\documentclass[draft,tightenlines,nofootinbib,preprint,aps,eqsecnum,amsmath,amssymb]{revtex4}

\newcommand{\beq}{\begin{equation}}
\newcommand{\eeq}{\end{equation}}
\newcommand{\bea}{\begin{eqnarray}}
\newcommand{\eea}{\end{eqnarray}}

\newcommand{\sgn}{\epsilon}

\begin{document}

\title{Relativistic Quantum Mechanics in the Rest-Frame Instant Form of Dynamics}

\author{David Alba}
\affiliation{Dipartimento di Fisica\\
Universita' di Firenze\\
Polo Scientifico, via Sansone 1\\
50019 Sesto Fiorentino, Italy\\
E-mail ALBA@FI.INFN.IT}
\author{Horace W. Crater}
\affiliation{The University of Tennessee Space Institute \\
Tullahoma, TN 37388 USA \\
E-mail: hcrater@utsi.edu}
\author{Luca Lusanna}
\affiliation{Sezione INFN di Firenze\\
Polo Scientifico\\
Via Sansone 1\\
50019 Sesto Fiorentino (FI), Italy\\
E-mail: lusanna@fi.infn.it}

\begin{abstract}

A new formulation of relativistic quantum mechanics is proposed in
the framework of the rest-frame instant form of dynamics with its
instantaneous Wigner 3-spaces and with its descrption of the
particle world-lines by means of derived non-canonical predictive
coordinates. In it we quantize the frozen Jacobi data of the
non-local 3-center of mass and the Wigner-covariant relative
variables in an abstract (frame-independent) internal space h1whose
existence is implied by Wigner-covariance. The formalism takes care
of the properties of both relativistic bound states and scattering
ones. There is a natural solution to the {\it relativistic
localization problem}. The non-relativistic limit leads to standard
quantum mechanics but with a frozen Hamilton-Jacobi description of
the center of mass. Due to the non-locality of the Poincare'
generators the resulting theory of relativistic entanglement is both
kinematically non-local and spatially non-separable, properties
absent in the non-relativistic limit.

\end{abstract}

\maketitle

\medskip

\today

\vfill\eject

\section{Introduction}

Atomic physics is an approximation to QED, in which the atoms are
described as non-relativistic particles in quantum mechanics (QM)
with a coupling to the electro-magnetic field of order  $1/c$
\cite{1,2,3}. For all the applications in which the energies
involved do not cross the threshold of pair production, this
description with a fixed number of particles is enough. Therefore
atomic physics and the theory of entanglement are formulated in the
absolute Euclidean 3-space and use Newton absolute time, namely they
are formulated in Galilei space-time. The main drawback is that, due
to the coupling to the electro-magnetic field there is not a
realization of the kinematical Galilei group  connecting
non-relativistic inertial frames. On the other hand, if we want to
arrive at an understanding of relativistic entanglement, we must
reformulate the theory in Minkowski space-time with a well defined
realization of the kinematical Poincare' group connecting
relativistic inertial frames. This would lead to {\it relativistic
atomic physics} as the quantization of a fixed number of classical
relativistic charged scalar (or spinning) particles interacting with
the classical electro-magnetic field.

\bigskip

In the papers in Refs.\cite{4,5,6,7} it was shown that it is
possible to describe any isolated relativistic system (particles,
fields, strings, fluids) admitting a Lagrangian formulation
(allowing one to define the energy-momentum tensor of the system) in
arbitrary non-inertial frames in Minkowski space-time by means of
{\it parametrized Minkowski theories}. The transition among
different non-inertial frames (with their different clock
synchronization conventions identifying the instantaneous, in
general non-Euclidean, 3-spaces) is performed by frame-preserving
diffeomorphisms, i.e. by suitable {\it gauge transformations}. As a
consequence the freedom in the choice of the clock synchronization
convention (needed to formulate a Cauchy problem for classical field
theories) becomes a choice of gauge. If we restrict ourselves to
inertial frames, the {\it inertial rest frame} is automatically
selected as the only one which can be identified in an intrinsic
geometric way: its instantaneous Euclidean 3-spaces are orthogonal
to the conserved 4-momentum of the isolated system.\medskip

This allows us to define the {\it rest-frame instant form of
dynamics for arbitrary isolated systems}: a complete exposition of
all its properties has been done in Ref.\cite{8} and extended to
non-inertial rest frames in Ref.\cite{9}. The study of relativistic
collective variables replacing the non-relativistic center of mass
leads to the description of the isolated system as a {\it decoupled
globally-defined non-local (and therefore un-observable)
non-covariant canonical external (Newton-Wigner) center of mass
carrying a pole-dipole structure} (the invariant mass $M$ and the
rest spin ${\vec {\bar S}}$ of the system) and an {\it external}
realization of the Poincare' group with generators $P^{\mu}$,
$J^{\mu\nu}$. $Mc$ and ${\vec {\bar S}}$ are the energy and angular
momentum of a {\it unfaithful internal} realization of the Poincare'
group, with generators $M\, c$, ${\vec {\cal P}}_{(int)}$, ${\vec
{\cal J}}_{(int)} = {\vec {\bar S}}$, ${\vec {\cal K}}_{(int)}$,
built with the energy-momentum tensor of the system and acting
inside the instantaneous Wigner 3-spaces where all the 3-vectors are
Wigner covariant. The vanishing of the internal 3-momentum and of
the internal Lorentz boosts eliminate the internal 3-center of mass
inside the Wigner 3-spaces, so that at the end the isolated system
is described only by {\it Wigner-covariant canonical internal
relative variables}, and imply that the Fokker-Pryce covariant
non-canonical 4-center of inertia has to be chosen as the inertial
observer origin of the 3-coordinates inside each Wigner 3-space. The
external 4-center of mass and the Fokker-Pryce 4-center of inertia
are parametrized in terms of canonical non-covariant frozen Jacobi
data $\vec z$, $\vec h$.\medskip

In particular, due to Refs.\cite{10,11}, it was shown in
Ref.\cite{8} that it is possible to define the inertial rest-frame
instant form of a semi-classical version of relativistic atomic
physics in which the electric charges of the positive-energy scalar
particles are Grassmann-valued (so that the Coulomb self-energies
are regularized) and in which the electro-magnetic potential is in
the radiation gauge (all the fields are transverse). Therefore the
isolated system is composed of {\it N positive-energy charged scalar
particles with mutual Coulomb interaction plus a transverse
electro-magnetic field}. The effect of this (both ultraviolet and
infrared) regularization is such that in the final Hamiltonian only
the potentials coming from the one-photon- exchange Feynman diagrams
appear, since all the radiative corrections and production diagrams
are eliminated. Therefore, our particles describe consistently the
semi-classical limit of a fixed- particle -number sector of some
matter QFT \footnote{The matter part of QED; at level of quark
sub-constituent it is the QFT of the standard model of particle
physics; for an atom of protons, neutrons and electrons it is an
effective QFT.}. Moreover the main features of the treatment of {\it
relativistic bound states} in the framework of QED are taken into
account, since the Darwin potential is emerging from the
Lienard-Wiechert solution \cite{10} \footnote{As shown in
Ref.\cite{11}, in this framework it is possible also to describe
positive-energy spinning particles (with Grassmann-valued spin) and
to identify the Salpeter potential instead of the Darwin one. The
Grassmann-valued 3-spins of the particles are all defined in the
same instantaneous 3-space, the Wigner hyper-planes, and therefore
transform as Wigner spin-1 3-vectors.  In Ref.\cite{12} the
positive-energy spinning particles are quantized in a special family
of non-inertial frames.}.

\medskip

The covariant world-lines of the particles are reconstructed in
terms of the covariant non-canonical external Fokker-Pryce center of
inertia, of the external 4-momentum and of the internal
Wigner-covariant relative variables: they are {\it covariant but not
canonical}, so that they correspond to the predictive coordinates of
predictive mechanics \cite{13}. Since they are not canonical, their
quantum version are operators which do not commute so that, in
general, the only covariant statements about them concern their
expectation values in given quantum states.
\medskip

In Ref.\cite{14} we showed how to determine a collective variable
associated with the internal 3-center of mass on the instantaneous
3-spaces, to be eliminated with the constraints ${\vec {\cal
K}}_{(int)} \approx 0$. Here ${\vec {\cal K}}_{(int)}$ is the
Lorentz boost generator in the unfaithful internal realization of
the Poincare' group and its vanishing is the gauge fixing to the
rest-frame conditions ${\vec {\cal P}}_{(int)} \approx 0$. We showed
how to find this collective variable for the following isolated
systems: a) charged particles with a Coulomb plus Darwin mutual
interaction; b) transverse radiation field; c) charged particles
with a mutual Coulomb interaction plus a transverse electro-magnetic
field.
\medskip

Moreover in Ref.\cite{8} it is shown that there is a canonical
transformation which allows one to describe the isolated system of
"N positive-energy charged scalar particles with mutual Coulomb
interaction plus a transverse electro-magnetic field" as a set  of
{\it N Coulomb-dressed charged particles interacting through a
Coulomb plus Darwin potential plus a free transverse radiation
field}: these two subsystems are not mutually interacting (the
internal Poincare' generators are a direct sum of the two
components) and are interconnected only by the rest-frame conditions
${\vec {\cal P}}_{(int)} \approx 0$ and the elimination of the
internal 3-center of mass with the gauge fixings ${\vec {\cal
K}}_{(int)} \approx 0$. Therefore in this framework with a fixed
number of particles there is a way out from the Haag theorem, at
least at the classical level.
\medskip

After the canonical transformation the two "non-interacting"
subsystems are only kinematically coupled by the rest-frame
conditions ${ \vec {\mathcal{P}}}_{(int)} \approx 0$ and ${\vec
{\mathcal{K}}}_{(int)} \approx 0$ (3 pairs of second class
constraints eliminating the spurious internal 3-center of mass):
only {\it internal relative variables} survive \footnote{Among them
${\hat {\vec \pi}}_{(12)3}$ is defined as a constant of motion,
describing the relative motion of the matter subsystem with respect
the radiation field subsystem.} with the exception of {\it two
collective variables} of the radiation field, i.e. the constant of
motion $p^{\tau}_{rad} = M_{rad}\, c$ (the energy of the radiation
field) and $X^{\tau}_{rad}\, {\buildrel \circ \over {=}}\, - \tau +
const.$ (an internal time discriminating the various symplectic
sub-manifolds of a surface of constant energy of the radiation
field).\medskip

The two papers of Refs\cite{8,14} will be quoted as I and II and
their formulas will be denoted (I.2.5) or (II.1.13).\medskip

As a consequence we now have a formalism which, {\it for the first
time}, takes into account all the known aspects of relativistic
kinematics and dynamics of point particles by means of 3+1
splittings of Minkowski space-time, parametrized Minkowski theories
and the rest-frame instant form of dynamics. One still open problem
is the possibility of defining a consistent {\it relativistic
statistical mechanics} by evaluating the relativistic
micro-canonical ensemble in the rest-frame instant form of dynamics.
Maybe other formulations are possible, but they have not yet been
developed.
\medskip

We refer to Subsection F of Section I of paper I for a review of the
other approaches to relativistic mechanics, in particular of those
with first-class constraints, which were the precursors of the
present formulation. However all these approaches suffered from some
problems. For instance it was too complicated to get a Lagrangian
description. See also the bibliography of the review part of
Ref.\cite{7}.

\medskip

Everyone of these approaches to relativistic mechanics tried to
perform the quantization and to define a consistent relativistic QM.
See Ref.\cite{15} and its bibliography for the attempt to quantiza
the two-particle models with two first-class constraints. However in
these models there was not a 3+1 splitting of Minkowski space-time.
Instead the problem of the instantaneous 3-space (a space-like
hyperplane) is present in the papers of Fleming in Ref.\cite{16}
(see also Refs.\cite{17,18}): however these papers did not succeeded
in giving an acceptable description of the comparison of the
dynamics on different space-like hyper-planes connected by Lorentz
transformations.

\bigskip

All the previous attempts to define relativistic QM employ the
so-called {\it zeroth postulate of QM} (see Zurek in Ref.\cite{19}).
According to it a composite system of two spatially separated
subsystems is described by the {\it tensor product} of the Hilbert
spaces of the subsystems. The notation ${\cal H}_t = ({\cal H}_1
\otimes {\cal H}_2)_t = ({\cal H}_{com} \otimes {\cal H}_{rel})_t$
means that the quantum 2-body isolated system can be imagined to be
constituted either by the two single particle subsystems with masses
$m_1$ and $m_2$ or as the tensor product of a decoupled
center-of-mass particle of mass $m = m_1 + m_2$ carrying an internal
space with an internal relative motion of reduced mass $\mu = m_1\,
m_2/m$. The second description is implied by the separation of
variables in the Schroedinger equation when the mutual interaction
respects the Galilei covariance of the isolated system. The two
descriptions are connected by a unitary transformation and
correspond to different choices of bases in ${\cal H}_t$
\footnote{Let us remark that in non-relativistic QM the Hilbert
space $\otimes_{i=1}^N\, {\cal H}_i$ for a N-body system could be
replaced with many other Hilbert spaces by means of unitary
transformations coming from the quantization of the canonical
transformations defining the possible canonical bases of Jacobi
coordinates for the N-particle system. This is possible because the
non-relativistic center of mass is a {\it local} notion not knowing
the whole Newton absolute 3-space: therefore we can consider the
center of mass of a 2-particle subsystem $(ab)$, then to couple it
with particle $(c)$ to get a 3-particle center of mass $((ab)c)$ and
so on. This is impossible in Minkowski space-time \cite{5,6} because
the non-relativistic notions of center of mass, reduced masses,
inertia tensors are not tensorial under the action of the Poincare'
group. The only possible non-relativistic global center of mass plus
relative variables which admit a relativistic extension are the {\it
canonical spin bases} of Ref.\cite{20} based on the coupling of the
angular momenta (and not of the centers of mass) of the subsystems.
The spin bases can be defined in Minkowski space-time \cite{6}, but
since the defining canonical transformations are {\it not point} it
is not yet clear whether they are unitarily implementable at the
quantum level. Moreover, since the canonical spin bases contain
angles, it is not clear whether they can be quantized.}.
\medskip

The zeroth postulate, i.e. ${\cal H}_t = ({\cal H}_1 \otimes {\cal
H}_2)_t$, is based on a notion of {\it separability} independent
from the Galilei group, which instead is at the basis of the
decomposition ${\cal H} = ({\cal H}_{com} \otimes {\cal H}_{rel})_t$
emphasizing that the center-of-mass momentum is a constant of motion
for an isolated system. This notion of separability goes back to
Einstein (see the EPR paper \cite{21} and Ref.\cite{22}): according
to him {\it proper separability} means that separate objects have
their independent real states, since for him it should be possible
to divide the world up into pieces about which statements can be
made (realism). The EPR argument leads to the statement that
non-relativistic QM is incomplete because realism and locality do
not coexist. Here {\it locality} means the real state of one system
remains unaffected by changes to a distant system (usually it is
said that it is locality which fails in orthodox QM with collapse,
even if in a benign way: it does not seem to make the testing of
predictions for isolated systems impossible, due to the presence of
the no-signalling theorem about the probabilities of the outcomes of
measurements; QM remains empirically testable despite violating
locality). The {\it no-signalling theorem} (ruling out the
possibility of signalling using entangled states) saves QM from
explicit non-locality conflicting with relativity. \bigskip

Given these notions and two subsystems A and B, we can introduce the
notions of a {\it separable pure state}  $| \Psi
>_{AB} = |\phi >_A \otimes | \psi >_B$ and of an {\it entangled non-separable state}
$| \Psi >_{AB} = \sum_i\, \sqrt{p_i}\, | {\bar \phi}_i >_A \otimes |
{\bar \psi}_i >_B$ ($\{ | {\bar \phi}_i \}$, $\{ | {\bar \psi}_i
> \}$ are  ortho-normal bases for subsystems A and B respectively
and $p_i$ are the non-zero eigenvalues of the reduced density matrix
of A). This is the starting point for the description of {\it
entanglement} in non-relativistic QM and for its foundational
problems connected to its probabilistic aspects and to its
non-locality (see Refs.\cite{23,24} for a review).
\medskip

Instead the attempts to define {\it relativistic entanglement} (see
for instance Ref.\cite{25}) usually start from quantum field theory
(QFT) and always use the notion of separability in the form of the
zeroth postulate. For a complete discussion of the state of the art
and for the open problems caused by Lorentz transformations and
massless particles see Refs. \cite{23,26}.\bigskip

The other source of problems in putting together QM and special
relativity is the notion of {\it localization}. This is connected
with the unusual properties of the non-covariant Newton-Wigner
operator \cite{27,28} and of its eigenvalues (absence of sharp
localization, an aspect of the non-locality present in special
relativity with self-adjoint position operators) and with the
connected problem of the instantaneous spreading of wave packets
(the Hegerfeldt theorem \cite{29,30}). As clearly shown in Ref.
\cite{31} in {\it local} QFT there is a notion of localization
deeply different from {\it Newton-Wigner localization}. Even if the
Reeh-Schlieder theorem \cite{32} says that the vacuum is
super-entangled (every state can be approximated with states
obtained from the vacuum by applying local operators defined in
bounded 4-regions (not 3-regions!) of space-time), the conclusion is
that the Newton-Wigner position operator cannot be described by
means of either local or quasi-local operators of algebraic QFT. The
accepted consequence is that this operators is {\it not measurable}.
Connected problems are the validity of micro-causality and the
relevance of un-sharp observables to try to define a theory of
measurement going beyond local QFT.

\bigskip

In this paper we propose a general scheme of quantization of
relativistic positive-energy scalar particles induced by the
rest-frame instant form of dynamics given by classical relativistic
mechanics \footnote{In  Refs.\cite{12} a first  attempt of
quantization of relativistic mechanics in inertial and non-inertial
frames (with the non-relativistic limit given in Ref.\cite{33}) was
done.}  and we will discuss in Section VI which of the quoted
problems are solved by our scheme.\medskip

The main result will be the non-validity of the zeroth postulate:
the quantization can be done only in a Hilbert space ${\cal
H}_{\tau}$ admitting the presentation ${\cal H}_{\tau} = \Big({\cal
H}_{com} \otimes {\cal H}_{rel}\Big)_{\tau}$ in each instantaneous
Wigner 3-space and in which the evolution is parametrized in terms
of the rest-frame time $\tau$. Actually after a 3+1 splitting of
Minkowski space-time it is not possible to define single-particle
Hilbert spaces ${\cal H}_1$, ${\cal H}_2$ ... : our basic operators
are the Jacobi data ${\hat {\vec z}}$, ${\hat {\vec h}}$ and
internal relative variables ${\hat {\vec \rho}}_a$, ${\hat {\vec
\pi}}_a$, $a=1,.., N-1$. The single particle (predictive) operators
${\hat x}^{\mu}_i$, ${\hat p}^{\mu}_i$, $i=1,..,N$, are derived
non-canonical quantities built in terms of the previous operators
and not independent variables like in the approaches considering the
tensor product $({\cal H}_1)_{x^o_1} \otimes ({\cal H}_2)_{x^o_2}
\otimes ...$ of free Klein-Gordon quantum particles. While ${\cal
H}_{\tau} = \Big({\cal H}_{com} \otimes {\cal H}_{rel}\Big)_{\tau}$
is the natural Hilbert space for the description of {\it
relativistic bound states} (and also of scattering states described
in terms of relative variables), in the tensor-product Hilbert space
$({\cal H}_1)_{x^o_1} \otimes ({\cal H}_2)_{x^o_2} \otimes ...$
there is no correlation among the times of the particles (their
clocks are not synchronized) so that in most of the states there are
some particles in the absolute future of the others. As a
consequence the two types of Hilbert spaces lead to inequivalent
descriptions.\medskip

Moreover the decoupled external non-covariant 4-center of mass is
{\it not measurable} (it is non-local in the sense of Newton-Wigner
localization) and evades Hegerfeldt's theorem being described by
frozen (non-evolving) Jacobi data.\medskip

In Section VI we will show what are the implications for
relativistic entanglement: since the dynamics is described by
relative variables in the Wigner 3-spaces, there is a {\it spatial
non-separability} and a {\it non-locality} of kinematical origin
besides the quantum non-locality.

\bigskip

The quantization scheme is defined initially for free particles and
then extended to particles with action-at-a-distance mutual
interactions. We will treat explicitly the two-body case and we will
show that there is no problem in the extension to N particles.
\medskip

In Subsection A of Section II we give a review of the rest-frame
instant form of dynamics for isolated systems of relativistic
positive-energy scalar particles living in the instantaneous Wigner
3-spaces. Then in Subsection B we study its non-relativistic limit.
Finally in Subsection C we define an abstract internal space of
relative variables independent of the orientation of the conserved
4-momentum of the isolated system: this is possible due to the
Wigner covariance of the relative variables.\medskip

In Subsection A of Section III we revisit the non-relativistic QM of
two particles, while in Subsection B we reformulate it in a form
suitable to be extended to the relativistic level ( which uses a
Hamilton-Jacobi description of the decoupled center of
mass).\medskip

In Section IV we introduce our quantization scheme. In Subsection A
we emphasize that we do not quantize  the non-covariant canonical
4-center of mass but only its frozen Jacobi data (it is a derived
quantity); instead for the particles we can either quantize  the
relative variables or the original variables ${\vec \eta}_i(\tau)$,
${\vec \kappa}_i(\tau)$ but with the supplementary requirements $<
{\hat {\vec {\cal P}}}_{(int)} > = < {\hat {\vec {\cal K}}}_{(int)}
> = 0$ (in both case the particle world-lines are derived
quantities). In Subsection B we describe the quantization of the
relative variables, while in Subsection C we delineate the
quantization before the elimination of the internal 3-center of
mass.\medskip

In Section V we give examples of quantization of two-particle
systems with action-at-a-distance interactions.
\medskip

In Section VI we show which problems connected to relativistic
localization are solved by our quantization scheme and its
implications for relativistic entanglement. The new  quantization
scheme contains a non-locality and a spatial non-separability
originating from the Lorentz signature of Minkowski space-time and
from the properties of the Poincare' group besides the standard
quantum non-locality. These new features disappear in the
non-relativistic limit due to the absolute nature of time and
3-space in Galileo space-time and due to the fact that Galilei
boosts are interaction independent.\medskip

In the Conclusions  we make some comments on the open problem of
quantizing the free transverse radiation field with the added
rest-frame requirements $< {\hat {\vec {\cal P}}}_{(int)}
> = < {\hat {\vec {\cal K}}}_{(int)} > = 0$.\medskip

In Appendix A we give the form of the Darwin potential in the
unequal mass case.

\medskip

In Appendix B there is the quantization of  two equal mass scalar
particles with mutual Coulomb plus Darwin interaction by means of
Weyl ordering.

\vfill\eject

\section{Review of the Rest-Frame Instant Form of Dynamics for
Relativistic Particles}

In this Section we review the rest-frame instant form of dynamics
for isolated systems developed in I using a two-particle system as
an example. $\eta_{\mu\nu} = \sgn\, (+---)$ is the flat metric
($\sgn = \pm 1$ according to either the particle physics $\sgn = 1$
or the general relativity $\sgn = - 1$ convention).

\subsection{The Rest-Frame Instant Form}

Let us consider an arbitrary inertial frame, centered on an inertial
observer whose world-line is the time axis, in Minkowski space-time.
If $P^{\mu} = Mc\, h^{\mu} =\, Mc\,  \Big(\sqrt{1 + {\vec h}^2};
\vec h\Big) $ ($\vec h = {{\vec v/c}\over {\sqrt{1 - (\vec
v/c)^2}}}= \vec P/Mc $ is an a-dimensional 3-velocity) is the
conserved total 4-momentum of the isolated particle system in this
inertial frame, the 3+1 splitting of Minkowski space-time associated
with the inertial rest-frame instant form description of the
isolated system has the instantaneous Wigner 3-spaces orthogonal to
$P^{\mu}$ (the 3-vectors inside them are Wigner spin-1 3-vectors;
the 3-metric inside the Euclidean Wigner 3-spaces is taken to be
positive definite, i.e. $\delta_{rs}$ with signature $(+++)$, so
that for the Wigner 3-vectors we have $V^r = V_r$). Their embedding
in Minkowski space-time is

\bea z_W^{\mu}(\tau ,\vec \sigma ) &=& Y^{\mu}(\tau ) +
\epsilon^{\mu}_r(\vec h)\, \sigma^r,\nonumber \\
&& h^{\mu} = \epsilon^{\mu}_{\tau}(\vec h) = (\sqrt{1 + {\vec h}^2};
\vec h),\qquad \epsilon^{\mu}_r(\vec h) = \Big( h_r; \delta^i_r +
{{h^i\, h_r}\over {1 + \sqrt{1 + {\vec h}^2}}}\Big), \label{2.1}
\eea

\medskip
\noindent where $Y^{\mu}(\tau ) = Y^{\mu}(0) + h^{\mu}\, \tau  =
z^{\mu}_W(\tau ,\vec 0)$ is the world-line of the external
Fokker-Pryce 4-center of inertia with $\eta_{\mu\nu}\,
\epsilon^{\mu}_A(\vec h)\, \epsilon^{\nu}_B(\vec h) =
\eta_{AB}$.\bigskip

In these rest frames there are only three notions of collective
variables, which can be built by using {\it only} the Poincare'
generators (they are {\it non-local} quantities knowing the whole
$\Sigma_{\tau}$): the canonical non-covariant Newton-Wigner 4-center
of mass (or center of spin) ${\tilde x}^{\mu}(\tau)$, the
non-canonical covariant Fokker-Pryce 4-center of inertia
$Y^{\mu}(\tau)$ and the non-canonical non-covariant M$\o$ller
4-center of energy $R^{\mu}(\tau)$. All of them tend to the
Newtonian center of mass in the non-relativistic limit.\medskip

As shown in I, these three variables can be expressed as known
functions of the Lorentz-scalar rest time $\tau = c\, T_s = h \cdot
\tilde x = h \cdot Y = h \cdot R$, of canonically conjugate Jacobi
data (frozen Cauchy data) $\vec z = Mc\, {\vec x}_{NW}(0)$ ($\{ z^i,
h^j\} = \delta^{ij}$; ${\vec x}_{NW}(\tau )$ is the standard
Newton-Wigner non-covariant 3-position, classical counterpart of the
corresponding position operator \cite{27}) and $\vec h = \vec P/Mc$,
of the invariant mass $Mc = \sqrt{\sgn\, P^2}$ of the system and of
its rest spin ${\vec {\bar S}}$:

\bigskip

1) the pseudo-world-line of the canonical non-covariant external
4-center of mass is

\bea {\tilde x}^{\mu}(\tau ) &=& \Big({\tilde x}^o(\tau ); {\tilde
{\vec x}}(\tau )\Big) = \Big(\sqrt{1 + {\vec h}^2}\, (\tau + {{\vec
h \cdot \vec z}\over {Mc}}); {{\vec z}\over {Mc}} + (\tau + {{\vec h
\cdot \vec z}\over {Mc}})\, \vec h\Big) =\nonumber \\
&=& z^{\mu}_W(\tau ,{\tilde {\vec \sigma}}) = Y^{\mu}(\tau ) +
\Big(0; {{- \vec S \times \vec h}\over {Mc\, (1 + \sqrt{1 + {\vec
h}^2})}}\Big), \label{2.2} \eea

\noindent so that we get $Y^{\mu}(0) = \Big(\sqrt{1 + {\vec h}^2}\,
{{\vec h \cdot \vec z}\over {Mc}}; {{\vec z}\over {Mc}} + {{\vec h
\cdot \vec z}\over {Mc}}\, \vec h + {{\vec S \times \vec h}\over
{Mc\, (1 + \sqrt{1 + {\vec h}^2})}}\Big)$ (we have used \cite{6}
${\tilde {\vec \sigma}} = {{- \vec S \times \vec h}\over {Mc\, (1 +
\sqrt{1 + {\vec h}^2})}}$);\medskip

2) the world-line of the non-canonical covariant external
Fokker-Pryce 4-center of inertia is

\bea Y^{\mu}(\tau ) &=& \Big({\tilde x}^o(\tau ); \vec Y(\tau )\Big)
= \Big(\sqrt{1 + {\vec h}^2}\, (\tau + {{\vec h \cdot \vec z}\over
{Mc}});  {{\vec z}\over {Mc}} + (\tau + {{\vec h \cdot \vec z}\over
{Mc}})\, \vec h + {{\vec S \times \vec h}\over {Mc\, (1 + \sqrt{1 +
{\vec h}^2})}} \Big) =\nonumber \\
&=& z_W^{\mu}(\tau ,\vec 0); \label{2.3} \eea
\medskip

3) the pseudo-world-line of the non-canonical non-covariant external
M$\o$ller 4-center of energy is

\bea R^{\mu}(\tau ) &=& \Big({\tilde x}^o(\tau ); \vec R(\tau )\Big)
= \Big(\sqrt{1 + {\vec h}^2}\, (\tau + {{\vec h \cdot \vec z}\over
{Mc}});\nonumber \\
&& {{\vec z}\over {Mc}} + (\tau + {{\vec h \cdot \vec z}\over
{Mc}})\, \vec h - {{ \vec S \times \vec h}\over {Mc\, \sqrt{1 +
{\vec h}^2}\, (1 + \sqrt{1 + {\vec
h}^2})}} \Big) =\nonumber \\
&=& z^{\mu}_W(\tau ,{\vec \sigma}_R) = Y^{\mu}(\tau ) + \Big(0;
{{-\, \vec S \times \vec h}\over {Mc\, \sqrt{1 + {\vec h}^2}}}\Big),
\label{2.4} \eea

\noindent (we have used \cite{6} ${\vec \sigma}_R = {{-\, \vec S
\times \vec h}\over {Mc\, \sqrt{1 + {\vec h}^2}}}$).\medskip

While $Y^{\mu}(\tau )$ is a 4-vector, ${\tilde x}^{\mu}(\tau )$ and
$R^{\mu}(\tau )$ are not 4-vectors. See Ref.\cite{5} for the
M$\o$ller non-covariance world-tube around the Fokker-Pryce 4-vector
identified by these collective variables. Their transformation
properties under Poincare' transformations $(a ,\Lambda )$ can be
deduced from those for $\vec h$, $\vec z$ and $\tau$ (see Appendix B
of Ref.\cite{15})

\bea h^{\mu}\, &\mapsto& h^{{'}\, \mu} =
\Lambda^{\mu}{}_{\nu}\, h^{\nu},\nonumber \\
z^i\, &\mapsto& z^{{'}\, i} = \Big(\Lambda^i{}_j -
{{\Lambda^i{}_{\mu}\, h^{\mu}}\over {\Lambda^o{}_{\nu}\, h^{\nu}}}\,
\lambda^o{}_j\Big)\, z^j + \Big(\Lambda^i{}_{\mu} -
{{\Lambda^i{}_{\nu}\, h^{\nu}}\over {\Lambda^o{}_{\rho}\,
h^{\rho}}}\, \Lambda^o{}_{\mu}\Big)\,
(\Lambda^{-1}\, a)^{\mu},\nonumber \\
\tau\, &\mapsto& \tau^{'} + h_{\mu}\, (\Lambda^{-1}\, a)^{\mu},
\nonumber \\
&&{}\nonumber \\
{\vec h}^{'} \cdot {\vec z}^{'} &=& \vec h \cdot \vec z +
{{\Lambda^o{}_j\, z^j}\over {\Lambda^o{}_{\mu}\, h^{\mu}}},\qquad
for\quad a^{\mu} = 0. \label{2.5} \eea

\bigskip

As said in I every isolated system (i.e. a closed universe) can be
visualized as a decoupled non-covariant collective (non-local)
pseudo-particle described by the frozen Jacobi data $\vec z$, $\vec
h$ carrying a {\it pole-dipole structure}, namely the invariant mass
$M\, c$ and the rest spin ${\vec {\bar S}}$ of the system, and with
an associated {\it external} realization  of the Poincare' group
(the last term in the Lorentz boosts induces the Wigner rotation of
the 3-vectors inside the Wigner 3-spaces):\medskip

\bea P^{\mu} &=& M\, c\, h^{\mu} = M\, c\, \Big(\sqrt{1 + {\vec
h}^2};
\vec h\Big),\nonumber \\
&&{}\nonumber \\
J^{ij} &=& z^i\, h^j - z^j\, h^i + \epsilon^{ijk}\, S^k,\qquad K^i =
J^{oi} = - \sqrt{1 + {\vec h}^2}\, z^i + {{(\vec S \times \vec
h)^i}\over {1 + \sqrt{1 + {\vec h}^2}}}, \label{2.6} \eea

\noindent satisfying the Poincare' algebra: $\{ P^{\mu}, P^{\nu} \}
= 0$, $\{ P^{\mu}, J^{\alpha\beta} \} = \eta^{\mu\alpha}\, P^{\beta}
- \eta^{\mu\beta}\, P^{\alpha}$, $\{ J^{\mu\nu}, J^{\alpha\beta} \}
= C^{\mu\nu\alpha\beta}_{\gamma\delta}\, J^{\gamma\delta}$,
$C^{\mu\nu\alpha\beta}_{\gamma\delta} = \delta^{\nu}_{\gamma}\,
\delta^{\alpha}_{\delta}\, \eta^{\mu\beta} + \delta^{\mu}_{\gamma}\,
\delta^{\beta}_{\delta}\, \eta^{\nu\alpha} - \delta^{\nu}_{\gamma}\,
\delta^{\beta}_{\delta}\, \eta^{\mu\alpha} - \delta^{\mu}_{\gamma}\,
\delta^{\alpha}_{\delta}\, \eta^{\nu\beta} $.

\medskip

The universal breaking of Lorentz covariance is connected to this
decoupled non-local collective variable and is irrelevant because
all the dynamics of the isolated system lives inside the Wigner
3-spaces and is Wigner-covariant. Inside these Wigner 3-spaces the
system is described by an internal 3-center of mass with a conjugate
3-momentum  and by relative variables and there is an {\it
unfaithful internal} realization of the Poincare' group: the
internal 3-momentum, conjugate to the internal 3-center of mass
\footnote{As shown in Ref.\cite{6} the three internal collective
3-variables (canonical ${\vec q}_+(\tau)$, Fokker-Pryce $\vec
y(\tau)$, M$\o$ller ${\vec R}_+(\tau)$) coincide due to the
rest-frame conditions: ${\vec q}_+ \approx \vec y \approx {\vec
R}_+$}, vanishes due the rest-frame condition. To avoid a double
counting of the center of mass, i.e. an external one and an internal
one, the internal (interaction-dependent) internal Lorentz boosts
must also vanish. As shown in I the only non-zero internal
generators are the invariant mass $M\, c$ and the rest spin ${\vec
{\bar S}}$ and the dynamics is re-expressed only in terms of
internal Wigner-covariant relative variables. Moreover this
construction implies that the time-like observer, at the origin of
the 3-coordinates on the Wigner 3-spaces, must be identified with
the Fokker-Pryce inertial observer as it was done in Eq.(\ref{2.1}).

\bigskip

As shown in Eq.(4.2) of the second paper of Ref.\cite{6}, given the
external realization (\ref{2.6}) of the Poincare' generators the
spatial part of the external M$\o$ller center of energy (\ref{2.4})
is given by $\vec R(0) = - \vec K/P^o$. In that paper it is also
shown that the Jacobi data $\vec z$ can be written in the form $\vec
z = Mc\, \vec R + {{Mc\, {\vec {\bar S}} \times \vec P}\over {P^o\,
(Mc + P^o)}}$, with ${\vec {\bar S}} = \vec J - \vec z \times {{\vec
P}\over {Mc}}$, and that this implies $\vec z = - {{P^o}\over {Mc}}
\vec K + {{\vec P \times \vec K}\over {P^o\, (Mc + P^o)}}\, \vec P +
{{\vec J \times \vec P}\over {Mc + P^o}}$. Eq.(\ref{2.2}) then
allows us  to express the external 4-center of mass ${\tilde
x}^{\mu}(\tau )$ in terms of the external Poincare' generators. The
same can be done for $Y^{\mu}(\tau)$ by using Eq.(\ref{2.3}).
Therefore the three collective variables of an isolated relativistic
system are {\it non-local} quantities like the Poincare'
generators.\medskip

As shown in I and in Ref.\cite{5}, in each Lorentz frame one has
different pseudo-world-lines describing $R^{\mu}$ and ${\tilde
x}^{\mu}$: the canonical 4-center of mass ${\tilde x}^{\mu}$ {\it
lies in between} $Y^{\mu}$ and $R^{\mu}$ in every (non rest)-frame.
As discussed in Subsection IIF of paper I, this leads to the
existence of the {\it M$\o$ller non-covariance world-tube}, around
the world-line $Y^{\mu}$ of the covariant non-canonical Fokker-Pryce
4-center of inertia $Y^{\mu}$. The {\it invariant radius} of the
tube is $\rho =\sqrt{- W^2}/p^2 = |\vec S|/\sqrt{P^2}$ where ($W^2 =
- P^2\, {\vec S}^2$ is the Pauli-Lubanski invariant when $P^2 > 0$).
This classical intrinsic radius delimitates the non-covariance
effects (the pseudo-world-lines) of the canonical 4-center of mass
${\tilde x}^{\mu}$ \footnote{In the rest-frame the world-tube is a
cylinder: in each instantaneous 3-space there is a disk of possible
positions of the canonical 3-center of mass orthogonal to the spin.
In the non-relativistic limit the radius $\rho$ of the disk tends to
zero and we recover the non-relativistic center of mass.}. They are
not detectable because the M$\o$ller radius is of the order of the
Compton wave-length: an attempt to test its interior would mean to
enter in the quantum regime of pair production \footnote{The
M$\o$ller radius of a field configuration (think to the radiation
field studied in Section III of paper II) could be a candidate for a
physical (configuration-dependent) ultraviolet cutoff in QFT
\cite{5}.}. Finally the M$\o$ller radius $\rho$ is also a remnant of
the energy conditions of general relativity in flat Minkowski
space-time \cite{5} and is the classical background of the violation
of the weak energy condition of the renormalized stress-energy
tensor in QFT (the Epstein, Glaser, Jaffe theorem \cite{34} ).

\bigskip

The world-lines of the positive-energy particles are parametrized by
Wigner 3-vectors ${\vec \eta}_i(\tau)$, $i = 1,2, ..,N$, and are
given by

\beq x^{\mu}_i(\tau) = z^{\mu}_W(\tau, {\vec \eta}_i(\tau)) =
Y^{\mu}(\tau) + \epsilon^{\mu}_r(\tau)\, \eta^r_i(\tau). \label{2.7}
\eeq

\medskip

For N free particles we have the following form of the internal
Poincare' generators (${\vec \kappa}_i(\tau)$ are the canonical
momenta conjugate to ${\vec \eta}_i(\tau)$, $\{ \eta^r_i(\tau),
\kappa^s_j(\tau)\} = \delta_{ij}\, \delta^{rs}$; the usual particle
4-momenta are the derived quantities $p^{\mu}_i = h^{\mu}\,
\sqrt{m_i^2\, c^2 + {\vec \kappa}^2_i} - \epsilon^{\mu}_r(\vec h)\,
\kappa_{ir}$ with $\sgn\, p^2_i = m_i^2\, c^2$)

\bea M\, c &=& {1\over c}\, {\cal E}_{(int)} = \sum_{i=1}^N\,
\sqrt{m_i^2\, c^2 + {\vec
\kappa}^2_i},\nonumber \\
{\vec {\cal P}}_{(int)} &=& \sum_{i=1}^N\, {\vec \kappa}_i \approx
0,\nonumber \\
\vec S &=& {\vec {\cal J}}_{(int)} = \sum_{i=1}^N\,
{\vec \eta}_i \times {\vec \kappa}_i,\nonumber \\
{\vec {\cal K}}_{(int)} &=& - \sum_{i=1}^N\, {\vec \eta}_i\,
\sqrt{m_i^2\, c^2 + {\vec \kappa}_i^2} \approx 0. \label{2.8} \eea

\bigskip

Instead of using the real internal 3-center-of-mass and relative
variables which can been obtained only with a non-linear non-point
canonical transformation as shown in the Appendix of the third paper
in Ref.\cite{6}, it is more convenient to use a naive linear point
canonical transformation. Therefore we will use the following
collective and relative variables which, written in terms of the
masses $m_i$ of the particles, make it easier to evaluate the
non-relativistic limit ($m = \sum_{i=1}^N\, m_i$)

\bea {\vec \eta}_+ &=& \sum_{i=1}^N\, {{m_i}\over m}\, {\vec
\eta}_i,\qquad {\vec \kappa}_+ = {\vec {\cal P}}_{(int)} =
\sum_{i=1}^N\, {\vec \kappa}_i,\nonumber \\
{\vec \rho}_a &=& \sqrt{N}\, \sum_{i=1}^N\, \gamma_{ai}\, {\vec
\eta}_i,\qquad {\vec \pi}_a = {1\over {\sqrt{N}}}\, \sum_{i=1}^N\,
\Gamma_{ai}\,
{\vec \kappa}_i,\qquad a = 1,..,N-1,\nonumber \\
&&{}\nonumber \\
{\vec \eta}_i &=& {\vec \eta}_+ + {1\over {\sqrt{N}}}\,
\sum_{a-1}^{N-1}\, \Gamma_{ai}\, {\vec \rho}_a,\qquad {\vec
\kappa}_i = {{m_i}\over m}\, {\vec \kappa}_+ + \sqrt{N}\,
\sum_{a=1}^{N-1}\, \gamma_{ai}\, {\vec \pi}_a, \label{2.9} \eea

\noindent with the following canonicity conditions
\footnote{Eqs.(\ref{2.9}) describe a family of canonical
transformations, because the $\gamma_{ai}$'s depend on
${\frac{1}{2}}(N-1)(N-2)$ free independent parameters.}

\bea &&\sum_{i=1}^{N}\, \gamma _{ai} = 0,\qquad  \sum_{i=1}^{N}\,
\gamma _{ai}\, \gamma _{bi} = \delta _{ab},\qquad \sum_{a=1}^{N-1}\,
\gamma _{ai}\, \gamma _{aj} = \delta _{ij} - {\frac{1}{N}},
\nonumber \\
&&{}\nonumber \\
&&\Gamma_{ai} = \gamma_{ai} - \sum_{k=1}^N\, {\frac{{m_k}}{ m}}\,
\gamma_{ak},\qquad \gamma_{ai} = \Gamma_{ai} -
{\frac{1}{ N}}\, \sum_{k=1}^N\, \Gamma_{ak},\nonumber \\
&&{}\nonumber \\
&&\sum_{i=1}^N\, {\frac{{m_i}}{m}}\, \Gamma_{ai} = 0,\qquad
\sum_{i=1}^N\, \gamma_{ai}\, \Gamma_{bi} = \delta_{ab},\qquad
\sum_{a=1}^{N-1}\, \gamma_{ai}\, \Gamma_{aj} =  \delta_{ij} -
{\frac{{m_i}}{ m}}.\nonumber \\
&&{} \label{2.10} \eea

\medskip

For $N=2$ we have $\gamma_{11} = - \gamma_{12} = {\frac{1}{
\sqrt{2}}}$, $\Gamma_{11} = \sqrt{2}\, {\frac{{m_2}}{ m}}$,
$\Gamma_{12} = - \sqrt{2}\, {\frac{{m_1}}{m}}$.\medskip

Therefore in the two-body case, by introducing the notation ${\vec
\eta}_{12} = {\vec \eta}_+$, ${\vec \kappa}_{12} = {\vec \kappa}_+ =
{\vec {\cal P}}_{(int)}$, we have the following collective and
relative variables

\bea {\vec \eta}_{12} &=& {\frac{{m_1}}{m}}\, {\vec \eta}_1 +
{\frac{{m_2}}{m}}\, {\vec\eta}_2,  \qquad
{\vec \rho}_{12} = {\vec \eta}_1 - {\vec \eta}_2,  \nonumber \\
{\vec \kappa}_{12} &=& {\vec \kappa}_1 + {\vec \kappa}_2 \approx 0,
\qquad {\vec \pi}_{12} =  {\frac{{m_2}}{m}}\, {\vec \kappa}_1 - {\
\frac{{m_1}}{m}}\, {\vec \kappa}_2,\nonumber \\
&&{}\nonumber \\
&&{}\nonumber \\
  {\vec \eta}_i &=& {\vec \eta}_{12} + (-)^{i+1}\, {\frac{{m_i}}{m}}\, {\vec \rho}_{12},
\qquad {\vec \kappa}_i = {\frac{{m_i}}{m}}\, {\vec \kappa}_{12} +
(-)^{i+1}\,  {\vec \pi}_{12},
  \label{2.11}
\eea

\noindent where we use the convention $m_3 \equiv m_1$.
\bigskip

The collective variable ${\vec \eta}_{12}(\tau )$ has to be
determined in terms of ${\vec \rho}_{12}(\tau )$ and ${\vec
\pi}_{12}(\tau )$ by means of the gauge fixings ${\vec
{\mathcal{K}}}_{(int)}\, {\buildrel {def}\over =}\, - M\, {\vec R}_+
\approx 0$. For two free particles Eqs.(\ref{2.8}) imply (${\vec
\eta}_{12}(\tau) \approx 0$ for $m_1 = m_2$)

\beq {\vec \eta}_{12}(\tau) \approx {{{{m_1}\over m}\, \sqrt{m_2^2\,
c^2 + {\vec \pi}^2_{12}(\tau)} - {{m_2}\over m}\, \sqrt{m_1^2\, c^2
+ {\vec \pi}^2_{12}(\tau)}}\over {\sqrt{m_1^2\, c^2 + {\vec
\pi}^2_{12}(\tau)} + \sqrt{m_2^2\, c^2 + {\vec
\pi}^2_{12}(\tau)}}}\, {\vec \rho}_{12}(\tau)\, \rightarrow_{c
\rightarrow \infty}\, 0. \label{2.12} \eeq

\bigskip

In the interacting case the rest-frame conditions ${\vec
\kappa}_{12} \approx 0$ and  the conditions eliminating the internal
3-center of mass ${\vec {\cal K}}_{(int)} \approx 0$ will determine
${\vec \eta}_{12}$ in terms of the relative variables ${\vec
\rho}_{12}$, ${\vec \pi}_{12}$ in an interaction-dependent way.
\medskip

Then the relative variables satisfy Hamilton equations with the
invariant mass $M({\vec \rho}_{12}, {\vec \pi}_{12})$ as Hamiltonian
and the particle world-lines $x^{\mu}_i(\tau )$ can be rebuilt
\cite{7}.
\medskip

The position of the two positive-energy particles in each
instantaneous Wigner 3-space is identified by the intersection of
the world-lines ($m_3 \equiv m_1$)

\bea
  x^{\mu}_i(\tau ) &=& Y^{\mu}(\tau ) + \epsilon^{\mu}_r(\vec h)\,
\eta^r_i(\tau ) \approx\, Y^{\mu}(\tau ) + \epsilon^{\mu}_r(\vec
h)\, \Big[\eta^r_{12}[{\vec \rho}_{12}(\tau ), {\vec \pi}_{12}(\tau
)] +
(-)^{i+1}\, {{m_{i+1}}\over m}\, \rho^r_{12}(\tau )\Big]\nonumber \\
&\approx_{free\, case}& Y^{\mu}(\tau ) + \epsilon^{\mu}_r(\vec h)\,
{{\sqrt{m_i^2\, c^2 + {\vec \pi}_{12}^2(\tau )}}\over {\sqrt{m_1^2\,
c^2 + {\vec \pi}_{12}^2(\tau )} + \sqrt{m_2^2\, c^2 + {\vec
\pi}_{12}^2(\tau )}}}\, \rho^r_{12}(\tau
),\nonumber \\
&&{}\nonumber \\
&& {\vec x}_i(\tau )\, \rightarrow_{c \rightarrow \infty}\, {\vec
x}_{(n)}(t) + (-)^{i+1}\, {{m_{i+1}}\over m}\, {\vec r}_{(n)}(t),\nonumber \\
&&{}\nonumber \\
&&{}\nonumber \\
p_i^{\mu}(\tau ) &=& h^{\mu}\, \sqrt{m_i^2\, c^2 + {\vec
\pi}_{12}^2(\tau )} + (-)^{i+1}\, \epsilon^{\mu}_r(\vec h)\,
\pi^r_{12}(\tau ),\qquad p^2_i(\tau ) = m_i^2\, c^2, \label{2.13}
\eea

\noindent with $Y^{\mu}(\tau)$ given in Eq.(\ref{2.3}) in terms of
$\vec z$, $\vec h$ and $\tau$. In the non-relativistic limit they
identify the Newton trajectories ${\vec x}_{(n) i}(t)$. The
covariant predictive world-lines $x^{\mu}_i(\tau)$ depend on the
relative position variables ${\vec \rho}_{12}$: a) if the
interaction among the particles is such that the relative position
variables have a compact support when $\tau$ varies (as happens with
the classical analogue of bound states) the world-lines will be
included in some finite time-like world-tube; b) instead, if the
interactions describe the classical analogue of scattering states,
the world-lines can diverge one from the other (cluster
decomposition property). This qualitative description has to be
checked in every system with a well defined action-at-a-distance
interaction.\medskip

They turn out to have a non-commutative (predictive) associated
structure since we have ($f^r_i = \eta^r_i({\vec \rho}_{12}(\tau ),
{\vec \pi}_{12}(\tau ))$)

\bea \{ x^{\mu}_i(\tau ), x^{\nu}_i(\tau ) \} &=& \{ Y^{\mu}(\tau ),
Y^{\nu}(\tau ) \} - \{ Y^{\mu}(\tau ), \epsilon^{\nu}_r(\vec h) \}\,
f^r_j + \{ Y^{\nu}(\tau ), \epsilon^{\mu}_r(\vec h) \}\, f^r_i +
\nonumber \\
&+& \epsilon^{\mu}_r(\vec h)\, \epsilon^{\nu}_s(\vec h)\, \{ f^r_i,
f^s_j \} \not= 0,\nonumber \\
&&{}\nonumber \\
&&\{ Y^o(\tau ), Y^i(\tau ) \} = - {{z^i\, \sqrt{1 + {\vec
h}^2}}\over {(Mc)^2}} + {{({\vec {\bar S}} \times \vec h)^i}\over
{(Mc)^2\, (1 +
\sqrt{1 + {\vec h}^2})}},\nonumber \\
&& \{ Y^i(\tau ), Y^j(\tau ) \} = {{\epsilon^{ijk}}\over {(Mc)^2}}\,
\Big[\Big(\vec z \times \vec h + {\vec {\bar S}}\Big)^k - {{h^k\,
\vec h \cdot {\vec {\bar S}}}\over
{(1 + \sqrt{1 + {\vec h}^2})^2}}\Big],\nonumber \\
&&\{ Y^o(\tau ), \epsilon^o_r(\vec h) \} = {{h^r\, \sqrt{1 + {\vec
h}^2}}\over {Mc}},\qquad \{ Y^i(\tau ), \epsilon^o_r(\vec h) \} =
{1\over {Mc}}\,
(\delta^{ir} + h^i\, h^r),\nonumber \\
&& \{ Y^o(\tau ), \epsilon^j_r(\vec h)
\} = {{h^j\, h^r}\over {Mc}},\nonumber \\
&&\{ Y^i(\tau ), \epsilon^j_r(\vec h) \} = {1\over {Mc\, (1 +
\sqrt{1 + {\vec h}^2})}}\, \Big(\delta^{ij}\, h^r + \delta^{ir}\,
h^j + {{2 + \sqrt{1 + {\vec h}^2}}\over {1 + \sqrt{1 + {\vec
h}^2}}}\, h^i\, h^j\, h^r\Big).
\nonumber \\
&&{} \label{2.14} \eea

\noindent Eqs.(\ref{2.3}) have been used to get these results

In the free case Eqs.(\ref{2.13}) imply $\{ f^r_i, f^j_s \} =
{{(m^2_1 - m^2_2)\, c^2\, (\sqrt{m_i^2\, c^2 + {\vec \pi}_{12}^2}\,
\pi_{12}^r\, \rho_{12}^s - \sqrt{m_j^2\, c^2 + {\vec \pi}_{12}^2}\,
\pi_{12}^s\, \rho_{12}^r)}\over {\sqrt{m_1^2\, c^2 + {\vec
\pi}_{12}^2}\, \sqrt{m_2^2\, c^2 + {\vec \pi}_{12}^2}\,
\sum_{k=1}^2\, \sqrt{m_k^2\, c^2 + {\vec \pi}_{12}^2}}}$.

\subsection{The Non-Relativistic Limit of the Rest-Frame Instant
Form}

Let us consider the non-relativistic limit of two positive-energy
scalar free particles, following I, where the kinematics is
described in Eq.(I-2.27) and the generators of the Galilei algebra
are given in Eq.(I-2.28).
\bigskip

The particles are described by the Newtonian canonical variables
${\vec x}_{(n)\, i}$, ${\vec p}_{(n)\, i}$, $i=1,2$, or by the
canonically equivalent center-of-mass and relative variables ${\vec
x}_{(n)}$, ${\vec p}_{(n)}$, ${\vec r}_{(n)}$, ${\vec q}_{(n)}$ (see
Ref.\cite{20} for the case of N particles)

\bea {\vec x}_{(n)} &=& {1\over m}\, \sum_{i=1}^2\, m_i\, {\vec
x}_{(n)\, i},\qquad {\vec p}_{(n)} = \sum_{i=1}^2\, {\vec p}_{(n)\,
i},\qquad m = m_1 + m_2,\nonumber \\
{\vec r}_{(n)} &=& {\vec x}_{(n)\, 1} - {\vec x}_{(n)\, 2},\qquad
{\vec q}_{(n)} = {1\over m}\, \Big(m_2\, {\vec p}_{(n)\, 1} - m_1\,
{\vec p}_{(n)2}\Big),\nonumber \\
&&{}\nonumber \\
{\vec x}_{(n)\, 1} &=& {\vec x}_{(n)} + {{m_2}\over m}\, {\vec
r}_{(n)},\qquad {\vec x}_{(n)\, 2} = {\vec x}_{(n)} - {{m_1}\over
m}\, {\vec r}_{(n)},\nonumber \\
{\vec p}_{(n)\, 1} &=& {{m_1}\over m}\, {\vec p}_{(n)} + {\vec
q}_{(n)},\qquad {\vec p}_{(n)\, 2} = {{m_2}\over m}\, {\vec p}_{(n)}
- {\vec q}_{(n)}. \label{2.15} \eea

The generators of the centrally extended Galilei algebra are (we
have changed the sign of the Galilei boosts with respect to
Refs.\cite{35})

\bea E_{Galilei} &=& \sum_{i=1}^2\, {{{\vec p}_{(n)\, i}^2}\over
{2m_i}} = {{{\vec p}^2_{(n)}}\over {2m}} + {{{\vec q}^2_{(n)}}\over
{2\mu}},\qquad {1\over
{\mu}} = {1\over {m_1}} + {1\over {m_2}},\nonumber \\
{\vec P}_{Galilei} &=& {\vec p}_{(n)} = \sum_{i=1}^2\, {\vec
p}_{(n)\,
i},\nonumber \\
{\vec J}_{Galilei} &=& \sum_{i=1}^2\, {\vec x}_{(n)\, i} \times
{\vec p}_{(n)\, i} = {\vec x}_{(n)} \times {\vec p}_{(n)} + {\vec
S}_{(n)},\qquad {\vec S}_{(n)} = {\vec r}_{(n)} \times {\vec
q}_{(n)},\nonumber \\
{\vec K}_{Galilei} &=& t\, {\vec p}_{(n)} - m\, {\vec
x}_{(n)},\nonumber \\
&&{}\nonumber \\
\{ E_{Galilei}, {\vec K}_{Galilei} \} &=& {\vec P}_{Galilei},\qquad
\{ P^i_{Galilei}, K^j_{Galilei} \} = m\, \delta^{ij},\qquad \{
K^i_{Galilei}, K^j_{Galilei} \} = 0,\nonumber \\
\{ A^i, J^j_{Galilei} \} &=& \epsilon^{ijk}\, A^k, \qquad \vec A =
{\vec P}_{Galilei}, {\vec J}_{Galilei}, {\vec K}_{Galilei}.
\label{2.16} \eea

The main property of the Galilei algebra is that the presence of
interactions changes the energy, $E_{Galilei}\, \rightarrow\,
E^{'}_{Galilei} = E_{Galilei} + V({\vec r}_{(n)})$ but not the
Galilei boosts \footnote{This is the reason why there is no
"No-Interaction Theorem" in Newtonian mechanics, so that Newtonian
kinematics is trivial. However, this theorem reappears when we make
a many-time reformulation of Newtonian mechanics \cite{36}.}.

\bigskip

Another property of the Galilei algebra, absent in the Poincare'
one, is that the energy generator is the sum of two distinct
constants of motion: the center-of-mass energy $E_{(n)com\,\, \vec
p} = {{{\vec p}^2}\over {2m}}$, $\vec p = {\vec p}_{(n)}$, and the
internal energy $\epsilon_{(n)} = {{{\vec q}^2_{(n)}}\over {2\mu}} +
V({\vec r}_{(n)})$ \footnote{Let us remark that this property is
preserved by the most general potential $V({\vec r}_{(n)}, {\vec
q}_{(n)}, E_{(n)com})$ admissible for an isolated two-particle
system.}. This justifies the separation of variables in the
Schroedinger equation. By comparison for two relativistic particles
we have $P^o = \sqrt{M^2\, c^2 + {\vec P}^2}$ with $Mc = \sum_i\,
\sqrt{m_i^2\, c^2 + {\vec \pi}^2_{12}} + V({\vec \rho}_{12})$ or $Mc
= \sum_i \sqrt{m_i^2\, c^2 + V({\vec \rho}_{12}) + {\vec
\pi}_{12}^2}$: $P^o$ is not a sum of two independent constants of
motion \cite{7}.

\bigskip

At the classical level the non-relativistic canonical transformation
separating the center of mass from the relative variables is {\it
point both} in the coordinate and in the momenta \footnote{  Its
relativistic version on the Wigner hyper-plane for the internal
motions is not point \cite{6,7} (in absence of interactions it is
point only in the momenta).}. The non-relativistic point canonical
transformation from the canonical basis ${\vec x}_{(n)i}$, ${\vec
p}_{(n)i}$, $i=1,2$, to the one ${\vec x}_{(n)} = \sum_i\,
{{m_1}\over m}\, {\vec x}_{(n)1} + {{m_2}\over m}\, {\vec
x}_{(n)2}$, ${\vec p}_{(n)} = {\vec p}_{(n)1} + {\vec p}_{(n)2}$,
${\vec r}_{(n)} = {\vec x}_{(n)1} - {\vec x}_{(n)2}$, ${\vec
q}_{(n)} = {{m_2}\over m}\, {\vec p}_{(n)2} - {{m_1}\over m}\, {\vec
p}_{(n)2}$ can be obtained from the sequence of the two following
canonical transformations connected with the identity  $e^{\{ ., S_2
\}}\, e^{\{ ., S_1 \}}$ with generating functions $S_1 = {{m_1}\over
m}\, {\vec x}_{(n) 1} \cdot {\vec p}_{(n) 2}$ and $S_2 = - {\vec
x}_{(n) 2} \cdot {\vec p}_{(n) 1}$ ($m = m_1 + m_2$)

\bea &&{\vec x}_{(n) 1}\, {\buildrel e^{\{ ., S_1 \}}\over
\rightarrow}\, {\vec x}_{(n) 1}\, {\buildrel e^{\{ ., S_2 \}}\over
\rightarrow}\, {\vec r}_{(n)} = {\vec x}_{(n) 1} - {\vec x}_{(n)
2},\nonumber \\
&&{\vec x}_{(n) 2}\, \rightarrow\, {\vec x}_{(n) 2} + {{m_1}\over
m}\, {\vec x}_{(n) 1}\, \rightarrow\, {\vec x}_{(n)} = {{m_1}\over
m}\, {\vec x}_{(n) 1} + {{m_2}\over m}\, {\vec x}_{(n) 2},
\nonumber \\
&&{\vec p}_{(n) 1}\, \rightarrow\, {\vec p}_{(n) 1} - {{m_1}\over
m}\, {\vec p}_{(n) 2}\, \rightarrow\, {\vec q}_{(n)} = {{m_2}\over
m}\, {\vec p}_{(n) 1} - {{m_1}\over m}\, {\vec p}_{(n) 2},\nonumber \\
&&{\vec p}_{(n) 2}\, \rightarrow\, {\vec p}_{(n) 2}\, \rightarrow\,
{\vec p}_{(n)} = {\vec p}_{(n) 1} + {\vec p}_{(n) 2}. \label{2.17}
\eea

\bigskip

Also at the non-relativistic level the 2-body system can be
presented as a decoupled particle, the external center of mass
${\vec x}_{(n)}(t)$ with momentum ${\vec p}_{(n)}$, of mass $m$ in
the absolute Euclidean 3-space carrying an internal space of
relative variables (${\vec r}_{(n)}(t)$, ${\vec q}_{(n)}(t)$) with
Hamiltonian $H_{rel} = {{{\vec q}_{(n)}^2}\over {2\, \mu}}$ and rest
spin ${\vec S}_{(n)}$.\medskip

The external center of mass  is associated with an external
realization of the Galilei group with generators $E_{Galilei} =
{{{\vec p}_{(n)}^2}\over {2m}} + H_{rel}$, ${\vec P}_{Galilei} =
{\vec p}_{(n)}$, ${\vec J}_{Galilei} = {\vec x}_{(n)} \times {\vec
p}_{(n)} + {\vec S}_{(n)}$, ${\vec K}_{Galilei} = t\, {\vec p}_{(n)}
- m\, {\vec x}_{(n)}(t)$.\medskip

The internal space can be identified with the rest frame (${\vec
p}_{(n)} \approx 0$) if we choose the origin of 3-coordinates in the
external center of mass (${\vec x}_{(n)}(t) \approx 0$): in it the
particles variables are ${\vec \eta}_{(n)i}(t) = {\vec
x}_{(n)i}(t){|}_{{\vec x}_{(n)} = {\vec p}_{(n)} = 0}$, ${\vec
\kappa}_{(n)i}(t) = {\vec p}_{(n)i}(t){|}_{{\vec x}_{(n)} = {\vec
p}_{(n)} = 0}$ (they are the non-relativistic counterpart of the
variables ${\vec \eta}_i(\tau )$, ${\vec \kappa}_i(\tau )$ on the
instantaneous Wigner 3-spaces). With this identification we get an
unfaithful internal realization of the Galilei group with generators
${\cal E}_{Galilei} = H_{rel}$, ${\vec {\cal P}}_{Galilei} = {\vec
p}_{(n)} \approx 0$ (the rest-frame conditions), ${\vec {\cal
J}}_{Galilei} = {\vec S}_{(n)}$, ${\vec {\cal K}}_{Galilei} = t\,
{\vec p}_{(n)} - m\, {\vec x}_{(n)}(t) \approx 0$ (the gauge fixings
to the rest-frame conditions implying ${\vec x}_{(n)}(t) \approx
0$). \medskip

Inside the internal space we have ${\vec x}_{(n)1} \approx {\vec
\eta}_{(n)1} = {{m_2}\over m}\, {\vec r}_{(n)}$, ${\vec x}_{(n)2}
\approx {\vec \eta}_{(n)2} = - {{m_1}\over m}\, {\vec r}_{(n)}$,
${\vec p}_{(n)1} \approx {\vec \kappa}_{(n)1} = {\vec q}_{(n)}$,
${\vec p}_{(n)2} \approx {\vec \kappa}_{(n)2} = - {\vec q}_{(n)}$
and we can introduce the following auxiliary variables (having an
obvious relativistic counterpart) ${\vec \rho}_{(n)12} = {\vec
\eta}_{(n)1} - {\vec \eta}_{(n)2} = {\vec r}_{(n)}$, ${\vec
\pi}_{(n)12} = {{m_2}\over m}\, {\vec \kappa}_{(n)1} - {{m_1}\over
m}\, {\vec \kappa}_{(n)2} = {\vec q}_{(n)}$, ${\vec \eta}_{(n)12} =
{{m_1}\over m}\, {\vec \eta}_{(n)1} + {{m_2}\over m}\, {\vec
\eta}_{(n)2} \approx 0$, ${\vec \kappa}_{(n)12} = {\vec
\kappa}_{(n)1} + {\vec \kappa}_{(n)2} \approx 0$.

\bigskip

In the relativistic rest-frame instant form the two-particle system
is described by\medskip

1) the external center-of-mass frozen Jacobi data $\vec z$, $\vec
h$, carrying the internal mass $M\, c = \sum_{i=1}^2\, \sqrt{m_i^2\,
c^2 + {\vec \kappa}_i^2}$ and the spin $\vec S = \sum_{i=1}^2\,
{\vec \eta}_i \times {\vec \kappa}_i$;\medskip

2) the two pairs of Wigner 3-vectors ${\vec \eta}_i$, ${\vec
\kappa}_i$, $i=1,2$, or by the canonically equivalent variables
(\ref{2.11}).

\bigskip

Since in the non-relativistic limit we have $\vec P = {\vec
p}_{(n)}$, $\vec h = {{\vec P}\over {M\, c}}\, \rightarrow_{c
\rightarrow \infty}\, 0$, implying $h^{\mu}\, \rightarrow_{c
\rightarrow \infty}\, \Big( 1; \vec 0\Big)$ and
$\epsilon^{\mu}_r(\vec h)\, \rightarrow_{c \rightarrow \infty}\,
\Big( 0; \delta^i_r\Big)$, it turns out that $\tau/c$, ${\tilde
x}^o/c$, $Y^o/c$, $R^o/c$  and $x^o_i/c$ all become the absolute
Newton time $t$.
\medskip

Moreover  we have the following results:\medskip

A) In the reference inertial system we get ${\tilde {\vec x}}(\tau
),\, \vec Y(\tau ),\, \vec R(\tau )\, \rightarrow_{c \rightarrow
\infty}\, {\vec x}_{(n)}(t)$, ${\vec x}_{NW} = {{\vec z}\over
{Mc}}\, \rightarrow_{c \rightarrow \infty}\, {\vec x}_{(n)}(0)$
because we have $\vec z\, = M\, c\, {\vec x}_{NW}(0)\,
\rightarrow_{c \rightarrow \infty}\, \infty$ and $\vec h \cdot \vec
z \, \rightarrow_{c \rightarrow \infty}\, {\vec p}_{(n)} \cdot
\Big({\vec x}_{(n)}(t) - {{{\vec p}_{(n)}}\over m}\, t\Big) = {\vec
p}_{(n)} \cdot {\vec x}_{(n)}(0)$ (it is a Jacobi data of the
non-relativistic theory).
\medskip

B) In the inertial rest frame, ${\vec p}_{(n)} \approx 0$, we get
${\vec \eta}_i(\tau )\, \rightarrow_{c \rightarrow \infty}\, {\vec
\eta}_{(n)i}(t)$, ${\vec \kappa}_i(\tau )\, \rightarrow_{c
\rightarrow \infty}\, {\vec \kappa}_{(n)i}(t)$, ${\vec x}_i(\tau )\,
\rightarrow_{c \rightarrow \infty}\, {\vec x}_{(n)}(t ) + {\vec
\eta}_{(n)i}(t )$, ${\vec p}_i(\tau )\, \rightarrow_{c \rightarrow
\infty}\, {\vec \kappa}_{(n)i}(t )$, $p_i^o\, \rightarrow_{c
\rightarrow \infty}\, m_i\, c + {{{\vec \kappa}_{(n)i}^2(t )}\over
{2m_i}}$.\bigskip

The  internal Poincare' generators (\ref{2.8}) have the limits
(modulo the rest mass $m\, c$ they are the internal Galilei
generators)

\bea M\, c &\rightarrow_{c \rightarrow \infty}& m\, c +
\sum_{i=1}^2\, {{{\vec \kappa}_{(n)i}^2}\over {2\, m_i}} \approx m\,
c + {{{\vec \pi}_{(n)12}^2}
\over {2\, \mu}} = m\, c + H_{rel},\nonumber \\
{\vec {\cal P}}_{(int)} &\rightarrow_{c \rightarrow \infty}\,&
{\vec \kappa}_{(n)12} \approx 0,\nonumber \\
\vec S &\rightarrow_{c \rightarrow \infty}& \sum_{i=1}^2\, {\vec
\eta}_{(n)i} \times {\vec \kappa}_{(n)i} \approx {\vec \rho}_{(n)12}
\times
{\vec \pi}_{(n)12} = {\vec S}_{(n)},\nonumber \\
{\vec {\cal K}}_{(int)} &\rightarrow_{c \rightarrow \infty}\,& -
\sum_{i=1}^2\, m_i\, {\vec \eta}_{(n)i} = - m\, {\vec \eta}_{(n)12}
\approx 0, \label{2.18} \eea

\noindent while the limits of the external Poincare' generators
(\ref{2.6}) are

\bea
\vec P &=& {\vec p}_{(n)} = {\vec P}_{Galilei},\nonumber \\
P^o &\rightarrow_{c \rightarrow \infty}& m\, c + {{{\vec
p}^2_{(n)}}\over {2m}} + \sum_{i=1}^2\, {{{\vec
\kappa}_{(n)i}^2}\over {2m_i}} \approx
  m\, c + {{{\vec p}^2_{(n)}}\over {2m}}
+ {{{\vec \pi}_{(n)12}^2}\over {2m_i}} = m\, c + E_{Galilei},\nonumber \\
\vec J &\rightarrow_{c \rightarrow \infty}& {\vec x}_{(n)} \times
{\vec p}_{(n)} + {\vec S}_{(n)} = {\vec J}_{Galilei},\nonumber \\
{\vec K}/c &\rightarrow_{c \rightarrow \infty}& t\, {\vec p}_{(n)} -
m\, {\vec x}_{(n)} = {\vec K}_{Galilei}. \label{2.19} \eea

\bigskip

Therefore the non-relativistic limit of the rest-frame instant form
leads to the following presentation of the Newton 2-body problem:

1) we have a decoupled external center of mass described by the
canonical variables ${\vec x}_{(n)}$, ${\vec p}_{(n)}$ and carrying
an internal space of relative variables coinciding with the
non-relativistic rest frame centered on the center of mass, ${\vec
p}_{(n)} \approx 0$ and ${\vec x}_{(n)}(t) \approx 0$ with the
Hamiltonian $H_{rel}$ and the rest spin ${\vec S}_{(n)}$;

2) in the internal space we have two pairs of variables ${\vec
\eta}_{(n)i}$, ${\vec \kappa}_{(n)i}$, or the canonically equivalent
${\vec \eta}_{(n)12} \approx 0$, ${\vec \kappa}_{(n)12} \approx 0$,
${\vec \rho}_{(n)12}$, ${\vec \pi}_{(n)12}$, and, as a consequence
from Eqs. (\ref{2.11}) and (\ref{2.13}) we have the following
identifications

\bea
  {\vec \rho}_{12}(\tau ) &=& {\vec \eta}_1(\tau ) - {\vec
  \eta}_2(\tau )\, \rightarrow_{c \rightarrow \infty}\, {\vec
  \rho}_{(n)12}(t) = {\vec \eta}_{(n)1}(t) - {\vec \eta}_{(n)2}(t) =
  {\vec r}_{(n)}(t),\nonumber \\
{\vec \pi}_{12}(\tau ) &=&  {{m_2}\over m}\, {\vec \kappa}_1(\tau )
-
  {{m_1}\over m}\, {\vec \kappa}_2(\tau)\, \rightarrow_{c \rightarrow \infty}\,
  {\vec \pi}_{(n)12}(t) = {{m_2}\over m}\, {\vec \kappa}_{(n)1}(\tau ) -
  {{m_1}\over m}\, {\vec \kappa}_{(n)2}(\tau) =  {\vec q}_{(n)}(t),\nonumber \\
  &&{}\nonumber \\
  &&\Downarrow\nonumber \\
  &&{}\nonumber \\
  {\vec x}_1(\tau ) &\rightarrow_{c \rightarrow \infty}& {\vec x}_{(n)}(t) + {\vec
  \eta}_{(n)1}(t) =  {\vec x}_{(n)1}(t),\nonumber \\
   {\vec x}_2(\tau ) &\rightarrow_{c \rightarrow \infty}& {\vec x}_{(n)}(t) + {\vec
  \eta}_{(n)2}(t ) {\vec x}_{(n)2}(t).
\label{2.20} \eea
\medskip

Let us remark that, while at the relativistic level the rest-frame
world-lines (\ref{2.8}) depend upon the 4-momentum $P^{\mu}$ of the
external 4-center of mass (because it identifies the instantaneous
Wigner 3-space in every inertial frame, being orthogonal to it), the
non-relativistic  trajectories ${\vec x}_{(n)\, i}(t)$ do not depend
upon ${\vec p}_{(n)}$, but only on ${\vec x}_{(n)}$ (the
non-relativistic definitions of center of mass and relative
variables do not mix coordinates and momenta).

\subsection{The Abstract Internal Space of Relative Variables}

In the fixed inertial frame chosen for the description of the
isolated two-body system, to each value of its constant 4-momentum
$P^{\mu} = Mc\, h^{\mu}$, i.e. to each value of the a-dimensional
3-velocity $\vec h = \vec v/c$, is associated a different rest-frame
3+1 splitting of Minkowski space-time, whose Wigner hyper-planes
$\Sigma_{\tau}^{(\vec h)}$ are orthogonal to the given $P^{\mu}$. In
the chosen inertial system the natural rest frame, with Wigner
3-spaces $\Sigma_{\tau}^{(\vec 0)}$, is associated with the
4-momentum ${\buildrel \circ \over P}^{\mu} = Mc\, (1; \vec 0)$,
i.e. to $\vec h = 0$. Let us denote ${\vec \rho}_{12}^{(\vec
h)}(\tau )$ and ${\vec \pi}_{12}^{(\vec h)}(\tau )$ the relative
variables living inside $\Sigma_{\tau}^{(\vec h)}$.

\bigskip

Since we have $P^{\mu} = Mc\, (\sqrt{1 + {\vec h}^2}; \vec h) =
L^{\mu}{}_{\nu}(P, {\buildrel \circ \over P})\, {\buildrel \circ
\over P}_{\nu} = Mc\, L^{\mu}{}_o(P, {\buildrel \circ \over P})$
\footnote{The standard Wigner boost for time-like Poincare' orbits
is $L^{\mu}{}_{\nu}(P, {\buildrel \circ \over P}) = \eta^{\mu}_{\nu}
- 2\, u^{\mu}(P)\, u_{\nu}({\buildrel \circ \over P}) -
{{[u^{\mu}(P) + u^{\mu}({\buildrel \circ \over P})]\, [u_{\nu}(P) +
u_{\nu}({\buildrel \circ \over P})]}\over {1 + u^o(P)}}$. We have
$L^{\mu}{}_{\nu}({\buildrel \circ \over P}, {\buildrel \circ \over
P}) = \eta^{\mu}_{\nu}$.}, we get $h^i = L^i{}_o(P, {\buildrel \circ
\over P})$. Therefore, since ${\vec \rho}_{12}^{(\vec h)}$ and
${\vec \pi}_{12}^{(\vec h)}$ are Wigner spin-1 3-vectors
transforming under Wigner rotations \footnote{To each Lorentz
transformation $\Lambda^{\mu}{}_{\nu}$ is associated the Wigner
rotation $R^{\mu}{}_{\nu}(\Lambda ,P) = [L({\buildrel \circ \over
P}, P)\, \Lambda^{-1}\, L(\Lambda P, {\buildrel \circ \over
P})]^{\mu}{}_{\nu}$, with $R^o{}_o(\Lambda ,P) = 1$,
$R^o{}_i(\Lambda ,P) = R^i{}_o(\Lambda ,P) = 0$, $R^i{}_j(\Lambda
,P) = (\Lambda^{-1})^i{}_j - {{(\Lambda^{-1})^i{}_o\, u_{\beta}(P)\,
(\Lambda^{-1})^{\beta}{}_j}\over {1 + u_{\rho}(P)\,
(\Lambda^{-1})^{\rho}{}_o}} - {{u^i(P)}\over {1 + u^o(P)}}\,
\Big[(\Lambda^{-1})^o{}_j - {{[(\Lambda^{-1})^o{}_o - 1]\,
u_{\beta}(P)\, (\Lambda^{-1})^{\beta}{}_j}\over {1 + u_{\rho}(P)\,
(\Lambda^{-1})^{\rho}{}_o}}\Big]$.}, we have that the 3-vectors
inside $\Sigma_{\tau}^{(\vec h)}$ can be obtained from those inside
$\Sigma_{\tau}^{(\vec 0)}$ by means of the Wigner rotation
$R^{\mu}{}_{\nu}(L(P, {\buildrel \circ \over P}), {\buildrel \circ
\over P}) = [L({\buildrel \circ \over P}, {\buildrel \circ \over
P})\, L^{-1}(P, {\buildrel \circ \over P})\, L(L(P, {\buildrel \circ
\over P})\, {\buildrel \circ \over P}, {\buildrel \circ \over
P})]^{\mu}{}_{\nu} = [L^{-1}(P, {\buildrel \circ \over P})\, L(P,
{\buildrel \circ \over P})]^{\mu}{}_{\nu} = \eta^{\mu}_{\nu}$
associated to the Wigner boosts $L(P, {\buildrel \circ \over P})$
sending $\vec h = 0$ into $\vec h$. As a consequence, we can make
the identifications

\bea {\vec \rho}_{12}^{(\vec h)}(\tau ) &=& {\vec \rho}_{12}^{(\vec
0)}(\tau ) \equiv {\vec \rho}_{12}(\tau ),\nonumber \\
{\vec \pi}_{12}^{(\vec h)}(\tau ) &=& {\vec \pi}_{12}^{(\vec
0)}(\tau ) \equiv {\vec \pi}_{12}(\tau ). \label{2.21} \eea

Therefore, there is an {\it abstract internal space of relative
variables}, living on an abstract Wigner 3-space $\Sigma_{\tau}\,
{\buildrel {def}\over =}\, \Sigma_{\tau}^{(\vec 0)}$, independent
from the rest-frame foliation, i.e. independent from $\vec h$. Both
the internal mass $M$ and the internal spin $\vec S$ depend only on
these abstract relative variables living in an abstract Wigner
3-space $\Sigma_{\tau}$: as a consequence there is a universal
pole-dipole structure carried by the external center of
mass.\bigskip

These identifications can be done also for the internal
3-center-of-mass variables ${\vec \eta}_{12} \approx {\vec
\eta}_{12}[{\vec \rho}_{12}, {\vec \pi}_{12}]$, ${\vec \kappa}_{12}
\approx 0$  before solving the Wigner-covariant constraints ${\vec
{\cal P}}_{(int)} \approx 0$, ${\vec {\cal K}}_{(int)} \approx 0$
and therefore also for the variables ${\vec \eta}_i(\tau)$, ${\vec
\kappa}_i(\tau)$.

\bigskip

This abstract internal relative space is carried by the external
3-center of mass, which is described by the Jacobi data $\vec z$,
$\vec h$  (the time-independent Cauchy data). While the Jacobi data
$\vec z$, $\vec h$, can be quantized independently from the
eigenvalues of the internal mass operator $\hat M\, c$, the
Newton-Wigner 3-position ${\vec x}_{NW} = \vec z / Mc$  and the
3-momentum $\vec P = Mc\, \vec h$, depend on these eigenvalues.
\bigskip

\vfill\eject

\section {Non-Relativistic Quantum Mechanics of Two Particles}

Let us review the standard QM description of a two-particle system
in Galilei space-time with the notation of I.

\subsection{Non-Relativistic Quantum Mechanics}

In non-relativistic QM  there is a Hilbert space ${\cal H}_t =
({\cal H}_1 \otimes {\cal H}_2)_t = ({\cal H}_{com} \otimes {\cal
H}_{rel})_t$ associated with each instant t of the absolute Newton
time: it contains wave functions $\psi_t({\vec x}_{(n)i})$ or
$\psi_t({\vec x}_{(n)}, {\vec r}_{(n)})$ depending upon the
3-coordinates of the particles in the absolute Euclidean 3-space.
The Galilei group acts in the Hilbert space ${\tilde {\cal H}} =
U_{t = -\infty}^{\infty}\,\, {\cal H}_t$ sending ${\cal H}_t
\rightarrow {\cal H}_t$ except for the time translations for which
we have ${\cal H}_t \rightarrow {\cal H}_{t + \delta\, t}$. However
all the Hilbert spaces ${\cal H}_t$ are isomorphic to an abstract
Hilbert space ${\cal H} = {\cal H}_1 \otimes {\cal H}_2 = {\cal
H}_{com} \otimes {\cal H}_{rel}$, so that the time evolution can be
described as a unitary transformation with parameter $t$   inside
${\cal H}$: in ${\cal H}$ we have the wave functions $\psi (t, {\vec
x}_{(n)i})$ or $\psi(t, {\vec x}_{(n)}, {\vec r}_{(n)})$ connected
by a unitary transformation.

\bigskip

By quantization of the sequence of the two canonical transformations
(\ref{2.17}) ($e^{ \{ ., S_i \} }\, \rightarrow\, e^{i\, {\hat
S}_i}$, ${\hat S}_i = {\hat S}_i^{\dagger}$) we get an explicit
unitary transformation connecting the description ${\cal H}_1
\otimes {\cal H}_2$ to the one ${\cal H}_{com} \otimes {\cal
H}_{rel}$. In ${\cal H}$ it corresponds to a {\it change of basis}:
it sends the position basis $\psi_1({\vec x}_{(n) 1})\, \psi_2({\vec
x}_{(n) 2})$ labeled by the eigenvalues of the maximal set ${\hat
{\vec x}}_{(n) 1}$, ${\hat {\vec x}}_{(n) 2}$ \footnote{${\hat A}$
denotes the operator corresponding to the classical variable $A$.}
of commuting operators to the position basis $\psi_{com}({\vec
x}_{(n)})\, \psi_{rel}({\vec r}_{(n)})$ labeled by the eigenvalues
of the maximal set ${\hat {\vec x}}_{(n)}$, ${\hat {\vec r}}_{(n)}$
of commuting operators.

\bigskip

When there is an interaction between the particles of an isolated
system, the separation of variables implies that the Schroedinger
equation is written in the coordinate representation associated with
the preferred basis in ${\cal H}_{com} \otimes {\cal H}_{rel}$ ($\vec L =
{\vec x}_{(n)} \times {\vec p}_{(n)} + \vec S$, $\vec S = {\vec
r}_{(n)} \times {\vec q}_{(n)}$)
\medskip

\begin{eqnarray*}
i\, {{\partial}\over {\partial\, t}}\, \psi_{(n)}(t, {\vec x}_{(n)},
{\vec r}_{(n)}) &=& \Big({{{\hat {\vec p}}_{(n)}^2}\over {2m}} +
{\hat H}_{rel}\Big)\, \psi_{(n)}(t, {\vec x}_{(n)}, {\vec
r}_{(n)}),\nonumber \\
&&{}\nonumber \\
&&{{{\hat {\vec p}}^2_{(n)}}\over {2m}}\, \psi_{(n)\vec p}({\vec
x}_{(n)}) = {{{\vec p}^2}\over {2m}}\, \psi_{(n)\vec p}({\vec
x}_{(n)}),\qquad \psi_{(n)\vec p}({\vec x}_{(n)}) = const. e^{i\,
{\vec p}\cdot {\vec
x}_{(n)}},\nonumber \\
&&{}\nonumber \\
&&{\hat H}_{rel}\, \phi_{(n) nlm}({\vec r}_{(n)}) = \epsilon_{(n)
n}\,
\phi_{(n) nlm}({\vec r}_{(n)}),\nonumber \\
&&{\hat {\vec S}}^2\, \phi_{(n) nlm}({\vec r}_{(n)}) = l\, (l +
1)\, \phi_{(n) nlm}({\vec r}_{(n)},\nonumber \\
&& {\hat S}_3\, \phi_{(n)
nlm}({\vec r}_{(n)}) = m\, \phi_{(n) nlm}({\vec r}_{(n)}),\nonumber \\
&&{}\nonumber \\
&&E_{(n) n \vec p} = {{{\vec p}^2}\over {2m}} + \epsilon_{(n) n} =
E_{(n) com\, \vec p} + \epsilon_{(n) n},
\end{eqnarray*}

\bea \Big({{{\hat {\vec p}}_{(n)}^2}\over {2m}} &+& {\hat
H}_{rel}\Big)\, \psi_{(n)}(t, {\vec x}_{(n)}, {\vec r}_{(n)}) =
E_{(n) n \vec p}\,
\psi_{(n) \vec p nlm}({\vec x}_{(n)}, {\vec r}_{(n)}),\nonumber \\
  &&{}\nonumber \\
\psi_{(n)\vec pnlm}(t, {\vec x}_{(n)}, {\vec r}_{(n)}) &=& e^{- i\,
E_{(n) n \vec p}\, t}\, \psi_{(n) nlm \vec p}({\vec x}_{(n)},
{\vec r}_{(n)}) =\nonumber \\
&=& e^{-i\, E_{(n) n \vec p}\, t}\, \psi_{(n) \vec p}({\vec
x}_{(n)})\, \phi_{(n) nlm}({\vec r}_{(n)}) =\nonumber \\
&=& \Big(e^{-i {{{\vec p}^2}\over {2m}}\, t}\, \psi_{(n) \vec
p}({\vec x}_{(n)})\Big)\, \Big(e^{-i\, \epsilon_{(n) n}\, t}\,
\phi_{(n) nlm}({\vec r}_{(n)})\Big),
\nonumber \\
&&{}\nonumber \\
&&i\, {{\partial}\over {\partial t}}\, \Big(e^{-i {{{\vec p}^2}\over
{2m}}\, t}\, \psi_{(n) \vec p}({\vec x}_{(n)})\Big) = {{{\hat {\vec
p}}^2_{(n)}}\over {2 m}}\, \Big(e^{-i {{{\vec p}^2}\over {2m}}\,
t}\, \psi_{(n) \vec p}({\vec x}_{(n)})\Big),\nonumber \\
&&i\, {{\partial}\over {\partial t}}\, \Big(e^{-i\, \epsilon_{(n)
n}\, t}\, \phi_{(n) nlm}({\vec r}_{(n)})\Big) = {\hat H}_{rel}\,
\Big(e^{-i\, \epsilon_{(n) n}\, t}\, \phi_{(n) nlm}({\vec
r}_{(n)})\Big). \label{3.1} \eea

\bigskip

Therefore the separation of variables implies that the Schroedinger
equation can be replaced by two separate Schroedinger equations, one
for the center of mass and one for the relative motion.

\bigskip

At the quantum level we have that the centrally extended Galilei
group is implemented with a projective realization. A Galilei boost
$\vec x \rightarrow \vec x - \vec v\, t$, $t \rightarrow t$ in
Galilei space-time (so that ${\vec x}_{(n)} \rightarrow {\vec
x}_{(n)} - \vec v\, t$ and ${\vec r}_{(n)} \rightarrow {\vec
r}_{(n)}$) is implemented as a projective unitary transformation:
$\psi (t, {\vec x}_{(n)}, {\vec r}_{(n)}) \rightarrow \, e^{- i\,
m\, \vec v \cdot {\vec x}_{(n)} + {i\over 2}\, m\, {\vec v}^2\, t}\,
\psi (t, {\vec x}_{(n)}, {\vec r}_{(n)}) = \psi^{'}(t, {\vec
x}_{(n)} - \vec v\, t, {\vec r}_{(n)})$.

\bigskip

Therefore, in the presence of mutual interactions, the bases of Hilbert
space corresponding to ${\cal H}_{1} \otimes {\cal H}_{2}$ is not
a natural one for isolated systems. Its use, for instance in the theory
of entanglement, is a realistic one only in the free case (see also footnote
42 in the conclusion).

\subsection{The Hamilton-Jacobi Description of the Center of Mass as
the Non-Relativistic Limit of the Rest-Frame Instant Form}

Since we want to make a comparison of the non-relativistic limit of
the rest-frame instant form of paper I with the standard
non-relativistic theory, let us define the quantum dynamics in
${\cal H} = {\cal H}_{com} \otimes {\cal H}_{rel}$ in the
representation arising after the transition to the Hamilton-Jacobi
form for the motion of the decoupled center of mass at the classical
level. Given the Hamiltonian $H = H_{com} + H_{rel}$, $H_{com} =
{{{\vec p}^2_{(n)}}\over {2m}}$, the transition to the
Hamilton-Jacobi description of the center of mass is usually done
with a time-dependent canonical transformation whose generating
function is the solution $\tilde S(t, {\vec x}_{(n)}(t), {\vec
p}_{(n)}(0)) = {\vec p}_{(n)} \cdot {\vec x}_{(n)}(t) - {{{\vec
p}_{(n)}^2}\over {2m}}\, t$ (${\vec p}_{(n)}$ is time-independent
being a constant of motion) of the Hamilton-Jacobi equation
$H_{com}({\vec x}_{(n)}(t), {{\partial\, \tilde S}\over {\partial\,
{\vec x}_{(n)}}}) + {{\partial\, \tilde S}\over {\partial t}} = 0$.
This canonical transformation can be implemented in the form $e^{\{
., S \}}$ if we choose the generating function $S = - {{{\vec
p}_{(n)}^2}\over {2m}}\, t = - H_{com}\, t$

\bea {\vec x}_{(n)}(t) &{\buildrel e^{\{ ., S \}}\over \rightarrow}&
{\vec x}_{(n)}(t) - {{{\vec p}_{(n)}}\over m}\, t = {\vec x}_{(n)}(0),\nonumber \\
{\vec p}_{(n)} &\rightarrow& {\vec p}_{(n)},\nonumber \\
&&{}\nonumber \\
H_{com} = {{{\vec p}^2_{(n)}}\over {2m}} &\rightarrow& e^{\{ ., S
\}}\, H_{com} + {{\partial S}\over {\partial t}} = 0. \label{3.2}
\eea
\medskip

Therefore ${\vec x}_{(n)}(0)$ and ${\vec p}_{(n)}$ are a set of {\it
Jacobi data} for the decoupled non-relativistic center of mass. As
already said, the classical isolated system is represented as a
decoupled {\it frozen} point particle, the center of mass ${\vec
x}_{(n)}(0)$ with conjugate momentum ${\vec p}_{(n)}$, carrying an
internal space spanned by the relative variables ${\vec
r}_{(n)}(t)$, ${\vec q}_{(n)}(t)$, with total angular momentum
(spin) ${\vec S}_{(n)} = {\vec r}_{(n)} \times {\vec q}_{(n)}$ and
Hamiltonian $H_{rel}$: namely the center of mass carries some kind
of pole-dipole structure (with the mass of the monopole replaced
with $H_{rel}$). {\it This internal space can be identified with the
rest frame description of the isolated system}, i.e. with the
inertial frame where ${\vec p}_{(n)} \approx 0$ (the rest-frame
conditions) and where the center of mass is chosen as the origin
(gauge fixings to the rest-frame conditions) ${\vec x}_{(n)}(0)
\approx 0$ as shown after Eqs.(\ref{2.16})

\bigskip

At the quantum level the associated unitary transformation to get
the center-of-mass Hamilton-Jacobi description is $e^{ i\, {{{\hat
{\vec p}}_{(n)}^2}\over {2m}}\, t}$ (it eliminates the evolution of
the center of mass): it sends the center-of-mass wave functions
$\psi_{(n)}(t, {\vec x}_{(n)}) = e^{- i\, {{{\vec p}^2}\over {2m}}\,
t}\, \psi_{(n) \vec p}({\vec x}_{(n)})$ for plane waves into the
{\it frozen} wave functions $\psi_{(n) \vec p}({\vec x}_{(n)}(0)) =
const. e^{i\, \vec p \cdot {\vec x}_{(n)}(0)}$ with the
identification ${\vec x}_{(n)} = {\vec x}_{(n)}(0)$. \medskip

In this basis the Schroedinger equation of Eqs.(\ref{3.1}) becomes

\bea i\, {{\partial}\over {\partial t}}\, {\tilde \psi}_{(n)}({\vec
x}_{(n)}(0)| t, {\vec r}_{(n)}) &=& {\hat H}_{rel}\,
{\tilde \psi}_{(n)}({\vec x}_{(n)}(0)| t, {\vec r}_{(n)}),\nonumber \\
&&{}\nonumber \\
&&{\tilde \psi}_{(n)}({\vec x}_{(n)}(0)| t, {\vec r}_{(n)}) =
\psi_{(n) \vec p}({\vec x}_{(n)}(0))\, \phi_{(n)}(t, {\vec
r}_{(n)}). \label{3.3} \eea
\medskip

If we go to the momentum basis for the frozen center of mass, we get
that the wave functions $\psi_{{\vec k}_{(n)}}({\vec p}_{(n)} | t,
{\vec r}_{(n)}) = \psi_{{\vec k}_{(n)}}({\vec p}_{(n)})\,
\phi_{(n)}(t, {\vec r}_{(n)})$ of a basis for ${\cal H}_{com}
\otimes {\cal H}_{rel}$ satisfy the following form of the last two
equations in Eqs.(\ref{3.1})

\bea {\hat {\vec p}}_{(n)}\, \psi_{{\vec k}_{(n)}}({\vec p}_{(n)} |
t, {\vec r}_{(n)}) = {\vec k}_{(n)}\, \psi_{{\vec k}_{(n)}}({\vec
p}_{(n)} | t, {\vec r}_{(n)}),\qquad or\,\, \psi_{{\vec
k}_{(n)}}({\vec p}_{(n)}) = \delta^3({\vec p}_{(n)} - {\vec k}_{(n)}),\nonumber \\
i\, {{\partial}\over {\partial\, t}}\, \psi_{{\vec k}_{(n)}}({\vec
p}_{(n)} | t, {\vec r}_{(n)}) = {\hat H}_{rel}\, \psi_{{\vec
k}_{(n)}}({\vec p}_{(n)} | t, {\vec r}_{(n)}). \label{3.4} \eea

\vfill\eject

\section{Quantization of Relativistic Particles in the Rest-Frame Instant Form
of Dynamics}

\subsection{Quantization}

Let us now study the quantization of the isolated 2-body problem in
the rest-frame instant form.\bigskip

We have to quantize the frozen Jacobi data $\vec z$, $\vec h$, of
the external 3-center of mass in the {\it preferred momentum basis}
$\vec h$ or $\vec P = Mc\, \vec h$, needed to define the foliations
and the abstract internal relative space, and the relative variables
${\vec \rho}_{12}$, ${\vec \pi}_{12}$ of the decoupled internal
space, whose evolution in the rest time $\tau = c\, T_s$ is governed
by the internal mass $M$, i.e. the energy of the internal Poincare'
group acting in the abstract Wigner 3-space $\Sigma_{\tau}$
\footnote{As shown in Refs.\cite{5,6,7}, before the reduction of the
external 4-center of mass ${\tilde x}^{\mu}$, $P^{\mu}$ to the
external 3-center of mass $\vec z$, $\vec h$, the variables $\tau =
c\, T_s = h \cdot \tilde x$ and $\epsilon_s = \sqrt{\sgn\, P^2}
\approx Mc$ are canonically conjugate variables. This is just the
same situation like with the Galilei energy $E$ and Newton time $t$
in ordinary quantum mechanics, where to get the time-dependent
Schroedinger equation 1) one sends $E$ into the operator $i\,
{{\partial}\over {\partial t}}$; 2) one uses $E = H$ to write $i\,
{{\partial}\over {\partial t}}\, \psi = \hat H\, \psi$ with the
Hamiltonian operator $\hat H$ depending on the canonically conjugate
particle variables.}.\medskip

The external canonical non-covariant 4-center of mass ${\tilde
x}^{\mu}(\tau )$ (the function of $\vec z/Mc = {\vec x}_{NW}$, $\vec
h$ and $\tau$ given in Eq.(\ref{2.2})) and its conjugate momentum
$P^{\mu} = Mc\, (\sqrt{1 + {\vec h}^2}; \vec h)$ are {\it derived
quantities}. The evolution in $\tau$ governed by $M$ will be shown
to imply  an evolution of the external 4-center of mass ${\tilde
x}^{\mu}(\tau )$ in terms of the time variable ${\tilde x}^o$:
consistently this evolution is governed by the Hamiltonian $P^o =
\sqrt{M^2\, c^2 + {\vec P}^2} = Mc\, \sqrt{1 + {\vec h}^2}$ (the
energy of the external Poincare' group) due to the relation $\tau =
c\, T_s = h \cdot \tilde x$ \footnote{We will get a positive-energy
Klein-Gordon equation for each eigenvalue $M_n$ of the internal mass
(like for a scalar positive-energy particle of mass $M_n$).}. Also
the covariant non-canonical world-lines $x^{\mu}_i(\tau ) =
z^{\mu}_W(\tau ,{\vec \eta}_i(\tau ))$ of the particles are {\it
derived quantities}, which becomes non-commuting operators, as
implied by Eqs.(\ref{2.14}), depending on the Jacobi data of the
external center of mass and on the relative variables.

\bigskip

For more complicated systems, like the ones of the standard
semi-relativistic atomic physics (see papers I and II), for which we
do not know how to solve the rest-frame conditions ${\vec
{\mathcal{P}}}_{(int)} \approx 0$, ${\vec { \mathcal{K}}}_{(int)}
\approx 0$, we must define a more general quantization scheme for
the internal space including also the conjugate variables describing
the internal 3-center of mass (${\vec \eta}_{12}$ and ${\vec
\kappa}_{12} \approx 0$ in the two-body case). Namely we have to
quantize the Jacobi data for the external center of mass and the
redundant variables ${\vec \eta}_i(\tau )$, ${\vec \kappa}_i(\tau )$
and then impose the quantum version of the second class constraints
${\vec {\mathcal{P}}}_{(int)} = {\vec \kappa}_{12} \approx 0$,
${\vec { \mathcal{K}}}_{(int)} \approx 0$ with some prescription
(the Gupta-Bleuler one if possible). Differently from the
non-relativistic case, where ${\vec { \mathcal{K}}}_{(int)} \approx
0$ becomes ${\vec \eta}_{12}(\tau ) \approx 0$, this is a
non-trivial task.

\bigskip

The extension of the previous quantization to isolated N-body
systems is automatic if we know the explicit form of the generators
of the internal Poincare' algebra and we use the relative variables
of Eqs.(II-2.1). As said in I, this requires the knowledge of the
energy-momentum tensor of the N-body system. This is known
explicitly only for a limited number of systems \cite{7,8}.

\subsection{Quantization after the Elimination of ${\vec
\eta}_{12}$}

\subsubsection{The Hilbert Space}

To quantize  we must consider a Hilbert space ${\cal H} = {\cal
H}_{com} \otimes {\cal H}_{rel}$ with the following constraints
implying the use of wave functions $\Psi (\vec h| \tau , {\vec
\rho}_{12})$ (i.e. in the center-of-mass momentum representation and
in the coordinate representation for the relative
variables):\bigskip

1) ${\cal H}_{com}$ is the Hilbert space of  a positive energy
frozen 3-center-of-mass particle described by the quantum Jacobi
data ${\hat {\vec z}}$, ${\hat {\vec h}}$. We must use the preferred
$\vec h$-basis in the momentum representation because it is needed
for the kinematical definition of the rest frame, i.e. of the Wigner
3-space. Consistently with the frozen nature of the external
3-center of mass, instead of an evolution equation we have

\bea &&{\hat {\vec h}}\, \Psi_{\vec k}(\vec h| \tau , {\vec
\rho}_{12}) = \vec k\,
\Psi_{\vec k} (\vec h| \tau , {\vec \rho}_{12}),\nonumber \\
&&{}\nonumber \\
&&or\nonumber \\
&&{}\nonumber \\
&&\Psi_{\vec k} (\vec h| \tau , {\vec \rho}_{12}) = \delta^3(\vec h
- \vec k)\, \phi (\tau ,{\vec \rho}_{12})\, {\buildrel {def}\over
=}\, \psi_{\vec k}(\vec h)\, \phi(\tau ,{\vec \rho}_{12}).
\label{4.1} \eea
\medskip

Let us assume that ${\hat {\vec z}}$ (and therefore also ${\hat
{\vec x}}_{NW} = {\hat {\vec z}}/\hat M$) is a self-adjoint
operator. Since this is a problematic assumption (see Section VI),
the preferred $\vec h$-basis  is also useful to avoid facing these
problems at this level. With ${\hat {\vec z}}$ self-adjoint we can
go to the $\vec z$-representation and use plane waves as elementary
solutions for the external 3-center of mass: $\Psi_{\vec k}(\vec z|
\tau, {\vec \rho}_{12}) = const. e^{i\, \vec k \cdot \vec z}\, \phi
(\tau ,{\vec \rho}_{12}) = \psi_{\vec k}(\vec z)\, \phi (\tau ,{\vec
\rho}_{12})$.
\bigskip

Let us remark that, as shown in Ref.\cite{12}, in the momentum
representation we have $z^i\, \rightarrow\, i\, \hbar\,
{{\partial}\over {\partial\, h^i}} - i\, \hbar\, {{h^i}\over {1 +
{\vec h}^2}}$ and the time-independent scalar product in this frozen
Hilbert space has the form

\beq < \Psi_1, \Psi_2>\, = \int {{d^3h}\over {2\, \sqrt{1 + {\vec
h}^2}}}\, \psi_1^*(\vec h)\, \psi_2(\vec h). \label{4.2} \eeq

\noindent It is a Lorentz scalar.

\bigskip

2) ${\cal H}_{rel}$ is the abstract internal rest-frame Hilbert
space, corresponding to the abstract internal relative space on the
abstract Wigner 3-space $\Sigma_{\tau}$, for the relative motions.
Its scalar product is

\beq < \phi_1, \phi_2 > = \int d^3\rho_{12}\, \phi_1^*(\tau ,{\vec
\rho}_{12})\,\, \phi_2 (\tau ,{\vec \rho}_{12}). \label{4.3} \eeq

\noindent It is conserved in the time $\tau$ and Lorentz scalar
\cite{12,15} \footnote{Strictly speaking the internal Poincare'
groups, acts in ${\tilde {\cal H}}_{rel} = U_{\tau = -
\infty}^{\infty}\,\, {\cal H}_{rel\, \tau}$.} with $\phi \in
L^2(R^3)$.\bigskip

At the classical level we have that the external Poincare' group has
the generators given in Eq.(\ref{2.6}) with  $M = M({\vec
\rho}_{12}, {\vec \pi}_{12})$ and ${\vec {\bar S}} = {\vec
\rho}_{12} \times {\vec \pi}_{12}$. In ${\cal H} = {\cal H}_{com}
\otimes {\cal H}_{rel}$ in the $\vec h$-basis they can be realized
as the following Hermitean operators

\bea &&{\hat P}^o = {\hat M}\, c\, \sqrt{1 + {\vec h}^2},\qquad
{\hat {\vec
P}} = {\hat M}\, c\, \vec h,\nonumber \\
&&{\hat {\vec J}} = {\hat {\vec z}} \times \vec h + {\hat {\vec
{\bar S}}},\nonumber \\
  &&{\hat {\vec K}} = - {1\over 2}\, \Big({\hat {\vec z}}\, \sqrt{1
  + {\vec h}^2} + \sqrt{1 + {\vec h}^2}\, {\hat {\vec z}}\Big) +
  {{{\hat {\vec {\bar S}}} \times \vec h}\over {1 + \sqrt{1 +
  {\vec h}^2}}},
\label{4.4} \eea

\noindent with $\hat M$ a self-adjoint suitably ordered operator
depending upon ${\hat {\vec \rho}}_{12}$, ${\hat {\vec \pi}}_{12}$
and with ${\hat {\vec {\bar S}}} = {\hat {\vec \rho}}_{12} \times
{\hat {\vec \pi}}_{12}$ ($[ \hat M, {\hat {\vec {\bar S}}}] = 0$,
$[{\hat {\bar S}}^r, {\hat {\bar S}}^s] = i\hbar \epsilon^{rsu}\,
{\hat {\bar S}}^u$). They satisfy the Poincare' algebra: $[{\hat
P}^{\mu}, {\hat P}^{\nu} ] = 0$, $[{\hat P}^o, {\hat {\vec J}}] =
0$, $[{\hat P}^o, {\hat {\vec K}}] = - i\hbar {\hat {\vec P}}$,
$[{\hat P}^i, {\hat J}^j] = i\hbar\, \epsilon^{ijk}\, {\hat P}^k$,
$[{\hat K}^i, {\hat J}^j] = i\hbar\, \epsilon^{ijk}\, {\hat K}^k$,
$[{\hat P}^i, {\hat K}^j] = i\hbar\, \delta^{ij}\, {\hat P}^o$,
$[{\hat K}^i, {\hat K}^j] = - i\hbar\, \epsilon^{ijk}\, {\hat J}^k$.
Therefore, as in Refs.\cite{15}, \cite{12}, there is a unitary
realization of the external Poincare' group.

\subsubsection{The Relativistic Schroedinger Equation}

As shown in Section III, in non-relativistic QM the Schroedinger
equation can be split in two separate Schroedinger equations for the
center of mass and the relative motion in the non-relativistic
Hilbert space ${\cal H}_{com} \otimes {\cal H}_{rel}$ due to the
special property of Galilei energy. In the center-of-mass
Hamilton-Jacobi description of the 2-body system these two equations
are given in Eqs.(\ref{3.4}).\medskip

Also, in the rest-frame instant form of dynamics the quantum
description of an isolated relativistic system is split in two
parts: \medskip

A) A non-evolving 3-center of mass described by frozen Jacobi data,
so that there is no relativistic Schroedinger equation as a
counterpart of the first equation in Eqs.(\ref{3.4});\medskip

B) An internal space of relative motions with a $\tau$-evolution
governed by the invariant mass $M$ (the energy generator of the
internal Poincare' group). This will lead to a Schroedinger equation
for the $\tau$-evolution of the internal motion, which is the
relativistic counterpart of the second of Eqs.(\ref{3.4}). The
eigenvalues $M_n$ of the invariant mass are determined by the
associated stationary Schroedinger equation, which will take also
into account the internal spin.

\bigskip

For the $\tau$-evolution of the internal motion inside the Wigner
3-space we have the following Schroedinger equations in ${\cal
H}_{com} \otimes {\cal H}_{rel}$ in the preferred $\vec h$-basis
with frozen 3-center-of-mass wave function $\psi_{\vec k}(\vec h)$,
[$\Psi_{\vec k} (\vec h | \tau, {\vec \rho}_{12}) = \psi_{\vec
k}(\vec h)\, \phi (\tau ,{\vec \rho}_{12})$]

\bea {\hat {\vec h}}\, \Psi_{\vec k} (\vec h | \tau, {\vec
\rho}_{12}) &=& \vec k\, \Psi_{\vec k} (\vec
h | \tau, {\vec \rho}_{12}),\nonumber \\
i\, {{\partial}\over {\partial\, \tau}}\, \Psi_{\vec k} (\vec h |
\tau, {\vec \rho}_{12}) &=&  {\hat M}({\vec \rho}_{12}, {\hat {\vec
\pi}}_{12})
\, c\,\,\, \Psi_{\vec k} (\vec h | \tau,  {\vec \rho}_{12}),\nonumber \\
&&{}\nonumber \\
&&\Downarrow\nonumber \\
&&{}\nonumber \\
  \psi_{\vec k}(\vec h)&& \Big[\Big(i\, {{\partial}\over
{\partial\, \tau}} - {\hat M}({\vec \rho}_{12}, {\hat {\vec
\pi}}_{12}) \, c\,\Big)\, \phi (\tau ,{\vec \rho}_{12})\Big] = 0,
\label{4.5} \eea

\noindent where $\hat M({\vec \rho}_{12}, {\hat {\vec \pi}}_{12})$
is the operator defined by the quantization of classical models with
either $M({\vec \rho}_{12}, {\vec \pi}_{12})\, c = \sum_{i=1}^2\,
\sqrt{m_i^2\, c^2 + V_1({\vec \rho}_{12}) +{\vec \pi}_{12}^2}$ or
$M({\vec \rho}_{12}, {\vec \pi}_{12})\, c = \sum_{i=1}^2\,
\sqrt{m_i^2\, c^2 + {\vec \pi}_{12}^2} + V_2({\vec \rho}_{12})$
($V_1$ or $V_2$ are a-a-a-d potentials) \footnote{In Ref.\cite{7}
there is the evaluation of the internal Poincare' generators for the
case in which the arbitrary potential $V_1({\vec \rho}_{12}^2)$ is
under the square root. For the more relevant case in which the
potential $V_2({\vec \rho}_{12}^2)$ is outside the square root the
form of the internal Lorentz boosts is not known except for the
Coulomb plus Darwin potential $V_2({\vec \rho}_{12}, {\vec
\pi}_{12})$ of II (in this case the knowledge of the energy-momentum
tensor of the system allows the determination). As shown in
Ref.\cite{7} and in II, they induce potential-dependent terms in the
internal Lorentz boosts, so that the solution ${\vec \eta}_{12} =
{\vec \eta}_{12}[{\vec \rho}_{12}, {\vec \pi}_{12}]$ of the
conditions ${\vec {\cal K}}_{(int)} \approx 0$ eliminating the
internal 3-center of mass are potential dependent.}.
\medskip

Let us put $\phi (\tau ,{\vec \rho}_{12}) = e^{- i\,
{{\epsilon}\over c}\, \tau}\, \phi ({\vec \rho}_{12})$. If we can
find the solutions $\phi_{nlm}({\vec \rho}_{12})$  of the stationary
equation in ${\cal H}_{rel}$

\bea {\hat M}\, c^2\, \phi_{nlm}({\vec \rho}_{12}) &=& \epsilon_n\,
\phi_{nlm}({\vec \rho}_{12}),\nonumber \\
{\hat {\vec S}}^2\, \phi_{nlm}({\vec \rho}_{12}) &=& l\, (l + 1)\,
\phi_{nlm}({\vec \rho}_{12}),\nonumber \\
{\hat S}_3\, \phi_{nlm}({\vec \rho}_{12}) &=& m\, \phi_{nlm}({\vec
\rho}_{12}), \label{4.6} \eea

\noindent then we have the elementary solutions ($P^{\mu}_n = \Big(
\epsilon_n\, \sqrt{1  + {\vec k}^2};  \epsilon_n\, c\, \vec k\Big)$,
$M_n\, c^2 = \epsilon_n\, {\buildrel {def}\over =}\, m\, c^2 +
{\tilde \epsilon}_n$ with ${\tilde \epsilon}_n\, \rightarrow_{c
\rightarrow \infty}\, \epsilon_{(n)n}$ \footnote{They are the
non-relativistic energy levels of the the relative Hamiltonian
${\hat H}_{rel}$ resulting from the non-relativistic limit of $\hat
M$ in  Eq.(\ref{4.5}). Let us remark that different relativistic
theories (potential either inside or outside the square roots) can
have the same non-relativistic potential as a limit.})

\beq \Psi_{ \vec k, nlm}(\vec h | \tau , {\vec \rho}_{12}) =
\delta^3(\vec h - \vec k)\, (2 \pi)^{-3/2}\, e^{ - i\, M_n\, c\,
\tau}\, \psi_{nlm}({\vec \rho}_{12}), \label{4.7} \eeq

\bigskip

In the $\vec z$-basis we have $ \Psi_{ \vec k, nlm}(\vec z | \tau ,
{\vec \rho}_{12}) = const. e^{i\, \vec k \cdot \vec z}\, (2
\pi)^{-3/2}\, e^{ - i\, M_n\, c\, \tau}\, \phi_{nlm}({\vec
\rho}_{12})$.\medskip

The wave packets for the internal motion are $\Psi_{ \vec k}(\vec z
| \tau , {\vec \rho}_{12}) = const. e^{i\, \vec k \cdot \vec z}\,
\sum_{nlm}\, F_{nlm}\, (2 \pi)^{-3/2}\, e^{ - i\, M_n\, c\, \tau}\,
\phi_{nlm}({\vec \rho}_{12})$.
\medskip

The wave packets also on the external 3-center of mass are $\Psi
(\vec z | \tau , {\vec \rho}_{12}) = \int {{d^3k}\over {2\, \sqrt{1
+ {\vec k}^2}}}\, G(\vec k)\, \Psi_{ \vec k}(\vec z | \tau , {\vec
\rho}_{12})$. These last wave packets correspond to superpositions
of different 3+1 rest-frame splittings. See Section VI for a
discussion on the self-adjointness of $\vec z$ and the status of
these wave packets.\medskip

If we can find a complete set of solutions of Eqs.(\ref{4.6}), then
the generic solutions of Eqs.(\ref{4.5}) will be the most general
square-integrable superposition of center-of-mass plane waves and
elementary solutions for the relative motion.
\bigskip

Let us remark that the non-relativistic limit of Eq.(\ref{4.7}) is

\bea && e^{i\, \vec k \cdot \vec z}\, e^{ - i\, M_n\, c\, \tau}\,
\phi_{nlm}({\vec \rho}_{12})\, \rightarrow_{c \rightarrow \infty}\,
e^{i\, M_nc\, \vec k \cdot {\vec x}_{(n)}(0)}\, e^{- i\, (mc^2 +
\epsilon_{(n)n})\, t}\, \phi_{(n)nlm}({\vec
r}_{(n)}) =\nonumber \\
&&\qquad e^{i\, {\vec p}_{(n)} \cdot {\vec x}_{(n)}(0)}\, e^{- i
mc^2\, t}\, \Big( non-relativistic\,
relative\, motion\, elementary\, solution\Big).\nonumber \\
&&{} \label{4.8} \eea

By comparison with Eq.(\ref{3.1}) we see that it corresponds to a
reformulation of non-relativistic quantum mechanics in a framework
in which the non-relativistic center of mass is described in terms
of the frozen Jacobi data $\vec k = {\vec p}_{(n)}/M_nc$ and $\vec z
= M_nc\, {\vec x}_{(n)}(0)$.

\subsubsection{The External 4-Center of Mass}

To recover the motion of the external 4-center of mass, carrying the
pole-dipole structure with mass $M$ and spin $\vec S$, we have to
replace the frozen $M$-independent plane wave $e^{i \vec k \cdot
\vec z}$, living in ${\cal H}_{com}$, with a wave function
$\psi_{M_n} ({\tilde x}^o, \vec P)$ knowing the levels $M_n$ of the
quantum invariant mass $\hat M$ (the internal wave function
$\phi_{nlm}({\vec \rho}_{12})$ in ${\cal H}_{rel}$ takes care of the
spin $\vec S$).\medskip

Its ${\tilde x}^o$- evolution is governed by the external Poincare'
energy $P^o = \sqrt{M_n^2\, c^2 + {\vec P}^2}$ corresponding to the
level. Therefore  we have to introduce as many new auxiliary Hilbert
spaces ${\cal H}_{ext com\,\, n}$ as mass levels $M_n$. For the
${\tilde x}^o$-evolution we have as Schroedinger equation the
positive-energy Klein-Gordon equation \footnote{It is obtained from
the Klein-Gordon equation by means of the Feshbach-Villars
transformation \cite{27,37}. With both signs of energy the scalar
product is the same as for a Klein-Gordon scalar particle of mass
$M_n$.}

\beq i \, {{\partial}\over {\partial \, {\tilde x}^o}}\,
\psi_{M_n}({\tilde x}^o, \vec P) = \sqrt{M^2_n\, c^2 + {\vec P}^2}\,
\psi_{M_n} ({\tilde x}^o, \vec P), \label{4.9} \eeq
\medskip

This is equivalent to undoing the Hamilton-Jacobi transformation on
the external center of mass independently for each level of the
internal motion: the non-relativistic limit of Eqs.(\ref{4.9}) is
the first equation in Eqs.(\ref{3.1}), because we have $M_n\,
\rightarrow_{c \rightarrow \infty}\, m + O(c^{-1})$, $\sqrt{M^2_n\,
c^2 + {\vec P}^2}\, \rightarrow_{c \rightarrow \infty}\, m c^2 +
{{{\vec p}^2_{(n)}}\over {2m}}$. The irrelevant phase factor $e^{-
i\, m\, c^2}$ has to be omitted.

\bigskip
If we take into account both positive- and negative-energies for the
external 4-center of mass, we have the Klein-Gordon equation  in the
{\it preferred} momentum basis

\beq \Big({\hat P}^2 - M^2_n\, c^2\,\Big) \,\psi_{M_n}(P^{\mu}) = 0,
\label{4.10} \eeq

\noindent whose solutions are ($\eta = sign\, P^o$)

\bea \psi_{M_n} (P^{\mu}) &=& const. e^{- i\, P\cdot \tilde x} =
const. e^{-i\,
M_n\, c\, \tau} = \nonumber \\
&=& const. e^{- i \Big(\eta\, \sqrt{M^2_n\, c^2 + {\vec P}^2}\,
{\tilde x}^o - \vec P \cdot {\tilde {\vec x}}\Big)} \,
\rightarrow_{c \rightarrow \infty,\,\, \eta = 1}\, const.\, e^{- i\,
\Big((mc^2 + {{{\vec p}^2_{(n)}}\over {2m}})\, t - {\vec p}_{(n)}
\cdot {\vec x}_{(n)}\Big)} =\nonumber \\
&=& e^{- i mc^2\, t}\, \Big(non-relativistic\, center-of-mass\,
plane\, wave\Big). \label{4.11} \eea

Consistently $\psi_{M_n} (P^{\mu})$ coincides with the piece
$e^{-i\, M_{nlm}\, c\, \tau}$ of Eq.(\ref{4.7}) due to the relation
$\tau = c\, T_s = h \cdot \tilde x$. Therefore, the auxiliary
Hilbert spaces ${\cal H}_{ext com\,\, n}$ are a byproduct of this
relation: due to it every elementary solution (\ref{4.7}) in ${\cal
H}_{com} \otimes {\cal H}_{rel}$ with fixed $\vec h$ and $M_n$ and
with the non-relativistic limit (\ref{4.8}) contains a phase
describing also a plane wave for the external 4-center of mass in
${\cal H}_{ext com\,\, n}$ as shown in Eq.(\ref{4.11})
\footnote{This is the relativistic description, which should be used
for the motion of an atom in atom interferometry instead of the
effective Schroedinger equation of Ref.\cite{38}, obtained by
extracting the positive-energy part of relativistic first-quantized
wave equations like Klein-Gordon, Dirac or Proca, whose second
quantization is assumed to describe an effective QFT for spin 0,
${\frac{1}{2}}$ or 1 (two-level) atoms.}.

\medskip
In the preferred momentum representation the plane wave solution is
$(2 \pi)^{-3/2}\, \delta^3(\vec P - \eta\, {{M_{nlm}}\over c}\, \vec
h)\, \delta (P^o - \eta\, {{M_n}\over c}\,  \sqrt{1 + {\vec h}^2})$.

\subsection{Quantization with ${\vec \eta}_{12}$}

If we cannot solve the rest-frame conditions and the conditions for
the elimination of the internal 3-center of mass, we must start with
an unphysical internal Hilbert space ${\cal H}_{{\vec \eta}_1}
\otimes {\cal H}_{{\vec \eta}_2} = {\cal H}_{{\vec \eta}_{12}}
\otimes {\cal H}_{{\vec \rho}_{12}}$ (its formal separability is
unphysical) with a unphysical scalar product, write Eq.(\ref{4.5})
in ${\cal H}_{com} \otimes {\cal H}_{{\vec \eta}_1} \otimes {\cal
H}_{{\vec \eta}_2}$ and then impose the 3 pairs of second class
constraints ${\vec {\cal P}}_{(int)} = {\vec \kappa}_{12} \approx
0$, ${\vec {\cal K}}_{(int)} \approx 0$ as restrictions on the
states. Therefore, besides Eq.(\ref{4.5}) with $\hat M$ function of
${\hat {\vec \eta}}_i$ and ${\hat {\vec \kappa}}_i$, $i=1,2$ there
will be the 6 equations

\beq < \Phi_{phys} |\, {\hat {\vec {\mathcal{P}}}}_{(int)} |\,
\Phi_{phys} >\,\, =\,\,  < \Phi_{phys} |\, { \hat {\vec
{\mathcal{K}}}}_{(int)} |\, \Phi_{phys} > = 0, \label{4.12} \eeq

\noindent which should lead to the identification of the physical
Hilbert space ${\cal H}_{rel}$ and of its physical scalar product.
But the second set of conditions (\ref{4.12}) are
interaction-dependent, so that the quantization is non-trivial and
could be unitarily inequivalent to the one of the previous
Subsection. An open problem is whether Eqs.(\ref{4.12}) can be
replaced by conditions of the type ${\hat {\vec {\cal A}}}\, |\,
\Phi_{phys} > = 0$ and $0 =  < \Phi_{phys} |\, {\hat {\vec {\cal
A}}}^{\dagger}$ corresponding to a generalized Gupta-Bleuer-like
approach.

\bigskip

In this case, besides writing the quantum external Poincare' algebra
with $\hat M$ and ${\hat {\vec S}}$ depending on the operators
${\hat {\vec \eta}}_i$ and ${\hat {\vec \kappa}}_i$, one should
check also the validity of the quantum internal Poincare' algebra by
using a suitable ordering.

\bigskip

Let us consider the case of two free particles as an example. From
Eqs.(\ref{2.8}) and (\ref{2.11}) we have the following two forms of
the internal Poincare' generators with the Poincare' algebra
trivially satisfied ($m_3 \equiv m_1$)

\bea M\, c &=& \sum_{i=1}^2\, \sqrt{m_i^2\, c^2 + {\vec \kappa}_i^2}
= \sum_{i=1}^2\, \sqrt{m_i^2\, c^2 + \Big({{m_i}\over m}\,
{\vec \kappa}_{12} - (-)^i\, {\vec \pi}_{12}\Big)^2} \approx\nonumber \\
&\approx& \sum_{i=1}^2\, \sqrt{m_i^2\, c^2 + {\vec \pi}_{12}^2},\nonumber \\
{\vec {\cal P}}_{(int)} &=& \sum_{i=1}^2\, {\vec \kappa}_i = {\vec
\kappa}_{12} \approx 0,\nonumber \\
{\vec {\cal J}}_{(int)} &=& \sum_{i=1}^2\, {\vec \eta}_i \times
{\vec \kappa}_i = {\vec \eta}_{12} \times {\vec \kappa}_{12} + {\vec
S} \approx {\vec S} = {\vec \rho}_{12} \times {\vec
\pi}_{12},\nonumber \\
{\vec {\cal K}}_{(int)} &=& - \sum_{i=1}^2\, {\vec \eta}_i\,
\sqrt{m_i^2\, c^2 + {\vec \kappa}_i^2} = - {\vec \eta}_{12}\,
\sum_{i=1}^2\, \sqrt{m_i^2\, c^2 + \Big({{m_i}\over m}\,
{\vec \kappa}_{12} - (-)^i\, {\vec \pi}_{12}\Big)^2} +\nonumber \\
&+& {\vec \rho}_{12}\, \sum_{i=1}^2\, (-)^i\, {{m_{i+1}}\over m}\,
\sqrt{m_i^2\, c^2 + \Big({{m_i}\over m}\, {\vec \kappa}_{12} -
(-)^i\, {\vec \pi}_{12}\Big)^2}
\approx\nonumber \\
&\approx& - {\vec \eta}_{12}\, \sum_{i=1}^2\, \sqrt{m_i^2\, c^2 +
{\vec \pi}^2_{12}} + {\vec \rho}_{12}\, \sum_{i=1}^2\, (-)^i\,
{{m_{i+1}}\over m}\, \sqrt{m_i^2\, c^2 + {\vec \pi}_{12}^2} \approx
0.\nonumber \\
&&{} \label{4.13} \eea
\medskip

The rest-frame conditions imply ${\vec \kappa}_{12} \approx 0$,
Eq.(\ref{2.12}) for ${\vec \eta}_{12}$ and Eqs.(\ref{2.13}) for
${\vec \eta}_i$, $x^{\mu}_i$ and $p^{\mu}_i$.\medskip

In the quantization without ${\vec \eta}_{12}$ in ${\cal H}_{rel}$
one uses the operators ${\hat {\cal M}}\, c = \sum_{i=1}^2\,
\sqrt{m_i^2\, c^2 + {\hat {\vec \pi}}^2_{12}}$ and ${\hat {\vec S}}
= {\hat {\vec \rho}}_{12} \times {\hat {\vec \pi}}_{12}$. For the
quantization of the derived quantities $x^{\mu}_i(\tau)$ and
$p^{\mu}_i(\tau)$ one must start from Eqs.(\ref{2.14}), as it will
be done in Eq.(\ref{5.2}) of the next Section.
\bigskip

Instead the quantization with ${\vec \eta}_{12}$ is done in the
unphysical Hilbert space ${\cal H}_{{\vec \eta}_1} \otimes {\cal
H}_{{\vec \eta}_2} = {\cal H}_{{\vec \eta}_{12}} \otimes {\cal
H}_{{\vec \rho}_{12}}$ with scalar product $ < {\tilde \phi}_1,
{\tilde \phi}_2 > = \int d^3\eta_{12}\, d^3\rho_{12}\, {\tilde
\phi}_1^*(\tau ,{\vec \eta}_{12}, {\vec \rho}_{12})\,\, {\tilde
\phi}_2 (\tau ,{\vec \eta}_{12}, {\vec \rho}_{12})$. In it we define
the quantum operators corresponding to the internal generators
(\ref{4.13}) (for the boosts we use a symmetrical ordering) and we
get that the quantum internal Poincare' algebra is trivially
satisfied.\medskip

It is still convenient to use as Hamiltonian ${\hat {\cal M}}\, c =
\sum_{i=1}^2\, \sqrt{m_i^2\, c^2 + {\hat {\vec \pi}}^2_{12}} = {\hat
H}_{rel}$, because it corresponds to a Hamilton-Jacobi description
of the internal 3-center of mass with frozen Jacobi data ${\hat
{\vec \eta}}_{12}$, ${\hat {\vec \kappa}}_{12}$. Therefore, in the
coordinate representation the  Schroedinger equation (\ref{4.5}) is
replaced by the following one

\beq i\, {{\partial}\over {\partial\, \tau}}\, \tilde \phi(\tau,
{\vec \eta}_{12}, {\vec \rho}_{12}) = \sum_{i=1}^2\, \sqrt{m_i^2\,
c^2 +  {\hat {\vec \pi}}_{12}^2} \,\,\, \tilde \phi(\tau, {\vec
\eta}_{12}, {\vec \rho}_{12}). \label{4.14} \eeq

\noindent The energy eigenfunctions $e^{- i \hbar\, E\, \tau}\,
\phi_E$ satisfy ${\hat H}_{rel}\, \phi_E = E\, \phi_E$ with $E =
\sum_{i=1}^2\, \sqrt{m_i^2\, c^2 + {\vec \pi}^2}$ if $\vec \pi$ is
the eigenvalue of ${\hat {\vec \pi}}_{12}$. By inversion we have
${\vec \pi}^2 = {1\over {4\, E^2}}\, [E^2 - (m_1 + m_2)^2\, c^2]\,
[E^2 - (m_1 - m_2)^2\, c^2]$.
\medskip

However the physical Hilbert space is identified by the following
conditions

\bea &&< \phi_{phys}| {\hat {\vec \kappa}}_{12} | \phi_{phys} >
= 0, \nonumber \\
&&< \phi_{phys}| {\hat {\vec \eta}}_{12} | \phi_{phys} > = {1\over
2}\, < \phi_{phys}| {\hat {\vec \rho}}_{12}\, {{{{m_1}\over m}\,
\sqrt{m_2^2\, c^2 + {\hat {\vec \pi}}^2_{12}} - {{m_2}\over m}\,
\sqrt{m_1^2\, c^2 + {\hat {\vec \pi}}^2_{12}}}\over {\sqrt{m_1^2\,
c^2 + {\hat {\vec \pi}}^2_{12}} + \sqrt{m_2^2\, c^2 + {\hat {\vec
\pi}}^2_{12}}}} +\nonumber \\
&+&  {{{{m_1}\over m}\, \sqrt{m_2^2\, c^2 + {\hat {\vec
\pi}}^2_{12}} - {{m_2}\over m}\, \sqrt{m_1^2\, c^2 + {\hat {\vec
\pi}}^2_{12}}}\over {\sqrt{m_1^2\, c^2 + {\hat {\vec \pi}}^2_{12}} +
\sqrt{m_2^2\, c^2 + {\hat {\vec \pi}}^2_{12}}}}\, {\hat {\vec
\rho}}_{12}  | \phi_{phys} >. \label{4.15} \eea
\medskip

Before studying the relativistic case let us look at the
non-relativistic one.

\subsubsection{The Non-Relativistic Case}

After Eq.(\ref{2.17}) we said that, given a two-body problem with
$E_{Galilei} = {{{\vec p}^2_{(n)}}\over {2\, m}} + {{{\vec
q}^2_{(n)}}\over {2\, \mu}} + V({\vec r}^2_{(n)}) = {{{\vec
p}^2_{(n)}}\over {2\, m}}  + H_{rel}$ at the classical level, the
identification of the non-relativistic internal space of relative
variables can be done by adding the second class constraints ${\vec
p}_{(n)} \approx 0$ (rest-frame condition) and ${\vec x}_{(n)}
\approx 0$ (elimination of the center of mass). As a consequence we
get $E_{Galilei} \approx H_{rel}$, i.e. a Hamilton-Jacobi
description of the 3-center of mass with frozen Jacobi data. At the
quantum level in the Hilbert space ${\cal H}_1 \otimes {\cal H}_2 =
{\cal H}_{com} \otimes {\cal H}_{rel}$ of section III, we quantize
the frozen center-of-mass variables and we use the Hamiltonian $\hat
H = {\hat H}_{rel} = {{{\hat {\vec q}}^2_{(n)}}\over {2\, \mu}} +
V({\hat {\vec r}}^2_{(n)})$. In the coordinate representation the
wave functions are $\psi(t, {\vec x}_{(n)}, {\vec r}_{(n)})$. Let us
restrict the Hilbert space to wave functions $\phi_{phys}(t, {\vec
x}_{(n)}, {\vec r}_{(n)})$ satisfying the requirements $<
\phi_{phys} | {\hat {\vec p}}_{(n)} | \phi_{phys} > = < \phi_{phys}
| {\hat {\vec x}}_{(n)} | \phi_{phys} > = 0$. If we define creation
and annihilation operators ${\hat {\vec a}}^{\dagger} = {\hat {\vec
x}}_{(n)} + i\, {{\hbar}\over {mc}}\, {\hat {\vec p}}_{(n)}$, ${\hat
{\vec a}} = {\hat {\vec x}}_{(n)} - i\, {{\hbar}\over {mc}}\, {\hat
{\vec p}}_{(n)}$, the wave functions $\phi_{phys}$ are identified by
the Gupta-Bleuler-like conditions ${\hat {\vec a}}^{\dagger}\, |
\phi_{phys} > = 0$, $< \phi_{phys} | {\hat {\vec a}} = 0$. In the
coordinate representation this implies the following form of the
wave functions: $\phi_{phys}(t, {\vec x}_{(n)}, {\vec r}_{(n)}) =
N\, e^{- \, {{{\vec x}^2_{(n)}}\over {2\, \beta^2}}}\, \tilde
\phi(t, {\vec r}_{(n)})$ ($\beta = {{\hbar^2}\over {mc}}$). Their
scalar product is $< \phi_{phys\, 1}, \phi_{phys\, 2}
>_{{\cal H}_{com} \otimes {\cal H}_{rel}} = \int d^3x_{(n)}\,
d^3r_{(n)}\, \phi^*_{phys 1}(t, {\vec x}_{(n)}, {\vec r}_{(n)})\,
\phi_{phys\, 2}(t, {\vec x}_{(n)}, {\vec r}_{(n)}) = \tilde N\, \int
d^3r_{(n)}\, {\tilde \phi}_1^*(t, {\vec r}_{(n)})\, {\tilde
\phi}_2(t, {\vec r}_{(n)}) = \tilde N\, < {\tilde \phi}_1, {\tilde
\phi}_2 >_{{\cal H}_{rel}}$.\medskip

Therefore we have a reduction from the Hilbert space ${\cal H}_1
\otimes {\cal H}_2 = {\cal H}_{com} \otimes {\cal H}_{rel}$ to the
Hilbert space ${\cal H}_{rel}$ with Hamiltonian ${\hat H}_{rel}$: we
have only to reabsorb the factor $e^{- \, {{{\vec x}^2_{(n)}}\over
{2\, \beta^2}}}$ in the normalization constant.

\subsubsection{The Relativistic Case}

In the relativistic case of two free particles the
Gupta-Bleuler-like conditions are not convenient because the
constraints ${\vec \chi} = {\vec \eta}_{12} - {{\sum_{i=1}^2\,
(-)^i\, {{m_i}\over m}\, \sqrt{m_i^2\, c^2 + {\vec
\pi}^2_{12}}}\over {\sum_{i=1}^2\, \sqrt{m_i^2\, c^2 + {\vec
\pi}^2_{12}}}}\, {\vec \rho}_{12} \approx 0$ do not have vanishing
Poisson bracket $\{ \chi^i, \chi^j \} \not= 0$. One should replace
the second class constraints $\vec \chi \approx 0$, ${\vec
\kappa}_{12} \approx 0$, with their suitable combinations ${\tilde
\chi}^i \approx 0$, ${\tilde \phi}^i \approx 0$, such that $\{
{\tilde \chi}^i, {\tilde \phi}^j \} = \delta^{ij}$, $\{ {\tilde
\chi}^i, {\tilde \chi}^j \} = \{ {\tilde \phi}^i, {\tilde \phi}^j \}
= 0$. Then, due to the resulting mixing of the canonical variables
${\vec \eta}_{12}$, ${\vec \kappa}_{12}$, ${\vec \rho}_{12}$, ${\vec
\pi}_{12}$, one should find a canonical transformation to a new base
${\vec \eta}^{'}_{12}$, ${\vec \kappa}^{'}_{12}$, ${\vec
\rho}^{'}_{12}$, ${\vec \pi}^{'}_{12}$, adapted to the second class
constraints, i.e. such that $\eta^{'\, i}_{12} = {\tilde \chi}^i
\approx 0$, $\kappa_{12}^{'\, i} = {\tilde \phi}^i \approx 0$.
Presumably the new Hamiltonian would weakly be function only of
${\vec \pi}^{'\, 2}_{12}$. A Gupta-Bleuler-like quantization of the
new canonical basis could then be done following the
non-relativistic pattern.
\medskip

Instead let us evaluate Eqs.(\ref{4.15}) by using the energy
eigenfunctions $\phi_E$ of Eq.(\ref{4.14}). We get $< \phi_E | i
\hbar\, {{\partial}\over {\partial\, {\vec \eta}_{12}}} | \phi_E > =
0$ and $< \phi_E | \Big({\vec \eta}_{12} - f(E)\, {\vec
\rho}_{12}\Big) | \phi_E > = 0$, where $f(E) = {{\sum_{i=1}^2\,
(-)^i\, {{m_i}\over m}\, \sqrt{m_i^2\, c^2 + {\vec \pi}^2(E)}}\over
{\sum_{i=1}^2\, \sqrt{m_i^2\, c^2 + {\vec \pi}^2(E)}}}$ with ${\vec
\pi}^2(E) = {1\over {4\, E^2}}\, [E^2 - (m_1 + m_2)^2\, c^2]\, [E^2
- (m_1 - m_2)^2\, c^2]$.\medskip

For each value of $E$ the conditions are satisfied by the following
energy eigenfunctions ${\tilde \phi}_E = e^{- {{({\vec \eta}_{12} -
f(E)\, {\vec \rho}_{12})^2}\over {2\, \beta^2}}}\, \psi_E(\tau,
{\vec \rho}_{12})$ ($\beta = {{\hbar^2}\over {mc}}$) with ${\hat
H}_{rel}\, \psi_E = e\, \psi_E$ and $<\psi_{E_1}, \psi_{E_2}>_{{\cal
H}_{rel}} = \delta(E_1 - E_2)$. As in the non-relativistic case we
get $<\phi_{phys\, E_1}, \phi_{phys\, E_2}>_{{\cal H}_{com} \otimes
{\cal H}_{rel}} = N\, <\psi_{E_1}, \psi_{E_2}>_{{\cal H}_{rel}} =
N\, \delta(E_1 - E_2)$. Therefore we can build the abstract physical
Hilbert space ${\cal H}_{rel}$ starting from its complete energy
basis $| \psi_E >$.\medskip

In presence of interactions the construction of the physical Hilbert
space ${\cal H}_{rel}$ is much more complex because the Hamiltonian
${\hat H}_{rel}$ depends also on the relative position operator
${\hat {\vec \rho}}_{12}$. Again one has to start from the energy
eigenfunctions.

\vfill\eject

\section{Examples of Two-Body Systems with Mutual
Action-a-a-Distance Interaction}

In this Section we analyze the two classes of models with
action-at-a-distance interaction of Refs.\cite{7,8}.

\subsection{Quantization of the Non-Trivial Interacting
Two-Particle System of Ref.\cite{7}.}

In Ref.\cite{7} we introduced the rest-frame instant form of a class
of positive-energy two-particle models with an arbitrary
action-at-a-distance potential \footnote{See Ref. \cite{7,15} and
its bibliography for the corresponding models with the two signs of
the energy and with mass-shell constraints.}. They were defined by
the following form of the internal Poincare' generators (use
Eq.(\ref{2.11}) and $m_3 \equiv m_1$)
\bigskip

\bea M\, c &=& \sum_{i=1}^2\, \sqrt{m_i^{2}\, c^2 + {\vec \kappa
}_i^{2} + \Phi ({\vec \rho }^2_{12}) } \approx \sum_{i=1}^2\,
\sqrt{m_i^{2}\,
c^2 + {\vec \pi}_{12}^2 + \Phi ({\vec \rho }^2_{12})},\notag \\
{\vec {\cal P}}_{(int)} &=& {\vec \pi}_{12} = {\vec \kappa }_{1} +
{\vec \kappa }_{2} \approx 0,  \notag \\
{\vec {\cal J}}_{(int)} &=& {\vec \eta }_{1} \times {\vec
\kappa}_{1} + {\vec \eta }_{2} \times {\vec \kappa }_{2} = {\vec
\eta}_{12} \times {\vec \kappa}_{12} + {\vec \rho}_{12} \times {\vec
\pi}_{12} \approx {\vec \rho}_{12} \times {\vec \pi}_{12} = \vec S,
  \notag \\
{\vec {\cal K}}_{(int)} &=& - \sum_{i=1}^2\, {\vec \eta }_i\,
\sqrt{m_i^{2}\, c^2 + {\vec \kappa }_i^{2} + \Phi ({\vec \rho
}_{12}^2)} \approx - {\vec \eta }_{12}\, \sum_{i=1}^2\,
\sqrt{m_i^{2}\,
c^2 + { \vec \pi}_{12}^{2} + \Phi ({\vec \rho }^2_{12})} +\nonumber \\
&+& {\vec \rho}_{12}\, \sum_{i=1}^2\, (-)^i\, {{m_{i+1}}\over m}\,
\sqrt{m_i^2\, c^2 + { \vec \pi}_{12}^{2} + \Phi ({\vec \rho
}^2_{12})} \approx 0. \label{5.1} \eea

The classical internal Poincare' algebra closes {\it only} using the
rest-frame condition ${\vec {\cal P}}_{(int)} \approx 0$.

\bigskip

The elimination of the internal 3-center of mass is done with the
conditions

\bea
 {\vec \eta}_{12} \approx {\vec \rho}_{12} \, {{\sum_{i=1}^2\,
 (-)^i\, {{m_{i+1}}\over m}\, \sqrt{m_i^2\, c^2 + { \vec
 \pi}_{12}^{2} + \Phi ({\vec \rho }^2_{12})}}\over {\sum_{i=1}^2\,
 \sqrt{m_i^{2}\, c^2 + { \vec \pi}_{12}^{2} + \Phi ({\vec \rho
 }^2_{12})}}},\qquad {\vec \kappa}_{12} \approx 0.
  \label{5.1a}
   \eea

The orbit reconstruction is given by Eqs.(\ref{2.12}) with ${\vec
\pi}_{12}^2\, \mapsto {\vec \pi}_{12}^2 + \Phi({\vec \rho}_{12}^2)$.

\subsubsection{Quantization without ${\vec \eta}_{12}$}

We have to quantize the Hamiltonian ${\cal M}\, c = \sum_{i=1}^2\,
\sqrt{m_i^2\, c^2 + {\vec \pi}_{12}^2 + \Phi ({\vec \rho }^2_{12})}$
together with the spin ${\vec S} = {\vec \rho}_{12} \times {\vec
\pi}_{12}$.\medskip

See Ref.\cite{40} for the definition of the pseudo-differential
operators connected with the quantization of quantities like
$\sqrt{m^2\, c^2 + {\vec \kappa}^2}$. When also the potential $ \Phi
({\vec \rho }^2_{12})$ appears under the square root, we follow
Ref.\cite{12}: in its Eq.(C7) the following definition was given
$\sqrt{m^2\, c^2 + \hat H} = m\, c\, \sum_{n=0}^{\infty}\, c_n\,
\Big({{\hat H}\over {mc}}\Big)^n$ where $c_n$ are the coefficients
of the Taylor expansion $\sqrt{1 + x} = \sum_{n=0}^{\infty}\, c_n\,
x^n$.\medskip

In our model we have the operator $\hat H = {\hat {\vec \pi}}_{12}^2
+ \Phi ({\hat {\vec \rho }}^2_{12})$ (${\hat {\cal M}}\, c =
\sum_{i=1}^2\, \sqrt{m_i^2\, c^2 + \hat H}$), which coincides with
the Hamiltonian of the relative motion of a non-relativistic
two-body problem with reduced mass $\mu = {1\over 2}$. Therefore if
a complete set of eigenfunctions of this problem is known ($\hat H\,
\psi_{nlm} = \epsilon_n\, \psi_{nlm}$, ${\hat {\vec S}}^2\,
\psi_{nlm} = s\, (s + 1)\, \psi_{nlm}$, ${\hat S}^3\, \psi_{nlm} =
m\, \psi_{nlm}$), then the relativistic mass levels will be $M_n\, c
= \sum_i\, \sqrt{m_i^2\, c^2 + \epsilon_n}$.
\bigskip

The derived (non-commuting) single particle self-adjoint operators
are obtained by quantizing Eq.(\ref{2.12}) and $Y^{\mu}(\tau)$ of
Eq.(\ref{2.3}) with a symmetric ordering

\bea {\hat x}^{\mu}_i(\tau) &=& {\hat Y}^{\mu}(\tau) + {1\over 2}\,
\epsilon _{r}^{\mu}({\hat {\vec h}})\, \Big[(-1)^{i+1}\,
{\hat\rho}^r_{12} -\nonumber \\
&-& {1\over 2}\, (m_1^2 - m_2^2)\, c^2\, \Big( {\hat \rho}^r_{12}\,
{1\over {\sum_{j=1}^{2}\, \sqrt{m_{j}^{2}\, c^2 + \hat H} }} +
{1\over {\sum_{j=1}^{2}\, \sqrt{m_{j}^{2}\, c^2 + \hat H} }}\, {\hat
\rho}^r_{12}
\Big)\Big],\nonumber \\
{\hat p}_i^{\mu} &=& {\hat h}^{\mu}\, \sqrt{m_i^2\, c^2 + {\hat
{\vec \pi}}_{12}^2} + (-)^{i+1}\, \epsilon^{\mu}_r({\hat {\vec
h}})\, {\hat
\pi}^r_{12},\nonumber \\
&&{}\nonumber \\
{\hat Y}^o(\tau) &=& {1\over 2}\, \Big({\hat {\cal M}}_{(int)}\,
c\Big)^{-1}\, \Big(\sqrt{1 + {\hat {\vec h}}^2}\, {\hat {\vec h}}
\cdot {\hat {\vec z}} + {\hat {\vec z}} \cdot {\hat {\vec h}}\,
\sqrt{1 + {\hat {\vec h}}^2} \Big)
+ \sqrt{1 + {\hat {\vec h}}^2}\, \tau,\nonumber \\
{\hat {\vec Y}}(\tau) &=& \Big({\hat {\cal M}}_{(int)}\,
c\Big)^{-1}\, \Big({\hat {\vec z}} + {1\over 2}\, ({\hat {\vec h}}\,
{\hat {\vec h}} \cdot {\hat {\vec z}} + {\hat {\vec z}} \cdot {\hat
{\vec h}}\, {\hat {\vec h}})\Big) + {\hat {\vec h}}\, \tau.
\label{5.2} \eea

\noindent Therefore ${\hat x}^{\mu}_i(\tau)$ depends both on the
quantum frozen Jacobi data ${\hat {\vec z}}$, ${\hat {\vec h}}$,
describing the external evolution, and on the quantum internal
relative variables ${\hat {\vec \rho}}_{12}$, ${\hat {\vec
\pi}}_{12}$, describing the mutual particle interaction.\medskip

\subsubsection{Quantization with ${\vec \eta}_{12}$}

As a consequence we cannot check the quantum internal Poincare'
algebra: to do it we should need a form of the internal Poincare'
generators satisfying the Poincare' algebra without using the
rest-frame conditions.\medskip

It cannot be done until one finds the form of the boosts ${\vec
{\cal K}}_{(int)}$ so that the internal Poincare' algebra closes
without using the rest-frame condition ${\vec {\cal P}}_{(int)}
\approx 0$.

\subsection{Quantization of the Two-Particle System with Coulomb plus
Darwin Mutual Interaction of Ref.\cite{8}.}

In Eq.(I-5.4) of I we found the following internal Poincare' algebra
for a system of two positive-energy charged scalar particles (with
Grassmann-valued electric charges) with a mutual Coulomb plus Darwin
potential

\begin{eqnarray*}
\mathcal{E}_{(int)} &=& M\, c^2 = c\, \sum_{i=1}^2\, \sqrt{m^2_i\,
c^2 +  {\vec \kappa}^2_i} + {\frac{{Q_1\, Q_2}}{{4\pi\, | { \vec
\eta }_1 - {\vec \eta}_2|}}} + V_{DARWIN}({\vec \eta}_1(\tau ) -
{\vec \eta}_2(\tau ); {\vec \kappa}_i(\tau )),  \nonumber \\
{\vec {\mathcal{P}}}_{(int)} &=& {\vec \kappa}_1 +
{\vec \kappa}_2 \approx 0,  \nonumber \\
{\vec {\mathcal{J}}}_{(int)} &=& \sum_{i=1}^2\, {\vec \eta}_i \times
{\vec \kappa}_i,
\end{eqnarray*}

\bea {\vec {\mathcal{K}}}_{(int)} &=& -\sum_{i=1}^2\,
\widetilde{\vec{\eta}_{i}} \, \Big[ \sqrt{m_{i}^{2}\, c^2
+ {\vec{\kappa}_{i}}^{2}}+ \nonumber \\
&+&\frac{{\vec{\kappa}_{i}}\cdot \sum_{j\neq
i}Q_{i}Q_{j}[\vec{\partial}_{{\eta}_{i}}\frac{1}{2} \,
{\mathcal{K}}_{ij}( {\vec{\kappa}_{i}}, {\vec{\kappa}_{j}},{\vec{
\eta}_{i}}- {\vec{\eta}_{j}})-2\vec{A}{{_{\perp Sj}(}}{
\vec{\kappa}_{j}},{\vec{\eta}_{i}} - {\vec{\eta}_{j}})]}{ 2\, c\,
\sqrt{ m_{i}^{2}\, c^2 + {\vec{\kappa}_{i}}^{2}}}\Big]
-\nonumber \\
&-& \frac{1}{2\, c}\, \sum_{i=1}^2\, \sum_{j\neq i}Q_i\, Q_j\,
\sqrt{ m_{i}^{2}\, c^2 + {\vec{\kappa}_{i}}^{2}} \vec{\partial}_{{
\kappa}_{i}}{ \mathcal{K}}_{ij}({\vec{\kappa}_{i}},
{\vec{\kappa}_{j}}, {\vec{\eta}_{i}} -
{\vec{\eta}_{j}}) -  \nonumber \\
&-& \sum_{i=1}^2\, \sum_{j\not=i}\, \frac{ Q_{i}\, Q_{j}}{ 4\pi\, c
}\int d^{3}\sigma \frac{{\vec{\pi}}_{\perp Sj}(\vec{ \sigma}-
{\vec{\eta}_{j}} ,{\vec{\kappa}_{j}})}{|\vec{
\sigma}- {\vec{\eta}_{i}}|}-  \nonumber \\
&-&{\frac{1}{2\, c}}\sum_{i=1}^2\, \sum_{j\neq i}Q_{i}Q_{j}\int
d^{3}\sigma \vec{ \sigma}[{\vec{\pi}_{\perp Si}}(\vec{\sigma}-
{\vec{\eta}_{i}} , {\vec{\kappa}_{i}})\cdot {\vec{\pi}}_{\perp
Sj}(\vec{\sigma}- {\vec{\eta}_{j}},
{\vec{\kappa}_{j}})+  \nonumber \\
&+&\vec{B}{_{Si}}(\vec{\sigma}- {\vec{\eta}_{i}},{\vec{
\kappa}_{i}})\cdot {\vec{B}}_{Sj}(\vec{\sigma} - {\vec{\eta}_j},
{\vec{\kappa}_j})] \approx 0. \label{5.3}
\end{eqnarray}

\noindent with the form of the Darwin potential and of the
Lienard-Wiechert quantities given in Appendix A.

\subsection{Quantization without ${\vec \eta}_{12}$}

By eliminating ${\vec \kappa}_{12} \approx 0$ and ${\vec \eta}_{12}$
we get for the invariant mass

\beq M\, c = {\cal M}\, c = \sum_{i=1}^2\, \sqrt{m_i^2\, c^2 + {\vec
\pi}^2_{12}} + {{Q_1\, Q_2}\over {4\pi\, |{\vec \rho}_{12}|}} +
{\tilde V}_{DARWIN}({\vec \rho}_{12}, {\vec \pi}_{12}). \label{5.4}
\eeq

The expression of the Darwin potential is given in Eq.(\ref{a7}).

\bigskip

In Eq.(6.37) of Ref.\cite{10} the following expression for the
Darwin potential was obtained in the case of equal masses $m_1 = m_2
= m$ (with $m = m_1 + m_2 \mapsto 2\, m$)

\bea &&{\tilde V}_{DARWIN}({\vec \rho}_{12}, {\vec \pi}_{12})
=\nonumber \\
&&{}\nonumber \\
&=&{{Q_1\, Q_2}\over {8\pi\, |{\vec \rho}_{12}|}}\,
  \Big(\,\, \Big(m^2\, c^2 + {\vec \pi}_{12}^2\Big)\, \Big[m^2\, c^2 +
\Big({\vec \pi}_{12} \cdot {{{\vec \rho}_{12}}\over {|{\vec
\rho}_{12}|}}\Big)^2\Big]\, \Big)^{-1}\nonumber \\
&&{}\nonumber \\
&&\Big( m^2\, \Big[3\, {\vec \pi}_{12}^2 + \Big({\vec \pi}_{12}
\cdot {{{\vec \rho}_{12}}\over {|{\vec \rho}_{12}|}}\Big)^2\Big] -
2\, {\vec \pi}_{12}^2\, \Big[{\vec \pi}_{12}^2 - 3\, \Big({\vec
\pi}_{12} \cdot {{{\vec \rho}_{12}}\over {|{\vec
\rho}_{12}|}}\Big)^2\Big]\, \sqrt{{{m^2\, c^2 + {\vec
\pi}_{12}^2}\over {m^2\, c^2 + \Big({\vec \pi}_{12} \cdot {{{\vec
\rho}_{12}}\over
{|{\vec \rho}_{12}|}}\Big)^2}}} -\nonumber \\
&-& 2\, \Big[{\vec \pi}_{12}^2 + \Big({\vec \pi}_{12} \cdot {{{\vec
\rho}_{12}}\over {|{\vec \rho}_{12}|}}\Big)^2\Big]\, \Big[m^2\, c^2
+ \Big({\vec \pi}_{12} \cdot {{{\vec \rho}_{12}}\over {|{\vec
\rho}_{12}|}}\Big)^2\Big]\Big). \label{5.5} \eea

\bigskip

In Appendix  B we obtain the Schroedinger equation corresponding to
the total Hamiltonian given in Eq. (\ref{5.4}).  Because of the
nontrivial momentum dependence in both its kinetic and potential
energy portions, we carry out its quantization by using Weyl
ordering \cite{41}. A noteworthy result of this quantization is that
not only do we obtain the expected nonlocal coordinate space form of
the kinetic energy term, but the Coulomb term itself, in the context
of the Darwin potential corresponding to Eq. (\ref{5.5}), takes on a
nonlocal coordinate space form.  Only in the limit of small  Compton
wavelength does it recover its local coordinate space form.

\bigskip
The Weyl ordering of the order $1/c^2$ Darwin potential below is
also carried out.  We demonstrate that the Weyl ordering leads to
the hermitian ordering given at the beginning of Appendix B. At the
order $1/c^2$, where the Darwin potential for unequal masses becomes

\beq {\tilde V}_{DARWIN}({\vec \rho}_{12}, {\vec \pi}_{12}) =
{{Q_1\, Q_2}\over {8\pi\, |{\vec \rho}_{12}|}}\, {{{\vec \pi}_{12}^2
- \Big({\vec \pi}_{12} \cdot {{{\vec \rho}_{12}}\over {|{\vec
\rho}_{12}|}}\Big)^2}\over {m_1m_2c^2}} + O(c^{- 4}), \label{5.6}
\eeq

\noindent as shown in Eq.(6.35) of Ref.\cite{10}, we recover the
effective stationary Schroedinger equation used for relativistic
bound states in Refs. \cite{42,43,44}

\bigskip

With the methods of Appendix A we can study the two-body problem for
positive-energy charged spinning particles \cite{11}.

\subsubsection{Quantization with ${\vec \eta}_{12}$}

Elsewhere, by using the Weyl ordering in, we will study the
implementation of the quantum internal Poincare' algebra, the
extended Schroedinger equation and its reduction to the previous
results.

\vfill\eject

\section{Implications for Relativistic Localization and Relativistic
Entanglement}

In this Section we will indicate which problems connected with
localization are solved with our rest-frame formulation of RQM.
Moreover we will delineate which are the implications for
relativistic entanglement.

\subsection{Relativistic Localization}

In non-relativistic QM a wave function strictly localized in a
finite volume at $t = 0$ will {\it spread instantaneously} to all
the 3-space with infinite tails as shown in Ref.\cite{45}. The
position operator ${\hat {\vec x}}$ is a self-adjoint operator with
a continuous spectrum, whose distributional eigenfunctions
corresponding to the localization associated to the eigenvalue $\vec
\xi$ are $\psi_{\vec \xi}(\vec x) = \delta^3(\vec x - \vec \xi )$.
These wave functions are mutually orthogonal: $< \psi_{{\vec
\xi}_1}, \psi_{{\vec \xi}_2} > = \delta^3({\vec \xi}_1 - {\vec
\xi}_2)$. Localization is invariant under the invariance group of
Galilei space-time, the Galilei group. The uncertainty relations
limit the sharpness with which a system's position can be determined
in certain circumstances. The only problems of {\it non-locality}
are connected with entanglement, see for instance the EPR argument
\cite{21,23,24}.

\subsubsection{The Newton-Wigner Position Operator and the
Hegerfeldt Theorem}

In relativistic QM, in a fixed inertial frame, to a scalar positive
-energy particle of mass $m$ is associated the {\it self-adjoint
Newton-Wigner (NW) position operator} \cite{46}, \cite{28}

\beq {\hat {\vec x}} = i\hbar {d\over {d\vec P}} - {{i\hbar \vec
P}\over {m^2\, c^2 + {\vec P}^2}}, \label{6.1} \eeq

\noindent in a Hilbert space with Lorentz-scalar scalar product

\beq < \psi_1, \psi_2 > = \int {{d^3P}\over {\sqrt{m^2\, c^2 + {\vec
P}^2}}}\, {\tilde \psi}^*_1(\vec P)\, {\tilde \psi}_2(\vec P),
\label{6.2} \eeq

\noindent in the momentum representation \footnote{The covariant
Fourier transform is $\psi (\vec x) = (2\pi)^{-3/2}\, \int\, e^{i
\vec P \cdot \vec x}\, \tilde \psi (\vec P)\, {{d^3P}\over
{\sqrt{m^2\, c^2 + {\vec P}^2}}}$.}. Its position eigenvectors at
time $t = x^o/c = 0$, corresponding to the eigenvalue $\vec \xi$,
are

\beq \psi_{\vec \xi}(\vec x, 0) = (2\pi)^{-3}\, \int {{d^3P}\over
{(m^2\, c^2 + {\vec P}^2)^{1/4}}}\, e^{i\, \vec P \cdot (\vec x -
\vec \xi)}, \label{6.3} \eeq

\noindent  with momentum representation ${\tilde \psi}_{\vec
\xi}(\vec P) = (2\pi)^{-3/2}\, (m^2\, c^2 + {\vec P}^2)^{1/4}\, e^{-
i \vec P \cdot \vec \xi}$. They are orthogonal, $< \psi_{{\vec
\xi}_1}, \psi_{{\vec \xi}_2} > = \delta^3({\vec \xi}_1 - {\vec
\xi}_2)$ but they {\it are spread out in $\vec x$}. Instead of a
delta function like in non-relativistic QM, they are proportional to
the Hankel functions of the first kind $H^{(1)}_{5/4}(\vec x - \vec
\xi )\, \rightarrow_{|\vec x| \rightarrow \infty}\, e^{- |\vec x -
\vec \xi|/\lambda_m}$, where $\lambda_m = \hbar/mc$ is the Compton
wavelength. Therefore there are infinite tails governed by the
Compton wavelength, even if at the classical level the associated
M$\o$ller radius is zero.\medskip

This {\it absence of sharp localization}, due to the form of the
scalar product and to the orthogonality requirement is an aspect of
the {\it non-locality} present in special relativity with
self-adjoint position operators \footnote{For a particle sharply
localized at $\vec \xi$ the non-relativistic wave function is $\psi
(\vec x) = \delta^3(\vec x - \vec \xi) = \psi_{\vec \xi}(\vec x)$.
Instead at the relativistic level we have $\psi (\vec x) = \int
d^3\xi\, G(\vec \xi )\, \psi_{\vec \xi}(\vec x)$ with $G(\vec \xi )
= (2\pi)^{-3}\, \int d^3x\, \int d^3P\, (m^2\, c^2 + {\vec
P}^2)^{1/4}\, e^{i\, \vec P \cdot (\vec \xi - \vec x)}\, \psi (\vec
x)$ for every wave function, also for those strongly peaked at some
${\vec \xi}_o$. }.\bigskip

This counterintuitive aspect of relativistic localization has the
following two inter-related implications:\medskip

A) {\it Newton-Wigner localization is not invariant under Lorentz
boosts} \cite{46}, \cite{28}, consistently with the classical
non-covariance of the 3-center of mass. If in the original inertial
frame we have the Newton-Wigner eigenstate $\psi_{\vec \xi = 0}(\vec
x)$ at $t = 0$, in a moving frame the boosted wave function is a
superposition of the Newton-Wigner eigenstates corresponding to
every value of $\vec \xi$. This means that the probability density
amplitude to be in a given eigenstate with eigenvalue $\vec \xi$ is
{\it frame-dependent}: if it is sharply localized in one frame, it
has infinite tails in a moving frame. {\it Frame-independent
objectivity of localization is lost}.

\bigskip

B) {\it Time evolution in a fixed inertial frame destroys sharp
localization}. At time $x^o = ct$ the Newton-Wigner eigenstate with
eigenvalue $\vec \xi = 0$ is

\bea \psi (x^o, \vec x) &=& (2\pi )^{-3}\, \int {{d^3P}\over {(m^2\,
c^2 + {\vec P}^2)^{1/4}}}\, e^{i(\vec P \cdot \vec x - \sqrt{m^2\,
c^2 + {\vec P}^2}\, x^o)} = \int d^3\xi\, G(x^o, \vec \xi )\,
\psi_{\vec
\xi}(0, \vec x),\nonumber \\
&&{}\nonumber \\
  &&G(x^o, \vec \xi ) = (2\pi
)^{-3}\, \int d^3P\, e^{i(\vec P \cdot \vec \xi - \sqrt{m^2\, c^2 +
{\vec P}^2}\, x^o)} \not= \delta^3(\vec \xi ). \label{6.4} \eea

\noindent The form of $G$ is due to the branch points at $|\vec P| =
\pm i\, mc$. Infinite tails in $\vec \xi$ develop and there is an
{\it apparent violation of Einstein causality}. $G(x^o, \vec \xi )$
is non-zero everywhere for arbitrarily small $x^o$ and this implies
the possibility of a non-local phenomenon.\medskip

This is the content of {\it Hegerfeldt theorem} \cite{29,30}, which
says that the requirement that the NW operator be a self-adjoint
operator implies the instantaneous super-luminal spreading of  wave
packets: only at the level of wave packets with power tails could
there be consistency with relativistic causality \footnote{In
Ref.\cite{30} it is also noted that the theorem does not create any
problem for the the interpretation of the Dirac equation due to the
presence of both positive- and negative-energy component as shown in
Ref.\cite{47}: but the same problems reappear if we restrict
ourselves to the positive-energy sector.}. As a consequence, the
requirement of relativistic causality implies {\it bad localization}
of the Newton-Wigner position, as already anticipated at the
classical level with the non-covariance M$\o$ller world-tube for the
relativistic canonical 3-center of mass. Since it is  impossible  to
explore the interior of the M$\o$ller world-tube (i.e. distances
less than the Compton wavelength of the isolated system) of the
isolated system \cite{5} without breaking manifest Lorentz
covariance, this would be compatible with a {\it non-self-adjoint}
Newton-Wigner position operator.

\bigskip

As clarified in Refs.\cite{30} with the hypotheses of the theorem
(Hilbert space and positive energy) it is not yet possible to show
that there is at least {\it weak causality}, namely that Einstein
causality holds only for the expectation values or the ensemble
averages of a projection operator $N(V)$ on a fixed 3-region V
\footnote{For such projectors there is  {\it Malament's theorem}
\cite{48} saying that the requirements of localizability,
translation covariance, energy bounded below and microcausality
imply that there is no chance that a particle will be detected in
any local region.}. If the position operator is not self-adjoint,
the operator $N(V)$ is not a projector but  a {\it positive
operator-valued measure} (POVM) \footnote{As shown by Peres in Ref.
\cite{26} (see also Ref.\cite{23}) POVM are complete sets of (in
general non-commuting) positive operators (more general than
projectors) describing {\it detectors} used to describe the {\it
measurement of an observable}. If the density matrix $\rho$
describes an {\it emitter}, then the probability that the detector
$\mu$ is excited is $Tr(\rho\, E_{\mu})$. According to Peres the
notion of particle has an operational meaning depending on the
context of experiments: particles are what is registered by
detectors localizing them (see Ref.\cite{26} for a review of the
localization of particles).}, but again the infinite tails spread
too fast.

\subsubsection{Quantum Field Theory}

In conclusion the localization problem in relativistic QM cannot be
solved without taking into account quantum field theory (QFT). In
Ref. \cite{49} it is claimed that {\it a relativistic QM of
localizable particles does not exist and that only relativistic QFT
makes sense} (the basic ontological objects are fields). It is
argued  that in QFT particle detection is an {\it approximately
local} measurement: for all practical purposes (FAPP) of
phenomenology non strictly localized objects will appear as strictly
localized (particles with localized mutual interactions) to local
finite observers. According to Haag \cite{50} the concept of
position at a given time is not a meaningful attribute of the
electron \footnote{In the basic Wightman axioms there is the {\it
time-slice axiom} (primitive causality) saying that there should be
a dynamical law which allows one to compute fields at an arbitrary
time in terms of the fields in a small time slice ${\cal O}_{t,
\epsilon} = \Big(x | | x^o - ct | < \epsilon \Big)$.}: rather it is
an attribute of the interaction between the electron and a suitable
detector.\medskip

Fraser \cite{51} shows that the particle concept (as elementary
quanta in Fock space) is meaningful only in the description of {\it
free fields} in QFT. Till now in interacting systems there is no
acceptable extension of this notion. The assumption that a particle
is localizable is not used in this exposition. Therefore the notion
of particle seems to be only an effective one to be used in
perturbative QFT. Let us note that in perturbative QFT one uses
Feynman diagrams as an intermediate tools to evaluate the S matrix.
These diagrams describe interacting particles by using the momentum
basis (they correspond to a Dirichlet problem  and not to a Cauchy
problem): in this way the problem of NW-localization is avoided.
Instead a well-posed Cauchy problem is needed for predictability in
classical field theory: only in this way (modulo integrability) can
we use the existence and uniqueness theorem for partial differential
equations. Only with the {\it non factual} 3+1 splitting of
Minkowski space-time and  the {\it non factual} definition of the
global Poincare' generators of an isolated system is it possible
define the instantaneous 3-spaces where to give the Cauchy data. In
this way, at least at the classical level, it is possible to avoid
the Haag theorem \cite{52} preventing the existence of interpolating
fields as shown in paper I.\medskip

See Ref.\cite{31} for {\it standard localization scheme in spatial
regions} of perturbative local QFT (where "localized in" means
"measurable in" and microcausality holds in 4-regions)
\footnote{However when one introduces the spectrum condition,
implying the positivity of the energy and that the velocity of light
is the upper bound for the propagation of physical effects, one
makes a {\it non local} statement on the global 4-momentum
operator.} and for the comparison with the {\it NW-localization
scheme} of Refs.\cite{53} and \cite{18} using the NW position
operator which cannot be described neither with local nor
quasi-local operators. See Ref.\cite{54} for the notion of {\it
unsharp} observables (if a local operator is not measurable with
local actions in a given 3-region) and for a criticism of the
request of microcausality, because sharp spatial localization is an
{\it operationally meaningless idealization} (it requires an
infinite amount of energy with unavoidable pair production; the
quantum nature of the constituents of the detectors should be taken
into account,...).

\medskip

Finally differently from perturbative QFT, in local algebraic QFT
{\it local (and quasi-local) operators} are introduced  having in
mind that they can be used to describe phenomena and measurements
confined in local bounded 4-regions of space-time with the vanishing
of the commutator of local operators  in disjoint
space-like-separated 4-regions (causality) implying the independence
of the disjoint measurements (no action-at-a-distance
communication).

\bigskip

However the relevance of relativistic QM against these attacks from
QFT, because in it particles are only {\it effective}
nearly-localized entities. See for instance Ref. \cite{55} where an
approximate notion of {\it effective localization in a 3-region G}
of radius $L$ of the order of the particle Compton wavelength (it is
an effective notion of NW-localization ) is given starting from an
analogy with a solid-state system on length-scales which are large
compared to the interatomic spacing.

\bigskip

Even if there is no agreement on the relevance of the notion of
particle in QFT, particles are effective tools for phenomenology and
for the S matrix. Moreover atomic and solid-state physics are
specific sectors of certain QFT's in which there is a wealth of
situations in which {\it particles} (electrons, atomic nuclei,...)
are strongly interacting and yet maintain their own particle
character.\bigskip

\subsubsection{The New Relativistic Quantum Mechanics}

The rest-frame instant form of relativistic QM developed in papers I
and II and its quantization done in this paper leads to an {\it
effective theory} for the description of relativistic atomic
physics, and hopefully quantum optics, below the threshold of pair
production. It can be interpreted as an approximation to QED in
which the particle number is fixed,  needed for going from quantum
optics with non-relativistic two-level atoms \cite{56}, used in the
experiments on non-relativistic entanglement where strictly speaking
photons do not exist (only their polarization and not their
world-line is described), to a relativistic theory in which both
atoms and photons can coexist. It will allow one to arrive at a
relativistic formulation of entanglement experiments with laser
beams with a fixed number of photons.
\bigskip

The quantization of the rest-frame instant form of relativistic
particle dynamics presented in this paper, which can be trivially
extended from two to N particles (see Ref.\cite{6} for the
kinematics), has the following advantages with respect to other
approaches to relativistic mechanics:\medskip

A) There is a complete solution to the problem of the {\it
non-objectivity of localization}, i.e. the dependence of the
particle position  also from the simultaneity hyperplane
(frame-dependence) of Refs.\cite{16,17,18}: it is avoided by using
the embedding of the Wigner 3-spaces (intrinsic rest frames) in
Minkowski space-time with the dynamics  described by
Wigner-covariant relative 3-variables living in an abstract  Wigner
3-space.

\medskip

B) We use a modified   NW-localization scheme: we do not quantize
the canonical non-covariant NW 3-center of mass. Instead we quantize
the non-covariant frozen Jacobi data of the external 3-center of
mass ${\hat {\vec z}}$, ${\hat {\vec h}}$ in the frozen Hilbert
space ${\cal H}_{com}$. Therefore {\it we do not have evolving wave
packets for the 3-center of mass so that we avoid the instantaneous
spreading of wave packets of Hegerfeldt theorem}. The Jacobi data
$\vec z$ are more fundamental than the center-of-mass Newton-Wigner
position ${\vec x}_{NW} = \vec z/M\, c$, because they do not depend
explicitly on the internal mass which is quantized at the quantum
level (so that the operators ${\hat {\vec x}}_{NW\, n} = {\hat {\vec
z}}/M_n\, c$ depend upon the mass eigenvalue $M_n$).\medskip

C) We quantize the Wigner covariant relative 3-variables inside the
Wigner 3-spaces with an abstract Hilbert space ${\cal H}_{rel}$. In
it there will be instantaneous spreading in the $\tau$-evolution of
initially localized (in the position relative variables) wave
packets for the relative motion. This Hilbert space, like its
non-relativistic counterpart, contains only relative variables with
action-at-a-distance interactions (the mutual Coulomb interaction
when the transverse electro-magnetic field is present), but, once
initial Cauchy data are given on an initial instantaneous 3-space,
then the evolution is compatible with Lorentz covariance (there is
no violation of Einstein causality or superluminal signalling). The
action of a Lorentz boost of the external Poincare' group on ${\cal
H}_{com} \otimes {\cal H}_{rel}$ induces a Wigner rotation (and not
a $\tau$-evolution) of the relative variables and leads to a
non-covariant transformation of the Jacobi data ${\hat {\vec z}}$
according to Eq.(\ref{2.5}).
\medskip

D) In terms of the quantum Jacobi data and of the quantum invariant
mass and rest-spin we can build the $\tau$-evolving position
operators for the external Fokker-Pryce 4-center of inertia ${\hat
Y}^{\mu}(\tau )$ (a 4-vector operator with non-commuting
components), the external 4-center of mass ${\hat {\tilde
x}}^{\mu}(\tau)$ (a pseudo-4-vector non-covariant operator whose
components require a suitable ordering to be commuting) and the
external M$\o$ller 4-center of energy ${\hat R}^{\mu}(\tau )$ (a
non-covariant non-commuting pseudo-4-vector operator). One should
study their mean value $< \phi | ... | \phi >$ and see whether some
form of  Eherenfest theorem holds for them. Since these collective
variables are {\it global non-local} quantities, they cannot be
localized with local means! This is our {\it answer the
NW-localization problem}. Moreover, when the spatial  region
containing the particles on a simultaneity Wigner instantaneous
3-space has a radius bigger than the M$\o$ller radius of the
particle configuration, then the classical energy density is
everywhere positive definite (weak energy condition; classical
version of the Epstein-Glaser-Jaffe theorem \cite{34}).\medskip

In $\mathcal{H}_{com} \otimes {\cal H}_{rel}$ the property of the
frozen Jacobi data $\vec h$ (or of the total 4-momentum $P^{\mu}$ of
the isolated system) of being a {\it constant of the motion} is made
explicit. Therefore, the description of the external non-covariant
center of mass carrying a pole-dipole structure fits with the point
of view that the isolated system is a {\it closed universe}. If we
use the Wigner-Araki-Yanase theorem on the constants of motion in QM
\cite{57} (see also p. 421 of Ref.\cite{23}), it turns out that the
conjugate variable, namely the Jacobi data $\vec z$  \footnote{As a
consequence of the canonical transformation of paper I, it turns out
that in this description of the isolated system "charged particles
with mutual Coulomb interaction plus a transverse electro-magnetic
field" there is another hidden constant of the motion, namely the
relative momentum ${\vec \pi}_{(12)3}$ of the particle subsystem
with respect to the center of phase of the transverse
electro-magnetic field. This implies that the relative variable
${\vec \rho}_{(12)3}$ is not measurable.} are {\it not measurable
quantities}. Therefore the same is true for the external
non-covariant 4-center of mass ${\tilde x}^{\mu}(\tau )$, the
decoupled pseudo-particle carrying the pole-dipole structure. The
conceptual problem is always the same: {\it Who will measure the
wave function of a closed universe?}.\medskip

This fact, together with the avoidance of the Hegerfeldt theorem due
to the frozen nature of ${\cal H}_{com}$, leads to the two following
{\it open problems}: a) do we take the Jacobi data ${\hat {\vec z}}$
self-adjoint?; b) if ${\hat {\vec z}}$ is chosen self-adjoint, is it
meaningful to consider superpositions of center of mass wave
functions with different eigenvalues of ${\hat {\vec h}}$ (one could
introduce a {\it superselection rule} forbidding them like it has
been proposed in canonical gravity \cite{58,59})? Actually a generic
non-factorizable wave function in ${\cal H}_{com} \otimes {\cal
H}_{rel}$ implies entanglement among different internal energy
levels and among the different rest-frame 3+1 splittings associated
with center-of-mass plane waves.

\medskip

E) The problem of NW-localization of the individual particles has a
different formulation, because the position 4-coordinates
$x^{\mu}_i(\tau )$ parametrizing the world-lines and the 4-momenta
$p^{\mu}_i(\tau )$ ($p^2_i = m_i^2\, c^2$) are derived quantities.
At the classical level the world-lines are obtained with the orbit
reconstruction of the 4-vectors $x^{\mu}_i(\tau )$ of
Eq.(\ref{2.13}). Therefore after quantization the information about
the individual particles is hidden in the quantum operators ${\hat
x}_i^{\mu}(\tau )$, ${\hat p}_i^{\mu}(\tau )$. The   position
operators ${\hat x}^{\mu}_i(\tau )$ have a non-commutative structure
(implied by Eqs.(\ref{2.14})) already at fixed time $[ {\hat
x}_1(\tau ), {\hat x}_1^{\nu}(\tau ) ] \not= 0$, $[ {\hat x}_1(\tau
), {\hat x}_2^{\nu}(\tau ) ] \not= 0$, $[ {\hat x}_2(\tau ), {\hat
x}_2^{\nu}(\tau ) ] \not= 0$ for $N=2$. Even if we have $[ {\hat
p}^{\mu}_i(\tau_1), {\hat p}^{\nu}_j(\tau_2) ] = 0$, also the
commutators $[ {\hat x}_i(\tau ), {\hat p}_j^{\nu}(\tau ) ]$ are
probably non trivial. Have these non-commutative properties any
connection with the existing non-commutative models for interactions
and/or space-time structure? One should study a version of the
Eherenfest theorem adapted to the rest-frame relativistic QM
\footnote{In Newton QM, by using ${\cal H}_1 \otimes {\cal H}_2$ we
can apply the Ehrenfest theorem to both ${\hat {\vec x}}_{(n)i}$ and
${\hat {\vec p}}_{(n)i}$.} for the recovering of the classical
world-lines $x^{\mu}_i(\tau )$ from the mean values $< \phi | {\hat
x}_i^{\mu}(\tau ) | \phi >$ on suitable {\it quasi-classical} states
$\phi$.
\medskip

As a consequence, {\it the operators of two space-like separated
particles do not satisfy microcausality} as happens in the
NW-localization scheme of Refs.\cite{18,31,53} {\it but without
implying superluminal signalling}. This supports the criticism to
the validity of the notion of local measurability associated to
local algebras and to the associated notion of microcausality (or
weak Einstein causality) of Ref.\cite{54}.\medskip

\subsection{Relativistic Entanglement}

As we have seen the absence of absolute simultaneity due to the
Lorentz signature of Minkowski space-time, the non-locality of
Poincare' generators, the non-covariance of the relativistic
canonical center of mass and the presence of interactions in the
Poincare' boosts (absent in the Galilei boosts) identify the tensor
product  $\mathcal{H}_{com} \otimes {\cal H}_{rel}$, ${\cal H}_{rel}
= \otimes_a\, \mathcal{H}_{rel\, a}$ as the relevant Hilbert space.
The Hilbert space $\Big({\cal H}_{com} \otimes {\cal
H}_{rel}\Big)_{\tau}$ cannot be presented in the form of the Hilbert
space $({\cal H}_1)_{x^o_1} \otimes ({\cal H}_2)_{x^o_2}$ of two
free Klein-Gordon quantum particles, even if these two Hilbert
spaces are isomorphic. In $\Big({\cal H}_{com} \otimes {\cal
H}_{rel}\Big)_{\tau}$ there is a frozen external center-of-mass wave
function\footnote{As said it can also be described by a Klein-Gordon
center-of-mass wave function with its conserved current implying the
independence of the external center-of-mass scalar product from
${\tilde x}^o$ in the auxiliary Hilbert spaces $\mathcal{H}_{ext\,
com\,\, n}$.} and a $\tau$-independent scalar product in ${\cal
H}_{rel}$ \footnote{${\cal H}_{rel}$ can be thought as the reduction
of the Hilbert space $\Big({\cal H}_{{\vec \eta}_1} \otimes {\cal
H}_{{\vec \eta}_2}\Big)_{\tau}$ by means of the conditions $< |
{\hat {\vec {\cal P}}}_{(int)} | > = < | {\hat {\vec {\cal
K}}}_{(int)} | > = 0$.}. In the Hilbert space $({\cal H}_1)_{x^o_1}
\otimes ({\cal H}_2)_{x^o_2}$ there are two conserved currents
implying that the scalar products in the Hilbert spaces $({\cal
H}_i)_{x^o_i}$ are independent from the times  $x^o_i$ as shown in
Ref.\cite{15}, but there is no correlation between $x^o_1$ and
$x^o_2$ \footnote{In Ref.\cite{15} there is also the quantization of
the first-class constraints $\sgn\, p_i^2 - m_i^2\, c^2 \approx 0$
after the introduction of suitable center-of-mass ($x^{\mu}$) and
relative ($r^{\mu}$) variables in place of the positions
$x^{\mu}_i$'s: in this way one gets a quantum model, adapted to the
sum and the difference of the two constraints, with a Hilbert space
$({\tilde {\cal H}})_{x^o} \otimes ({\cal H}_{rel})_{r^o}$ where
there are conserved currents implying that the new scalar product is
independent from the center-of-mass time $x^o$ and from the relative
time $r^o$. As a consequence also the presentation $({\tilde {\cal
H}})_{x^o} \otimes ({\cal H}_{rel})_{r^o}$ (a precursor of the
approach in this paper) is inequivalent to the one $({\cal
H}_1)_{x^o_1} \otimes ({\cal H}_2)_{x^o_2}$ with its single-particle
conserved currents.}. The problem is that in the tensor product
${\cal H}_1 \otimes {\cal H}_2$ the clocks of the two particles are
not synchronized: there are states in which one particle is in
absolute future of the other one, so that we cannot define a
well-posed Cauchy problem.

\bigskip

One relevant point of the definition of relativistic rest-frame QM
is that it selects a {\it preferred bases} for ${\cal H}_{com}$,
i.e. the momentum basis, because with each eigenvalue $\vec k$ of
${\hat {\vec h}}$ is associated an inertial 3+1 splitting of
Minkowski space-time with the Euclidean instantaneous Wigner
3-spaces orthogonal to $h^{\mu} = (\sqrt{1 + {\vec h}^2}; \vec h)$.
This preferred basis is therefore induced by the need of clock
synchronization for the identification of the instantaneous 3-space:
it is a consequence of Lorentz signature. Instead the selection of
preferred bases in ${\cal H}_{rel}$ has to be done with the methods
of decoherence \cite{24}. The derived momentum operators ${\hat
p}_i^{\mu}(\tau )$, needed for the description of the individual
particles, will depend on the preferred basis of ${\cal H}_{com}$
and on the chosen basis for ${\cal H}_{rel}$. The same holds for the
derived world-lines of the particles.\medskip

Let us remark that one could also study relativistic entanglement in
the unphysical Hilbert space ${\cal H}_{{\vec \eta}_1} \otimes {\cal
H}_{{\vec \eta}_2} \otimes {\cal H}_{{\vec \eta}_3} \otimes ...$,
where there is separability on the instantaneous Wigner 3-spaces.
However this type of separability is then destroyed by the quantum
version of the interaction-dependent second class constraints ${\vec
{\cal P}}_{(int)} \approx 0$, ${\vec {\cal K}}_{(int)} \approx 0$.
In the non-relativistic limit, where the interaction dependent terms
are at order $1/c^2$, this amounts to study the non-relativistic
entanglement in the rest frame with the center of mass put in the
origin of the coordinates.

\bigskip

In conclusion relativistic rest-frame QM has the following important
{\it kinematical} properties induced by the absence of absolute
simultaneity due to the Lorentz signature of Minkowski space-time
and to the  structure of the Poincare' group: {\it non-locality} of
the collective relativistic variables and {\it spatial
non-separability}. The fact that {\it a relativistic composite
system is never the tensor product of the elementary subsystems},
but is described by the Hilbert space ${\cal H}_{com} \otimes {\cal
H}_{rel}$, implies an \textit{intrinsic spatial non-separability}.
It is induced by the clock synchronization problem, which is not
present in Galilei space-time where time and space are separate
absolute notions, so that the {\it separability} of the subsystems
of a composite system is always assumed (the zeroth law of QM).
However, as shown in Section III, non-relativistic QM can be
presented in the same non-separable form as the rest-frame instant
form of relativistic QM if we emphasize the role of the Galilei
group in the separation of variables in the Schroedinger equation in
presence of interactions.
\bigskip

Let us remark that if we do not succeed to solve the
interaction-dependent constraints ${\vec {\cal K}}_{(int)} \approx
0$ (gauge fixings of the rest frame conditions ${\vec {\cal
P}}_{(int)} \approx 0$), so that the internal 3-center of mass
becomes an interaction-dependent function, ${\vec \eta}_+ \approx
{\vec \eta}_+[{\vec \rho}_a, {\vec \pi}_a]$, of the relative degrees
of freedom, we must work in the unphysical Hilbert space
$\otimes_{i=1}^N\, {\cal H}_{{\vec \eta}_i}$ and then make a
Gupta-Bleurer reduction to ${\cal H}_{rel}$ as said in Section IV.
The formal separability in subsystems inside the Wigner 3-spaces of
the unphysical Hilbert space is destroyed by the dependence upon the
interaction of the constraints. It is only in the non-relativistic
limit, where the solution of ${\vec {\cal K}}_{(int)} \approx 0$ is
${\vec \eta}_+(\tau ) \approx 0$ independently from the
interactions, that separability can be recovered (if wished) as
shown in Section III.

\bigskip

Since the non-separable physics is completely contained in the
relative variables of ${\cal H}_{rel}$, we can say that the absence
of an absolute notion of simultaneity in special relativity induces
a {\it weak-relationist point of view}: only relative motions are
locally accessible because the globally defined center of mass
motion cannot be locally determined. Therefore an isolated system (a
closed universe) composed by subsystems of the type {\it physical
system + observer 1 + observer 2 + (particles of the experimental
protocol) + environment} must be analyzed in terms of relative
variables after the separation of the global (not locally
accessible) center of mass (being decoupled its non-covariance is
irrelevant). In this respect there are some analogies with Rovelli's
relational QM \cite{60} (all systems and observers are equivalent
and all the observations are observer dependent), but Rovelli's
notions of locality and separability are completely different.

\bigskip

The previously described {\it kinematical} properties of {\it
non-locality}  and {\it spatial non-separability} derive from the
choice (required by predictability) of the instantaneous 3-space
with a clock synchronization convention which introduces a
correlation among all the particles. Therefore this {\it kinematical
property is independent of the distances between the particles like
the non-local aspects of quantum mechanics} connected with the
entanglement (the {\it fake} a-a-a-d implied by entanglement if we
accept Einstein notion of reality). Therefore {\it quantum
non-locality is superimposed to already existing relativistic
non-locality and spatial non-separability}.
\medskip

Let us remark that till now the approaches to relativistic
entanglement  have been based on Hilbert spaces of the type of
tensor product of the constituents (the type of separability
suggested by scattering theory but incompatible with relativistic
bound states) trying to analyze it using group theoretical methods
from the theory of representations of the Poincare' group. See Refs.
\cite{26} for the attempts to define relativistic entanglement.

\medskip

See Refs. \cite{23,24,61}  for the problems of entanglement, of what
is a measurement and for the discussion on the interpretations of
non-relativistic QM. For  the role of decoherence see
Refs.\cite{19,24,62}. The implications of our relativistic version
of entanglement for these problems and for the emergence of
classical properties will be investigated elsewhere.

\bigskip

Finally, to include Maxwell equations and their quantization in the
relativistic theory of entanglement (the great absent in
non-relativistic entanglement), we must either use Fock states with
fixed number of photons or make an eikonal approximation of
classical Maxwell equations to introduce rays of light (in both
cases we can use classical massless helicity 2 classical
relativistic particles \cite{63} and their first quantization
adapted to the rest-frame instant form \footnote{For the
positive-energy spinning particles in the rest-frame instant form
see the Appendix of Ref.\cite{12} and Ref.\cite{64}. For
positive-energy massless particle with helicity one (photons in the
eikonal approximation of Maxwell equations with light rays) see
Ref.\cite{65}.} ); see also Refs.\cite{26,66}. This will be needed
to study relativistic teleportation, before facing the problem of
gravity \footnote{See Refs.\cite{67} for an attempt to formulate
atom interferometry in the gravitational field of the Earth by
assuming that atoms follow time-like geodetics.} as in the proposed
teleportation experiments between Earth and the Space Station
\cite{68}.

\vfill\eject

\section{Conclusions}

In this paper we propose a new quantization scheme for
positive-energy relativistic particles in the inertial rest-frame
instant form of dynamics. The isolated system of N particles is
visualized as a non-local decoupled 4-center of mass, described by
canonical non-covariant frozen Jacobi data $\vec z$ and $\vec h$,
carrying a pole-dipole structure, i.e. a rest mass $Mc$ and a rest
spin ${\vec {\bar S}}$ functions of Wigner-covariant relative
variables ${\vec \rho}_a$, ${\vec \pi}_a$, $a=1,..,N-1$ lying in the
instantaneous Wigner 3-spaces centered on the Fokker-Pryce 4-center
of inertia. The internal 3-center of mass inside the Wigner 3-space
is eliminated with the rest-frame condition avoiding a double
counting of the center of mass. The Wigner 3-spaces are orthogonal
to the conserved 4-momentum of the isolated system, but the internal
relative variables are independent of its orientation due to their
Wigner covariance (abstract frame-independent internal space). The
particle world-lines are derived quantities described by
non-canonical 4-vectors (predictive coordinates): a well defined (in
general interaction-dependent) non-commutative structure
emerges.\medskip

The non-relativistic limit of this relativistic QM reproduces the
ordinary QM in the Hamilton-Jacobi description of the
non-relativistic center of mass.\medskip

The quantization scheme is applied to two classes of models with
mutual action-at-a-distance interaction among the particles. Besides
scattering states also the known properties of relativistic bound
states can be described by this quantization scheme. Included in 
Appendix B is the Weyl-ordered quantization of the classical 
two-body Hamiltonian including Coulomb plus Darwin interactions 
to all orders of $1\over c^2$

\medskip

After a review of the known problems with the notion of {\it
relativistic localization} in classical relativistic mechanics, in
relativistic QM and in QFT, we emphasize that the only open problem
in our quantization scheme is connected with the quantum Jacobi data
${\hat {\vec z}}$ : A) If we take them self-adjoint (like in  
non-relativistic QM), we may either allow superpositions of 
center-of-mass states or introduce superselection rules forbidding
them; B) if we take them to be non-self-adjoint, we need to introduce
a modified theory of measurement.  The non-observability of the 
center-of-mass gives rise to these global problems, whose solution
requires further study.

\medskip
Then we study the properties of the {\it
relativistic entanglement} implied by the new quantization scheme.
It turns out to be qualitatively different from {\it
non-relativistic entanglement} whose most relevant property is {\it
quantum non-locality} whichever attitude one takes about the
foundational interpretative problems. At the relativistic level the
prominent properties are the {\it kinematical non-locality and
spatial non-separability} induced by the non-local nature of the
relativistic 4-center of mass and by the use of relative variables
in the instantaneous Wigner 3-spaces and not quantum non-locality in
the absolute Euclidean 3-space of Galilei space-time. Both
properties are consequences of the Lorentz signature of Minkowski
space-time and of the structure of the Poincare' group whose
generators are non-local quantities knowing the whole instantaneous
3-space (moreover with the Lorentz boosts interaction-dependent
differently from the Galilei boosts). These properties of
relativistic entanglement disappear as $1/c$ effects in the
non-relativistic limit.
\bigskip

The future developments of the research will be:
\medskip

A) The extention of the calculations of Appendix B for the
quantization of charged particles with mutual Coulomb plus Darwin
interaction to include positive energy spin-one-half particles.
\bigskip

B) The standard quantization of the radiation field in the radiation
gauge (see paper I), in the transverse Fock space ${\cal H}$ with
creation and annhilation operators ${\hat
a}^{\dagger}_{\lambda}(\vec k)$, ${\hat a}_{\lambda}(\vec k)$,
$\lambda = 1,2$, followed to its reduction to the rest-frame instant
form of dynamics. The physical reduced Fock space ${\cal H}_{phys}$
has to be defined by imposing the conditions $<{\hat {\cal P}}^r> =
<{1\over c}\, \sum_{\lambda =1,2}\, \int d\tilde k\,  k^r\, {\hat
a}^{\dagger}_{\lambda}(\vec k)\, {\hat a}_{\lambda}(\vec k)> = 0$
and $<{\hat {\cal K}}^r> = <{i\over c}\, \sum_{\lambda = 1,2}\, \int
d\tilde k\, {\hat a}^{\dagger}_{\lambda}(\vec k)\, \omega(\vec k)\,
{{\partial}\over {\partial\, k^r}}\, {\hat a}_{\lambda}(\vec k) +
{i\over {2c}}\, \sum_{\lambda, \lambda^{'} = 1,2}\, \int d\tilde k\,
\Big[{\hat a}_{\lambda}(\vec k)\, {\hat
a}^{\dagger}_{\lambda^{'}}(\vec k) - {\hat
a}^{\dagger}_{\lambda}(\vec k)\, {\hat a}_{\lambda^{'}}(\vec k)\Big]
{\vec \epsilon}_{\lambda}(\vec k) \cdot \omega(\vec k)\,
{{\partial\, {\vec \epsilon}_{\lambda^{'}}(\vec k)}\over {\partial\,
k^r}}>= 0$ [see Eq.(II-3.2); $d\tilde k = d^3k/ 2 \omega(\vec k)\,
(2 \pi)^3$; $\omega(\vec k) = |\vec k|$].\medskip

If ${\cal H}_{phys}$ is well defined and can be explicitly
constructed, this method would be a first definition of the
quantization of the modulus-phase variables of II with the
elimination of the un-observable global phase (only relative phases
can be measured) described by the internal 3-center of mass (it is a
3-center of phase) of the field configuration on the instantaneous
Wigner 3-spaces \footnote{For fermion fields, which must be
Grassmann-valued to become anti-commuting fields after quantization,
it is still an open problem how to eliminate the internal 3-center
of mass, because action-angle variables cannot be defined for
fermion fields.}. See the reviews of Refs. \cite{69} and Ref.
\cite{3} for the obstruction to quantize angles and phases. If phase
could be quantized, then  we could quantize the relative variables
of Eq.(II-3.10) with Hamiltonian $M_{rad}\, c = {\cal
P}^{\tau}_{rad}$ of Eq.(II-3.2)\footnote{Without fixing the gauge
$X^{\tau}_{rad} \approx \pm \tau$ where $X^{\tau}_{rad}$ is the
phase center conjugate to ${\cal P}^{\tau}_{rad}$, i.e. conjugated
to the Hamiltonian.} and to get the quantum theory defined in the
Hilbert space ${\cal H}_{com} \otimes {\cal H}_{Fock\, rel}$.

\bigskip

C) If the previous quantization of the transverse radiation field
would work, then we could study the first quantization of the
positive-energy particles with Coulomb plus Darwin mutual
interaction together with a second quantized transverse radiation
field in the rest-frame instant form, i.e. of the system obtained in
I after the canonical transformation.

If the inverse (I-3.10) of the canonical transformation (I-3.6)
\footnote{It is neither a coordinate- nor momentum- point
transformation.} would be unitarily implementable after this
quantization, we would get a definition of positive-energy charged
quantum particles with mutual Coulomb interaction coupled to a
transverse (not radiation) electro-magnetic field in the radiation
gauge. Therefore by construction we would get that this fixed-
particle- number semi-classical approximation admits a quantum
interaction picture description unitarily equivalent to a QM of
mutually interacting dressed- particle system plus an "IN" second
quantized free radiation field kinematically connected by the
rest-frame conditions \footnote{A consequence of the clock
synchronization convention needed to formulate a Cauchy problem for
the isolated system.}.

\bigskip

D) Finally, as a preliminary step in the study of the properties of
protocols like teleportation from the space station to an earth
station requiring the theory of relativistic entanglement, we have
to rephrase non-relativistic entanglement in the rest-frame instant
form after the elimination of the center of mass so that the theory
depends only on relative variables \footnote{We can try to
reformulate the non-relativistic theory of entanglement for quantum
non-relativistic N-particles systems in the rest-frame framework
developed in Section III to mimic relativistic rest-frame QM. We
have to identify the constants of motion (whose conjugate variables
cannot be measured) and to understand which information is lost if
the center of mass is considered as a global not locally accessible
quantity. In a 2-body problem with canonical relative variables
$\vec \rho$, $\vec \pi$, the relative momentum $\vec \pi$ is a
constant of motion in the free case, but not in the interacting one.
In a 3-body problem with canonical relative variables ${\vec
\rho}_1$, ${\vec \pi}_1$, ${\vec \rho}_2$, ${\vec \pi}_2$, with 1
and 2 interacting and 3 free we must choose a canonical basis of
relative variables such that a) ${\vec \rho}_1 = {\vec \eta}_1 -
{\vec \eta}_2$ and b) ${\vec \pi}_{(12)3}$ is a constant of motion.
Therefore there will be preferred canonical bases of relative
variables selected by the kind of interactions existing among the
particles and including the maximal existing set of constants of
motion. There are analogies with molecular physics, where the
non-relativistic Jacobi bases of relative variables are used: one
chooses the Jacobi basis adapted to the dominating bonds and treats
the other bonds perturbatively  (see Refs.\cite{66}).}.\medskip

The limitation of the approach is that to get an isolated system we
must use a mixing of macroscopic and microscopic objects without
knowing which is the "relevant effective" description of the
macro-objects needed to describe the observers and their
instruments. In the isolated system {\it physical system + observer
1 + observer 2 + (particles of the experimental protocol) +
environment} the observers (measuring apparatuses or Alice and Bob)
have to be described as quasi-classical systems. However the spatial
non-separability implies that they must be described by relative
variables which interconnect them with the microscopic physical
system and with the environment. With macroscopic bodies the
constraints ${\vec {\cal K}}_{(int)} \approx 0$ are probably
dominated by the approximate solution ${\vec \eta}_+ \approx 0$
(with corrections depending on the interactions; ${\vec \eta}_+$ is
the internal 3-center of mass) so that the use of the separable
unphysical Hilbert space $\otimes_i\, {\cal H}_{{\vec \eta}_i}$
becomes an acceptable approximation.

\vfill\eject

\appendix

\section{Darwin Potential in the Unequal Mass Case}

From Eq.(I-4.5) the Darwin potential has the following expression

\begin{eqnarray}
&&V_{DARWIN}({{\vec{\eta}}}_{1}(\tau )-{{\vec{\eta}}}_{2}(\tau );{{
\vec{\kappa}}}_{i}(\tau ))
=\sum_{i\not=j}^{1,2}\,Q_{i}\,Q_{j}\,\Big({\frac{{{\
{\vec{\kappa}}}_{i}\cdot {{\vec{A}}}_{\perp Sj}({{\vec{ \eta
}}}_{i}(\tau) - {\vec \eta}_j(\tau), {\vec
\kappa}_j(\tau))}}{\sqrt{m_{i}^{2}\,c^{2}+{{\vec{\kappa}}}_{i}^{2}}}}
+  \nonumber \\
&+&\int d^{3}\sigma \,\Big[{\frac{1}{2}}\,\Big({{\vec{\pi} }}_{\perp
Si}(\vec{\sigma} - {\vec{\eta}_i}, {\vec{\kappa}_i})\cdot
{{\vec{\pi}}}_{\perp Sj}(\vec{\sigma} - {\vec{\eta}_j},
{\vec{\kappa}_j}) + {{\vec{B}}}_{Si}(\vec{\sigma} - {\vec{\eta}_i},
{\vec{\kappa}_i})\cdot {\ {\vec{ B}}}_{Sj}(\vec{\sigma} -
{\vec{\eta}_j}, {\vec{\kappa}_j})\Big)+  \nonumber \\
&+&({\frac{{{{\vec{\kappa}}}_{i}}}{\sqrt{m_{i}^{2}\,c^{2}+{{\vec{
\kappa}}}_{i}^{2}}}}\cdot {\frac{{\partial }}{{\partial
\,{{\vec{\eta}}} _{i}}}})\, \Big({{\vec{A}}}_{\perp Si}(\vec{\sigma}
- {\vec{\eta}_i}, {\vec{\kappa}_i})\cdot {{\vec{\pi}}} _{\perp
Sj}(\vec{\sigma} - {\vec{\eta}_j}, {\vec{\kappa}_j}) -\nonumber \\
  &-&{{\vec{\pi}}}_{\perp Si}(\vec{\sigma} -
{\vec{\eta}_i}, {\vec{\kappa}_i})\cdot {{\vec{A}}}_{\perp
Sj}(\vec{\sigma} -
{\vec{\eta}_j}, {\vec{\kappa}_j})\Big) \Big]\Big).  \nonumber \\
&&{}  \label{a1}
\end{eqnarray}

\noindent with the following form of the Lienard-Wiechert fields
[see  Eqs. (I-2.51), (I-2.52) and (I-2.53)]

\begin{eqnarray}
&&{\vec{A}}_{\perp S}(\tau ,\vec{\sigma})\, {\buildrel \circ \over
{{=}}}\, \sum_{i=1}^2\, Q_{i}{\vec{A}} _{\perp
Si}(\vec{\sigma}-{\vec{\eta}}_{i}(\tau
),{\vec{\kappa}}_{i}(\tau )),  \nonumber \\
&&{}  \nonumber \\
&&\vec{A}_{\perp Si}(\vec{\sigma} -
\vec{\eta}_{i},{\vec{\kappa}}_{i}) = {\ \frac{1}{4\pi |\vec{\sigma}
- \vec{\eta}_{i}|}} {\frac{1}{{\sqrt{m^2_i\, c^2 + {\vec
\kappa}^2_i} + \sqrt{ m_{i}^{2}\, c^2 + (\vec{\kappa}_{i} \cdot {\
\frac{{\vec{\sigma} - \vec{\eta}_{i}}}{{| \vec{\sigma} -
\vec{\eta}_{i}|}}}
)^{2}}}}}\times  \nonumber \\
&&\Big[{\vec \kappa}_i + {\frac{{[\vec{\kappa}_{i} \cdot
(\vec{\sigma} - \vec{\eta}_{i})]\, (\vec{ \sigma} -
\vec{\eta}_{i})}}{{|\vec{\sigma} - \vec{ \eta}_{i}|^{2}}}}\,
{\frac{\sqrt{ m_{i}^{2}\, c^2 + {{\vec{\kappa}_{i}}^{2}}}
}{\sqrt{m_{i}^{2}\, c^2 + (\vec{\kappa}_{i} \cdot {\
\frac{{\vec{\sigma} - \vec{\eta}_{i}}}{{|\vec{\sigma} -
\vec{\eta}_{i}|}}})^{2}}}} \Big], \label{a2}
\end{eqnarray}
\bigskip

\begin{eqnarray*}
\vec{E}_{\perp S}(\tau ,\vec{\sigma}) &=&{\vec{\pi}}_{\perp S}(\tau
,\vec{ \sigma}) = - {\frac{\partial \vec{A}_{\perp S}(\tau
,\vec{\sigma})}{\partial \tau }} =   \sum_{i=1}^2\, Q_{i}\,
{\vec{\pi}}_{\perp Si}(\vec{\sigma} - {\vec{\eta} }_{i}(\tau
),{\vec{\kappa}}_{i}(\tau )) =
\end{eqnarray*}

\begin{eqnarray*}
&=&\sum_{i=1}^2\, Q_{i}\, {\frac{{{\vec{\kappa}}_{i}(\tau ) \cdot
{\vec{\partial}} _{\sigma }}}{\sqrt{m_{i}^{2}\, c^2 +
{\vec{\kappa}}_{i}^{2}(\tau ) }}}\, {\vec{A}} _{\perp
Si}(\vec{\sigma}-{\vec{\eta}}_{i}(\tau
),{\vec{\kappa }}_{i}(\tau )) =  \nonumber \\
&=& - \sum_{i=1}^2\, Q_{i} \times  \nonumber \\
&&{\frac{1}{{4\pi |\vec{\sigma} - {\vec{\eta}}_{i}(\tau )|^{2}}}}\,
\Big[{\ \vec{ \kappa}}_{i}(\tau )\, [{\vec{\kappa}}_{i}(\tau ) \cdot
{\frac{{\vec{ \sigma} - {\ \vec{\eta}}_{i}(\tau )}}{{|\vec{\sigma} -
{\vec{\eta}}_{i}(\tau )|}}}]\, {\frac{ \sqrt{m_{i}^{2}\, c^2 +
{\vec{\kappa}}_{i}^{2}(\tau )}}{ [m_{i}^{2}\, c^2 + ({\vec{\kappa}}
_{i}(\tau ) \cdot {\frac{{\vec{\sigma} - { \ \vec{\eta}}_{i}(\tau
)}}{{|\vec{\sigma} - {\vec{\eta}}_{i}(\tau )|}}}
)^{2}]^{3/2}}}+  \nonumber \\
&+& {\frac{{\vec{\sigma} - {\vec{\eta}}_{i}(\tau )}}{{|\vec{\sigma}
- {\vec{ \eta}} _{i}(\tau )|}}}\, \Big({\frac{{\
{\vec{\kappa}}_{i}^{2}(\tau ) + ({\ \vec{\kappa}} _{i}(\tau ) \cdot
{\frac{{\vec{\sigma} - {\vec{\eta}}_{i}(\tau )}}{{|\vec{\sigma } -
{\vec{\eta}}_{i}(\tau )|}}})^{2}}}{{{\vec{\kappa}} _{i}^{2}(\tau ) -
({\vec{ \kappa}}_{i}(\tau ) \cdot {\frac{{\vec{\sigma} - { \
\vec{\eta}}_{i}(\tau )}}{{| \vec{\sigma} - {\vec{\eta}}_{i}(\tau
)|}}})^{2} }} }\, ({\frac{\sqrt{m_{i}^{2}\, c^2 +
{\vec{\kappa}}_{i}^{2}(\tau )}}{\sqrt{ m_{i}^{2}\, c^2 +
({\vec{\kappa}}_{i}(\tau ) \cdot {\frac{{\vec{\sigma} - {\
\vec{\eta}}_{i}(\tau )}}{{|\vec{\sigma} - {\vec{ \eta}}_{i}(\tau
)|}}})^{2}}}} - 1) +
\end{eqnarray*}

\bea &+& {\frac{{({\vec{\kappa}}_{i}(\tau ) \cdot
{\frac{{\vec{\sigma} - {\vec{ \eta}} _{i}(\tau )}}{{|\vec{\sigma} -
{\vec{\eta}}_{i}(\tau )|}}})^{2}\, \sqrt{m_{i}^{2}\, c^2 +
{\vec{\kappa}}_{i}^{2}(\tau )}}}{{[m_{i}^{2}\, c^2 +
({\vec{\kappa}}_{i}(\tau ) \cdot {\frac{{\vec{\sigma} - {\vec{\eta}}
_{i}(\tau )}}{{|\vec{\sigma} - {\vec{\eta}} _{i}(\tau
)|}}})^{2}\,]^{3/2}}}} \Big)\Big],
  \label{a3}
\end{eqnarray}

\bigskip

\begin{eqnarray}
\vec{B}_{S}(\tau ,\vec{\sigma}) &=& \vec \partial \times {\vec
A}_{\perp S}(\tau ,\vec \sigma ) =\sum_{i=1}^2\, Q_{i}\,
{\vec{B}}_{Si}( \vec{\sigma
} - {\vec{\eta}}_{i}(\tau ), {\vec{\kappa}}_{i}(\tau ))=  \nonumber \\
&=&\sum_{i=1}^2\, Q_{i}\, {\frac{1}{{4\pi |\vec{\sigma} -
{\vec{\eta}} _{i}(\tau )|^{2}}}}\, {\frac{{m_{i}^{2}\, c^2\,
{\vec{\kappa}}_{i}(\tau ) \times {\frac{{\vec{ \sigma} -
{\vec{\eta}}_{i}(\tau )}}{{|\vec{\sigma} - {\ \vec{\eta}}_{i}(\tau
)|}}}} }{{[m_{i}^{2}\, c^2 + ({\vec{\kappa}}_{i}(\tau ) \cdot
{\frac{{\vec{\sigma} - {\vec{\eta }}_{i}(\tau )}}{{|\vec{\sigma} -
{\ \vec{\eta}}_{i}(\tau )|}}})^{2}\,]^{3/2}}}}. \label{a4}
\end{eqnarray}
\medskip

\noindent From Eq.(I-3.5) we get the following form  of the function
${\cal K}_{ij}(\tau)$

\beq {\cal K}_{12}(\tau) = \int d^3\sigma\, \Big[{\vec A}_{\perp S1}
\cdot {\vec \pi}_{\perp S2} - {\vec \pi}_{\perp S1} \cdot {\vec
A}_{\perp S2}\Big](\tau, \vec \sigma). \label{a5} \eeq

\bigskip

The internal Poincare' algebra closes without using the rest-frame
condition ${\vec {\cal P}}_{(int)} \approx 0$.
\medskip

By using Eq.(II-2.4) of II, the vanishing of the internal boost in
Eq.(\ref{5.4}) gives the following form of ${\vec \eta}_{12}(\tau)$
(${\vec \kappa}_i \approx (-)^{i+1}\, {\vec \pi}_{12}$)

\begin{eqnarray*}
  {{\vec \eta}}_{12}\, &=& \Big[\sum_{i=1}^2\,
\sqrt{m_{i}^{2}\, c^2 + {\vec \pi}_{12}^2} +  \nonumber \\
&+& {\frac{{Q_1\, Q_2}}{c}}\, \Big({\frac{{\ { {\vec \kappa}}_1
\cdot \Big[{\frac{1}{2}}\, {\vec \partial}_{{ {\vec \rho}}_{12}}\, {
{ \mathcal{K}}}_{12}({\vec \pi}_{12}, - {\vec \pi}_{12},  {\vec
\rho}_{12}) - 2\, {\vec A}_{\perp S2}({\vec \rho}_{12}, - {\vec
\pi}_{12}) \Big]}}{{2\, \sqrt{m_1^2\, c^2 + {
\vec \pi}^2_{12}}}}} +  \nonumber \\
&+& {\frac{{\ { {\vec \kappa}}_2 \cdot \Big[{\frac{1}{2}}\,
{\vec\partial}_{{ {\vec \rho}}_{12}}\, { {\mathcal{K}}}_{12}( {\vec
\pi}_{12}, - {\vec \pi}_{12}, {\vec \rho}_{12}) - 2\, { \vec
A}_{\perp S1}( {\vec \rho}_{12}, {\vec \pi}_{12}) \Big] }}{{2\,
\sqrt{m_2^2\, c^2 + {\vec{\pi}_{12}}^2}}}}\Big) \Big]^{-1}
\end{eqnarray*}

\begin{eqnarray*}
&\times&\Big( - { {\vec \rho}}_{12}\, \Big({\frac{{m_2}}{m}}\,
\sqrt{m_1^2\, c^2 + { {\vec \pi}}_{12}^2} - {\frac{{m_1}}{m}}\,
\sqrt{m_2^2\, c^2 + { {\vec \pi}}_{12}^2} +  \nonumber \\
&+& {\frac{{Q_1\, Q_2}}{c}}\, \Big[{\frac{{\ m_2\, { {\vec
\pi}}_{12} \cdot \Big[{\frac{1}{2}}\, {\vec \partial}_{{ {\vec
\rho}}_{12}}\, {  {\mathcal{K}}}_{12}({\vec \pi}_{12}, - {\vec \pi}
_{12}, {\vec \rho}_{12}) - 2\, {\vec A}_{\perp S2}( {\vec
\rho}_{12}, - {\vec \pi}_{12}) \Big] }}{{2\, m\, \sqrt{m_1^2\,
c^2 + {\vec \pi}_{12}^2} }}} +  \nonumber \\
&+& {\frac{{\ m_1\, {\vec \pi}_{12} \cdot \Big[{\frac{1}{2}}\, {
\vec \partial}_{{ {\vec \rho}}_{12}}\, { {\mathcal{K}}}_{12}({\vec
\pi}_{12}, - {\vec \pi}_{1}2,  {\vec \rho} _{12}) - 2\, {\vec
A}_{\perp S1}( {\vec \rho }_{12}, {\vec \pi}_{12}) \Big] }}{{2\, m\,
\sqrt{m_2^2\, c^2 +  {\vec \pi}_{12}^2}}}} \Big]\Big) -
\end{eqnarray*}

\bea &-& {\frac{1}{{2\, c}}}\, Q_1\, Q_2\, \Big(\sqrt{m_1^2\, c^2 +
{ {\vec \pi}}_{12}^2}\, {\vec \partial}_{{ {\vec \kappa}}_1} +
\sqrt{m_2^2\, c^2 + { {\vec \pi}}_{12}^2}\, {\vec
\partial}_{{ {\vec \kappa}} _2}\Big)\, {
{\mathcal{K}}}_{12}({ {\vec \kappa}}_1, { { \vec \kappa}}_2, { {\vec
\rho}}_{12}){|}_{{\vec \kappa}_1
= - {\vec \kappa}_2 = {\vec \pi}_{12}} -  \nonumber \\
&-& {\frac{{Q_1\, Q_2}}{{4\pi\, c}}}\, \int d^3\sigma\,
\Big({\frac{{{  {\vec \pi}}_{\perp S1}(\vec \sigma -
{\frac{{m_2}}{m}}\, { { \vec \rho}}_{12}, { {\vec
\pi}}_{12})}}{{|\vec \sigma + {\frac{{m_1}}{m }}\, { {\vec
\rho}}_{12}|}}} + {\frac{{{ {\vec \pi}}_{\perp S2}(\vec \sigma +
{\frac{{m_1}}{m}}\, { {\vec \rho}}_{12},  - { \vec
\pi}_{12})}}{{|\vec \sigma - {\frac{{m_2}}{m}}\, { {\vec
\rho}}_{12}|}}} \Big) -  \nonumber \\
&-& {\frac{{Q_1\, Q_2}}{c}}\, \int d^3\sigma\, \vec \sigma\,\,
\Big[{ { \vec \pi}}_{\perp S1}(\vec \sigma - {\frac{{m_2}}{m}}\, {
{\vec \rho}} _{12}, {\vec \pi}_{12}) \cdot { {\vec \pi}}_{\perp
S2}(\vec \sigma + {\frac{{m_1}}{m}}\, { {\vec \rho}}_{12}, {\vec
\pi}_{12}) +  \nonumber \\
&+& { {\vec B}}_{S1}(\vec \sigma - {\frac{{m_2}}{m}}\, { {\vec
\rho}}_{12}, {\vec \pi}_{12}) \cdot { {\vec B}}_{S2}(\vec \sigma +
{\frac{{m_1}}{m}}\, { {\vec \rho}}_{12}, - {\vec \pi}_{12}) \Big]
\,\, \Big).
  \label{a6}
\end{eqnarray}

\bigskip

By eliminating ${\vec \eta}_{12}$ and ${\vec \kappa}_{12} \approx 0$
we get the following form of the Darwin potential and of the
Lienard-Wiechert quantities

\bea &&{\tilde V}_{DARWIN}({\vec \rho}_{12}, {\vec \pi}_{12})
=\nonumber \\
&&{}\nonumber \\
&=&Q_1\, Q_2\, \Big({{{\vec \pi}_{12}(\tau) \cdot {\vec A}_{\perp
S2}({\vec \rho}_{12}(\tau), - {\vec \pi}_{12}(\tau))}\over
{\sqrt{m_1^2\, c^2 + {\vec \pi}_{12}^2(\tau)}}} - {{{\vec
\pi}_{12}(\tau) \cdot {\vec A}_{\perp S2}(- {\vec \rho}_{12}(\tau),
{\vec \pi}_{12}(\tau))}\over
{\sqrt{m_2^2\, c^2 + {\vec \pi}_{12}^2(\tau)}}} +\nonumber \\
&+& \int d^3\sigma\, \Big[{{{m\over {m_2}}}\over {\sqrt{m_1^2\, c^2
+ {\vec \pi}_{12}^2(\tau)}}}\, \Big( \Big(\Big[{\vec \pi}_{12}(\tau)
\cdot {{\partial}\over {\partial\, {\vec \rho}_{12}}}\Big]\, {\vec
A}_{\perp S1}(\vec \sigma - {{m_2}\over m}\, {\vec \rho}_{12}(\tau),
{\vec \pi}_{12}(\tau))\Big) \cdot
\nonumber \\
&&\cdot {\vec \pi}_{\perp S2}(\vec \sigma + {{m_1}\over m}\, {\vec
\rho}_{12}(\tau),
- {\vec \pi}_{12}(\tau)) -\nonumber \\
&-& \Big(\Big[{\vec \pi}_{12}(\tau) \cdot {{\partial}\over
{\partial\, {\vec \rho}_{12}}}\Big]\, {\vec \pi}_{\perp S1}(\vec
\sigma - {{m_2}\over m}\, {\vec \rho}_{12}(\tau), {\vec
\pi}_{12}(\tau))\Big) \cdot {\vec A}_{\perp S2}(\vec \sigma +
{{m_1}\over m}\, {\vec \rho}_{12}(\tau),
- {\vec \pi}_{12}(\tau)) \Big) +\nonumber \\
&+& {{{m\over {m_1}}}\over {\sqrt{m_2^2\, c^2 + {\vec
\pi}_{12}^2(\tau)}}}\, \Big( \Big(\Big[{\vec \pi}_{12}(\tau) \cdot
{{\partial}\over {\partial\, {\vec \rho}_{12}}}\Big]\, {\vec
A}_{\perp S2}(\vec \sigma + {{m_1}\over m}\, {\vec \rho}_{12}(\tau),
- {\vec \pi}_{12}(\tau))\Big) \cdot
\nonumber \\
&&\cdot {\vec \pi}_{\perp S1}(\vec \sigma - {{m_2}\over m}\, {\vec
\rho}_{12}(\tau),
{\vec \pi}_{12}(\tau)) -\nonumber \\
&-& \Big(\Big[{\vec \pi}_{12}(\tau) \cdot {{\partial}\over
{\partial\, {\vec \rho}_{12}}}\Big]\, {\vec \pi}_{\perp S2}(\vec
\sigma + {{m_1}\over m}\, {\vec \rho}_{12}(\tau), - {\vec
\pi}_{12}(\tau))\Big) \cdot {\vec A}_{\perp S1}(\vec \sigma -
{{m_2}\over m}\, {\vec \rho}_{12}(\tau),
{\vec \pi}_{12}(\tau)) \Big)\,\,  +\nonumber \\
&+& {\vec \pi}_{\perp S1}(\vec \sigma - {{m_2}\over m}\, {\vec
\rho}_{12}, {\vec \pi}_{12}) \cdot {\vec \pi}_{\perp S2}(\vec \sigma
+ {{m_1}\over m}\,
{\vec \rho}_{12}, - {\vec \pi}_{12}) +\nonumber \\
&+& {\vec B}_{S1}(\vec \sigma - {{m_2}\over m}\, {\vec \rho}_{12},
{\vec \pi}_{12}) \cdot {\vec B}_{S2}(\vec \sigma + {{m_1}\over m}\,
{\vec \rho}_{12}, - {\vec \pi}_{12}) \Big]\Big)(\tau), \label{a7}
\eea

\bigskip

\begin{eqnarray*}
&&{\vec A}_{\perp S1}(- {\vec \rho}_{12}, {\vec \pi}_{12}) = {1\over
{4\pi\, |{\vec \rho}_{12}|}}\, {1\over {\sqrt{m_1^2\, c^2 + {\vec
\pi}_{12}^2} + \sqrt{m_1^2\, c^2 + \Big({\vec \pi}_{12} \cdot
{{{\vec \rho}_{12}}\over {|{\vec \rho}_{12}|}}\Big)^2}}}\nonumber \\
&&\Big[{\vec \pi}_{12} + {{({\vec \pi}_{12} \cdot {\vec
\rho}_{12})\, {\vec \rho}_{12}}\over {|{\vec \rho}_{12}|^2}}\,
{{\sqrt{m_1^2\, c^2 + {\vec \pi}_{12}^2}}\over {\sqrt{m_1^2\, c^2 +
\Big({\vec \pi}_{12} \cdot
{{{\vec \rho}_{12}}\over {|{\vec \rho}_{12}|}}\Big)^2}}} \Big],\nonumber \\
&&{}\nonumber \\
  &&{\vec A}_{\perp S2}( {\vec \rho}_{12}, - {\vec \pi}_{12}) = -
{1\over {4\pi\, |{\vec \rho}_{12}|}}\, {1\over {\sqrt{m_2^2\, c^2 +
{\vec \pi}_{12}^2} + \sqrt{m_2^2\, c^2 + \Big({\vec \pi}_{12} \cdot
{{{\vec \rho}_{12}}\over {|{\vec \rho}_{12}|}}\Big)^2}}}\nonumber \\
&&\Big[{\vec \pi}_{12} + {{({\vec \pi}_{12} \cdot {\vec
\rho}_{12})\, {\vec \rho}_{12}}\over {|{\vec \rho}_{12}|^2}}\,
{{\sqrt{m_2^2\, c^2 + {\vec \pi}_{12}^2}}\over {\sqrt{m_2^2\, c^2 +
\Big({\vec \pi}_{12} \cdot {{{\vec \rho}_{12}}\over {|{\vec
\rho}_{12}|}}\Big)^2}}} \Big],
\end{eqnarray*}

\begin{eqnarray*}
&&{\vec A}_{\perp S1}(\vec \sigma - {{m_2}\over m}\, {\vec
\rho}_{12}, {\vec \pi}_{12}) =
{1\over {4\pi\, |\vec \sigma - {{m_2}\over m}\, {\vec \rho}_{12}|}}\nonumber \\
&&{1\over {\sqrt{m_1^2\, c^2 + {\vec \pi}_{12}^2} + \sqrt{m_1^2\,
c^2 + \Big({\vec \pi}_{12} \cdot {{\vec \sigma - {{m_2}\over m}\,
{\vec \rho}_{12}}\over {|\vec \sigma - {{m_2}\over m}\, {\vec
\rho}_{12}|}}\Big)^2}}}\,
\Big[{\vec \pi}_{12} +\nonumber \\
&+& {{{\vec \pi}_{12} \cdot \Big(\vec \sigma - {{m_2}\over m}\,
{\vec \rho}_{12}\Big)\, \Big(\vec \sigma - {{m_2}\over m}\, {\vec
\rho}_{12}\Big)}\over {|\vec \sigma - {{m_2}\over m}\, {\vec
\rho}_{12}|^2}}\, {{\sqrt{m_1^2\, c^2 + {\vec \pi}_{12}^2}}\over
{\sqrt{m_1^2\, c^2 + \Big({\vec \pi}_{12} \cdot {{\vec \sigma -
{{m_2}\over m}\, {\vec \rho}_{12}}\over {|\vec \sigma - {{m_2}\over
m}\, {\vec \rho}_{12}|}}
\Big)^2}}}\Big], \nonumber \\
&&{}\nonumber \\
&&{\vec A}_{\perp S2}(\vec \sigma + {{m_1}\over m}\, {\vec
\rho}_{12}, - {\vec \pi}_{12}) =
- {1\over {4\pi\, |\vec \sigma + {{m_1}\over m}\, {\vec \rho}_{12}|}}\nonumber \\
&&{1\over {\sqrt{m_2^2\, c^2 + {\vec \pi}_{12}^2} + \sqrt{m_2^2\,
c^2 + \Big({\vec \pi}_{12} \cdot {{\vec \sigma + {{m_1}\over m}\,
{\vec \rho}_{12}}\over {|\vec \sigma + {{m_1}\over m}\, {\vec
\rho}_{12}|}}\Big)^2}}}\,
\Big[{\vec \pi}_{12} +\nonumber \\
&+& {{{\vec \pi}_{12} \cdot \Big(\vec \sigma + {{m_1}\over m}\,
{\vec \rho}_{12}\Big)\, \Big(\vec \sigma + {{m_1}\over m}\, {\vec
\rho}_{12}\Big)}\over {|\vec \sigma + {{m_1}\over m}\, {\vec
\rho}_{12}|^2}}\, {{\sqrt{m_2^2\, c^2 + {\vec \pi}_{12}^2}}\over
{\sqrt{m_2^2\, c^2 + \Big({\vec \pi}_{12} \cdot {{\vec \sigma +
{{m_1}\over m}\, {\vec \rho}_{12}}\over {|\vec \sigma + {{m_1}\over
m}\, {\vec \rho}_{12}|}} \Big)^2}}}\Big],
\end{eqnarray*}

\begin{eqnarray*}
&&{\vec \pi}_{\perp S1}(- {\vec \rho}_{12}, {\vec \pi}_{12}) =  {1
\over {4\pi\, |{\vec \rho}_{12}|^2}}\, \Big({\vec \pi}_{12}\, {\vec
\pi}_{12} \cdot {{{\vec \rho}_{12}}\over {|{\vec \rho}_{12}|}}\,
{{\sqrt{m_1^2\, c^2 + {\vec \pi}_{12}^2}}\over {\Big[m_1^2\, c^2 +
\Big({\vec \pi}_{12} \cdot {{{\vec \rho}_{12}}\over {|{\vec
\rho}_{12}|}}\Big)^2\Big]^{3/2}}} +\nonumber \\
&+&{{{\vec \rho}_{12}}\over {|{\vec \rho}_{12}|}}\, \Big[ {{{\vec
\pi}_{12}^2 + \Big({\vec \pi}_{12} \cdot {{{\vec \rho}_{12}}\over
{|{\vec \rho}_{12}|}}\Big)^2}\over {{\vec \pi}_{12}^2 - \Big({\vec
\pi}_{12} \cdot {{{\vec \rho}_{12}}\over {|{\vec
\rho}_{12}|}}\Big)^2}}\, \Big({{\sqrt{m_1^2\, c^2 + {\vec
\pi}_{12}^2}}\over {\sqrt{m_1^2\, c^2 + \Big({\vec \pi}_{12} \cdot
{{{\vec \rho}_{12}}\over {|{\vec \rho}_{12}|}}\Big)^2}}}  - 1\Big) +\nonumber \\
&+& {{\Big({\vec \pi}_{12} \cdot {{{\vec \rho}_{12}}\over {|{\vec
\rho}_{12}|}}\Big)^2\, \sqrt{m_1^2\, c^2 + {\vec \pi}_{12}^2}}\over
{\Big[m_1^2\, c^2 + \Big({\vec \pi}_{12} \cdot {{{\vec
\rho}_{12}}\over {|{\vec \rho}_{12}|}}\Big)^2\Big]^{3/2}}}
\Big]\Big),\nonumber \\
&&{}\nonumber \\
  &&{\vec \pi}_{\perp S2}( {\vec \rho}_{12}, - {\vec \pi}_{12}) = -  {1
\over {4\pi\, |{\vec \rho}_{12}|^2}}\, \Big({\vec \pi}_{12}\, {\vec
\pi}_{12} \cdot {{{\vec \rho}_{12}}\over {|{\vec \rho}_{12}|}}\,
{{\sqrt{m_2^2\, c^2 + {\vec \pi}_{12}^2}}\over {\Big[m_2^2\, c^2 +
\Big({\vec \pi}_{12} \cdot {{{\vec \rho}_{12}}\over {|{\vec
\rho}_{12}|}}\Big)^2\Big]^{3/2}}} +\nonumber \\
&+&{{{\vec \rho}_{12}}\over {|{\vec \rho}_{12}|}}\, \Big[ {{{\vec
\pi}_{12}^2 + \Big({\vec \pi}_{12} \cdot {{{\vec \rho}_{12}}\over
{|{\vec \rho}_{12}|}}\Big)^2}\over {{\vec \pi}_{12}^2 - \Big({\vec
\pi}_{12} \cdot {{{\vec \rho}_{12}}\over {|{\vec
\rho}_{12}|}}\Big)^2}}\, \Big({{\sqrt{m_2^2\, c^2 + {\vec
\pi}_{12}^2}}\over {\sqrt{m_2^2\, c^2 + \Big({\vec \pi}_{12} \cdot
{{{\vec \rho}_{12}}\over {|{\vec \rho}_{12}|}}\Big)^2}}}  - 1\Big) +\nonumber \\
&+& {{\Big({\vec \pi}_{12} \cdot {{{\vec \rho}_{12}}\over {|{\vec
\rho}_{12}|}}\Big)^2\, \sqrt{m_2^2\, c^2 + {\vec \pi}_{12}^2}}\over
{\Big[m_2^2\, c^2 + \Big({\vec \pi}_{12} \cdot {{{\vec
\rho}_{12}}\over {|{\vec \rho}_{12}|}}\Big)^2\Big]^{3/2}}}
\Big]\Big),
  \end{eqnarray*}

\begin{eqnarray*}
&&{\vec \pi}_{\perp S1}(\vec \sigma - {{m_2}\over m}\, {\vec
\rho}_{12}, {\vec \pi}_{12}) = - {1\over {|\vec \sigma - {{m_2}\over
m}\, {\vec \rho}_{12}|^2}}\, \Big({\vec \pi}_{12}\, {\vec \pi}_{12}
\cdot {{\vec \sigma - {{m_2}\over m}\, {\vec \rho}_{12}}\over {|\vec
\sigma - {{m_2}\over m}\, {\vec \rho}_{12}|}}\, {{\sqrt{m_1^2\, c^2
+ {\vec \pi}_{12}^2}}\over {\Big[m_1^2\, c^2\, + \Big({\vec
\pi}_{12} \cdot {{\vec \sigma - {{m_2}\over m}\, {\vec
\rho}_{12}}\over {|\vec \sigma - {{m_2}\over m}\, {\vec
\rho}_{12}|}}\Big)^2\Big]^{3/2}}} +\nonumber \\
&+& {{\vec \sigma - {{m_2}\over m}\, {\vec \rho}_{12}}\over {|\vec
\sigma - {{m_2}\over m}\, {\vec \rho}_{12}|}}\, \Big[ {{{\vec
\pi}_{12}^2 + \Big({\vec \pi}_{12} \cdot {{\vec \sigma - {{m_2}\over
m}\, {\vec \rho}_{12}}\over {|\vec \sigma - {{m_2}\over m}\, {\vec
\rho}_{12}|}}\Big)^2}\over {{\vec \pi}_{12}^2 - \Big({\vec \pi}_{12}
\cdot {{\vec \sigma - {{m_2}\over m}\, {\vec \rho}_{12}}\over {|\vec
\sigma - {{m_2}\over m}\, {\vec \rho}_{12}|}}\Big)^2}}\,
\Big({{\sqrt{m_1^2\, c^2 + {\vec\pi}_{12}^2}}\over {\sqrt{m_1^2\,
c^2 + \Big({\vec \pi}_{12} \cdot {{\vec \sigma - {{m_2}\over m}\,
{\vec \rho}_{12}}\over {|\vec \sigma -
{{m_2}\over m}\, {\vec \rho}_{12}|}}\Big)^2}}} - 1\Big) +\nonumber \\
&+& {{\Big({\vec \pi}_{12} \cdot {{\vec \sigma - {{m_2}\over m}\,
{\vec \rho}_{12}}\over {|\vec \sigma - {{m_2}\over m}\, {\vec
\rho}_{12}|}}\Big)^2\, \sqrt{m_1^2\, c^2 + {\vec \pi}_{12}^2}}\over
{\Big[m_1^2\, c^2 + \Big({\vec \pi}_{12} \cdot {{\vec \sigma -
{{m_2}\over m}\, {\vec \rho}_{12}}\over {|\vec \sigma - {{m_2}\over
m}\, {\vec \rho}_{12}|}}\Big)^2\Big]^{3/2}}}
\Big]\Big),\nonumber \\
&&{}\nonumber \\
  &&{\vec \pi}_{\perp S2}(\vec \sigma + {{m_1}\over m}\,
  {\vec \rho}_{12}, - {\vec \pi}_{12}) = -
{1\over {|\vec \sigma + {{m_1}\over m}\, {\vec \rho}_{12}|^2}}\,
\Big({\vec \pi}_{12}\, {\vec \pi}_{12} \cdot {{\vec \sigma +
{{m_1}\over m}\, {\vec \rho}_{12}}\over {|\vec \sigma + {{m_1}\over
m}\, {\vec \rho}_{12}|}}\, {{\sqrt{m_2^2\, c^2 + {\vec
\pi}_{12}^2}}\over {\Big[m_2^2\, c^2\, + \Big({\vec \pi}_{12} \cdot
{{\vec \sigma + {{m_1}\over m}\, {\vec \rho}_{12}}\over {|\vec
\sigma + {{m_1}\over m}\, {\vec
\rho}_{12}|}}\Big)^2\Big]^{3/2}}} +\nonumber \\
&+& {{\vec \sigma + {{m_1}\over m}\, {\vec \rho}_{12}}\over {|\vec
\sigma + {{m_1}\over m}\, {\vec \rho}_{12}|}}\, \Big[ {{{\vec
\pi}_{12}^2 + \Big({\vec \pi}_{12} \cdot {{\vec \sigma + {{m_1}\over
m}\, {\vec \rho}_{12}}\over {|\vec \sigma + {{m_1}\over m}\, {\vec
\rho}_{12}|}}\Big)^2}\over {{\vec \pi}_{12}^2 - \Big({\vec \pi}_{12}
\cdot {{\vec \sigma + {{m_1}\over m}\, {\vec \rho}_{12}}\over {|\vec
\sigma + {{m_1}\over m}\, {\vec \rho}_{12}|}}\Big)^2}}\,
\Big({{\sqrt{m_2^2\, c^2 + {\vec\pi}_{12}^2}}\over {\sqrt{m_2^2\,
c^2 + \Big({\vec \pi}_{12} \cdot {{\vec \sigma + {{m_1}\over m}\,
{\vec \rho}_{12}}\over {|\vec \sigma
+ {{m_1}\over m}\, {\vec \rho}_{12}|}}\Big)^2}}} - 1\Big) +\nonumber \\
&+& {{\Big({\vec \pi}_{12} \cdot {{\vec \sigma + {{m_1}\over m}\,
{\vec \rho}_{12}}\over {|\vec \sigma + {{m_1}\over m}\, {\vec
\rho}_{12}|}}\Big)^2\, \sqrt{m_2^2\, c^2 + {\vec \pi}_{12}^2}}\over
{\Big[m_2^2\, c^2 + \Big({\vec \pi}_{12} \cdot {{\vec \sigma +
{{m_1}\over m}\, {\vec \rho}_{12}}\over {|\vec \sigma + {{m_1}\over
m}\, {\vec \rho}_{12}|}}\Big)^2\Big]^{3/2}}} \Big]\Big),
\end{eqnarray*}

\bea &&{\vec B}_{S1}(\vec \sigma - {{m_2}\over m}\, {\vec
\rho}_{12}, {\vec \pi}_{12}) = {1\over {4\pi\, |\vec \sigma -
{{m_2}\over m}\, {\vec \rho}_{12}|^2}}\, {{m_1^2\, c^2\, {\vec
\pi}_{12} \times {{\vec \sigma - {{m_2}\over m}\, {\vec
\rho}_{12}}\over {|\vec \sigma - {{m_2}\over m}\, {\vec
\rho}_{12}|}}}\over {\Big[m_1^2\, c^2 + \Big({\vec \pi}_{12} \cdot
{{\vec \sigma - {{m_2}\over m}\, {\vec \rho}_{12}}\over
{|\vec \sigma - {{m_2}\over m}\, {\vec \rho}_{12}|}}\Big)^2\Big]^{3/2}}},\nonumber \\
&&{}\nonumber \\
&&{\vec B}_{S2}(\vec \sigma + {{m_1}\over m}\, {\vec \rho}_{12}, -
{\vec \pi}_{12}) = - {1\over {4\pi\, |\vec \sigma - {{m_2}\over m}\,
{\vec \rho}_{12}|^2}}\, {{m_2^2\, c^2\, {\vec \pi}_{12} \times
{{\vec \sigma + {{m_1}\over m}\, {\vec \rho}_{12}}\over {|\vec
\sigma + {{m_1}\over m}\, {\vec \rho}_{12}|}}}\over {\Big[m_2^2\,
c^2 + \Big({\vec \pi}_{12} \cdot {{\vec \sigma + {{m_1}\over m}\,
{\vec \rho}_{12}}\over {|\vec \sigma + {{m_1}\over m}\, {\vec
\rho}_{12}|}}\Big)^2\Big]^{3/2}}} \label{a8} \eea

\vfill\eject

\section{Weyl Ordering of the Invariant Mass of Two Equal Mass Particles with Mutual
Coulomb plus Darwin Interaction}

This appendix consists of two parts. \ In the first part we show
that the Weyl ordered form obtained from  the $O(1/c^{2})$ classical
Darwin interaction given in Eq. (\ref{5.6}) is identical to the self
adjoint version
\begin{equation}
\frac{Q_{1}Q_{1}}{8\pi m_{1}m_{2}c^{2}}[{\hat{\vec{\pi}}_{12}}\cdot \frac{1}{%
\left\vert {\hat{\vec{\rho}}_{12}}\right\vert }{\hat{\vec{\pi}}_{12}}+{\hat{%
\vec{\pi}}_{12}}\cdot {\hat{\vec{\rho}}_{12}}\frac{1}{\left\vert {\hat{\vec{%
\rho}}_{12}}\right\vert ^{3}}{\hat{\vec{\rho}}_{12}}\cdot {\hat{\vec{\pi}}%
_{12}].}
\end{equation}%
In the second part of this appendix we develop the Weyl quantization
of the exact equal mass Darwin Hamiltonian given in Eq. (\ref{5.5})
including the kinetic and Coulomb portions. An unusual and
unexpected result of that part of
this appendix is that the usual local form of the Coulomb potential is the $%
c\rightarrow \infty $ limit of the complete and nonlocal Coulomb
plus Darwin interactions.

\subsection{Weyl Ordering, the operator $\frac{1}{{|\hat{\vec{\protect%
\rho}}_{12}|}}$ and the Weyl Ordered Darwin Operators}

We wish to compare the Weyl ordered quantum operator corresponding to the $%
O(1/c^{2})$ classical Darwin interaction to the standard hermitian
form. \ Its classical form is from Eq. (\ref{5.6})
\begin{equation}
H_{D}=\frac{Q_{1}Q_{1}}{8\pi m_{1}m_{2}c^{2}}(\vec{\pi}_{12}^{2}\frac{1}{%
\rho _{12}}+\left( \vec{\pi}_{12}\cdot \vec{\rho}_{12}\right) ^{2}\frac{1}{%
\rho _{12}^{3}}).
\end{equation}%
(For simplicity of notation in this part of the appendix we use for
the
hatted quantum operators the abbreviations $\vec{\pi}={\hat{\vec{\pi}}_{12}},%
\vec{\rho}={\hat{\vec{\rho}}_{12}}$).$~$In order to use the Weyl
ordered
product for the Coulomb potential we replace its singular form with%
\begin{equation}
\frac{1}{\rho }\rightarrow \frac{1}{\mathfrak{r}},
\end{equation}%
where we define%
\begin{equation}
\mathfrak{r}=\sqrt{\rho ^{2}+\varepsilon ^{2}}=\sqrt{\rho
_{x}^{2}+\rho _{y}^{2}+\rho _{z}^{2}+\varepsilon ^{2}}.
\end{equation}%
This removes the singularity of this operator at the origin. \ For small $%
\varepsilon ,$ our results are independent of $\varepsilon $ and
reproduce
the known behaviors. \ Part of this follows from the form%
\begin{equation}
\nabla ^{2}\frac{1}{\mathfrak{r}}=\frac{1}{\rho }\frac{d^{2}}{d\rho ^{2}}%
\rho \frac{1}{\mathfrak{r}}=-\frac{3\varepsilon ^{2}}{\mathfrak{r}^{5}}=-%
\frac{3\varepsilon ^{2}}{(\rho ^{2}+\varepsilon ^{2})^{5/2}},
\end{equation}%
of the Laplacian. \ This equation is a particular form, for infinitesimal $%
\varepsilon ,$ of the Poisson equation for a point charge
\begin{equation}
\nabla ^{2}\frac{1}{\rho }=-4\pi \delta ^{3}(\vec{\rho}).
\end{equation}%
To see this notice that%
\begin{equation}
-3\varepsilon ^{2}\int \frac{d^{3}\rho }{(\rho ^{2}+\varepsilon ^{2})^{5/2}}%
=-4\pi .  \label{1e}
\end{equation}%
Thus,%
\begin{equation}
\underset{\varepsilon \rightarrow 0}{\lim }\frac{3\varepsilon
^{2}}{(\rho ^{2}+\varepsilon ^{2})^{5/2}}=4\pi \delta
^{3}(\vec{\rho}).  \label{1d}
\end{equation}

Now, we return to the determination of the Weyl ordered Darwin form
of the Hamiltonian. \ In rectangular coordinates the classical
Darwin interaction Eq. (\ref{5.6}) is
\begin{equation}
H_{D}=\frac{Q_{1}Q_{1}}{8\pi
m_{1}m_{2}c^{2}}[\frac{1}{\mathfrak{r}}\pi
_{x}^{2}+\frac{1}{\mathfrak{r}}\pi
_{y}^{2}+\frac{1}{\mathfrak{r}}\pi
_{z}^{2}+\frac{1}{\mathfrak{r}^{3}}(\rho _{x}\pi _{x}+\rho _{y}\pi
_{y}+\rho _{z}\pi _{z})^{2}].
\end{equation}%
We examine $\frac{1}{\mathfrak{r}}\pi _{z}^{2}$ first. The others in
the initial portion would be similarly treated. \ Let
\begin{equation}
\zeta ^{2}=\rho _{x}^{2}+\rho _{y}^{2}+\varepsilon ^{2}.
\end{equation}%
Then
\begin{equation}
\frac{1}{\mathfrak{r}}\pi _{z}^{2}=\frac{1}{\zeta }\sum_{n=0}^{\infty }%
\binom{-1/2}{n}\frac{\rho _{z}^{2n}}{\zeta ^{2n}}\pi _{z}^{2}.
\end{equation}%
The Weyl ordered quantum form of this is (see Eq. (2.41) in \cite{46} )%
\begin{equation}
\left( \rho _{z}^{2n}\pi _{z}^{2}\right) ^{W}=\frac{1}{2^{2n}}\sum_{m=0}^{2n}%
\binom{2n}{m}\rho _{z}^{2n-m}\pi _{z}^{2}\rho _{z}^{m}.
\end{equation}%
Use \footnote{%
In this Appendix we use the $\hbar =1$ convention.}%
\begin{equation}
\pi _{z}^{2}\rho _{z}^{m}=\rho _{z}^{m}\pi _{z}^{2}-2im\rho
_{z}^{m-1}\pi _{z}-m(m-1)\rho _{z}^{m-2},
\end{equation}%
and so%
\begin{equation}
\left( \frac{1}{\mathfrak{r}}\pi _{z}^{2}\right) ^{W}=\frac{1}{\zeta }%
\sum_{n=0}^{\infty }\frac{1}{(2\zeta )^{2n}}\binom{-1/2}{n}\sum_{m=0}^{2n}%
\binom{2n}{m}(\rho _{z}^{2n}\pi _{z}^{2}-2im\rho _{z}^{2n-1}\pi
_{z}-m(m-1)\rho _{z}^{2n-2}).
\end{equation}%
Perform the inner summations and we obtain%
\begin{eqnarray}
\left( \frac{1}{\mathfrak{r}}\pi _{z}^{2}\right) ^{W} &=&\frac{1}{\zeta }%
\sum_{n=0}^{\infty }\frac{1}{(\zeta )^{2n}}\binom{-1/2}{n}(\rho
_{z}^{2n}\pi _{z}^{2}-2ni\rho _{z}^{2n-1}\pi
_{z}-\frac{2n(2n-1)}{4}\rho _{z}^{2n-2})
\nonumber \\
&=&\frac{1}{\mathfrak{r}}\pi _{z}^{2}+\frac{i\rho
_{z}}{\mathfrak{r}^{3}}\pi
_{z}-\frac{1}{4\mathfrak{r}^{3}}+\frac{3}{4}\frac{\rho _{z}^{2}}{\mathfrak{r}%
^{5}}.
\end{eqnarray}%
By cyclic symmetry we thus have%
\begin{eqnarray}
\left( \frac{1}{\mathfrak{r}}\pi _{x}^{2}+\frac{1}{\mathfrak{r}}\pi _{y}^{2}+%
\frac{1}{\mathfrak{r}}\pi _{z}^{2}\right) ^{W} &=&\frac{1}{\mathfrak{r}}\vec{%
\pi}^{2}+i\frac{\vec{\rho}}{\mathfrak{r}^{3}}\cdot \vec{\pi}+\frac{3}{4%
\mathfrak{r}^{3}}-\frac{3}{4}\frac{\mathfrak{r}^{2}-\varepsilon ^{2}}{%
\mathfrak{r}^{5}}=\frac{1}{\mathfrak{r}}\vec{\pi}^{2}+i\frac{\vec{\rho}}{%
\mathfrak{r}^{3}}\cdot \vec{\pi}+\frac{3\varepsilon
^{2}}{4\mathfrak{r}^{5}}
\nonumber \\
&\rightarrow &\frac{1}{\rho }\vec{\pi}^{2}+i\frac{\vec{\rho}}{\rho ^{3}}%
\cdot \vec{\pi}+\pi \delta ^{3}(\vec{\rho}).
\end{eqnarray}%
On the other hand, we would obtain from the standard hermitean form%
\begin{equation}
\vec{\pi}\cdot \frac{1}{\rho }\vec{\pi}=\frac{1}{\rho }\vec{\pi}^{2}-i\frac{%
\vec{\rho}}{r^{3}}\cdot \vec{\pi}.
\end{equation}%
Thus,
\begin{equation}
\left( \frac{1}{\rho }\vec{\pi}^{2}\right) ^{W}-\vec{\pi}\cdot
\frac{1}{\rho
}\vec{\pi}=\frac{2i\vec{\rho}}{r^{3}}\cdot \vec{\pi}+\pi \delta ^{3}(\vec{%
\rho})).
\end{equation}

The remaining part of the Weyl ordered Darwin interaction has the
classical
form of%
\begin{equation}
\frac{\left( \rho _{z}\pi _{x}+\rho _{y}\pi _{y}+\rho _{z}\pi
_{z}\right) ^{2}}{\mathfrak{r}^{3}}=\frac{\rho _{x}^{2}\pi
_{x}^{2}+\rho _{y}^{2}\pi _{y}^{2}+\rho _{z}^{2}\pi _{z}^{2}+2\rho
_{x}\rho _{y}\pi _{x}\pi _{y}+2\rho
_{y}\rho _{z}\pi _{z}\pi _{y}+2\rho _{x}\rho _{z}\pi _{x}\pi _{z}}{\mathfrak{%
r}^{3}}.
\end{equation}%
It is sufficient to examine the two terms
\begin{equation}
\frac{\rho _{x}^{2}\pi _{x}^{2}+2\rho _{x}\rho _{y}\pi _{x}\pi _{y}}{%
\mathfrak{r}^{3}},
\end{equation}%
and the rest we determine by cyclic symmetry. \

Consider first $\frac{\rho _{x}^{2}\pi _{x}^{2}}{\mathfrak{r}^{3}}$.
\
\begin{equation}
\frac{\rho _{x}^{2}\pi _{x}^{2}}{\mathfrak{r}^{3}}=\frac{\rho
_{x}^{2}\pi _{x}^{2}}{\zeta ^{3}(1+\rho _{x}^{2}/\zeta
^{2})^{3/2}}=\frac{\rho
_{x}^{2}\pi _{x}^{2}}{\zeta ^{3}}\sum_{n=0}^{\infty }\binom{-3/2}{n}\frac{%
\rho _{x}^{2n}}{\zeta ^{2n}}.
\end{equation}%
The term that needs Weyl ordering is $\rho _{x}^{2n+2}\pi _{x}^{2}$.
\ In
analogy to above we find%
\begin{eqnarray}
\left( \frac{\rho _{x}^{2}\pi _{x}^{2}}{\mathfrak{r}^{3}}\right) ^{W} &=&%
\frac{1}{\zeta }\sum_{n=0}^{\infty }\frac{1}{(2\zeta )^{2n+2}}\binom{-3/2}{n}%
\sum_{m=0}^{2n+2}\binom{2n+2}{m}(\rho _{x}^{2n+2}\pi
_{x}^{2}-2im\rho
_{x}^{2n+1}\pi _{x}-m(m-1)\rho _{x}^{2n})  \nonumber \\
&=&\frac{\rho _{x}^{2}}{\mathfrak{r}^{3}}\pi _{x}^{2}-i\frac{2\rho _{x}}{%
\mathfrak{r}^{3}}\pi _{x}+i\frac{3\rho _{x}^{3}}{\mathfrak{r}^{5}}\pi _{x}-%
\frac{1}{2\mathfrak{r}^{3}}+\frac{15\rho _{x}^{2}}{4\mathfrak{r}^{5}}-\frac{%
15\rho _{x}^{4}}{4\mathfrak{r}^{7}}.
\end{eqnarray}%
\ \ Including cyclic terms we find

\begin{eqnarray}
&&\left( \frac{\rho _{x}^{2}\pi _{x}^{2}+\rho _{y}^{2}\pi
_{y}^{2}+\rho
_{z}^{2}\pi _{z}^{2}}{\mathfrak{r}^{3}}\right) ^{W}  \nonumber \\
&=&\frac{1}{\mathfrak{r}^{3}}(\rho _{x}^{2}\pi _{x}^{2}+\rho
_{y}^{2}\pi
_{y}^{2}+\rho _{z}^{2}\pi _{z}^{2})-i\frac{2}{\mathfrak{r}^{3}}\vec{\rho}%
\cdot \vec{\pi}+i\frac{3}{\mathfrak{r}^{5}}(\rho _{x}^{3}\pi
_{x}+\rho
_{y}^{3}\pi _{y}+\rho _{z}^{3}\pi _{z})  \nonumber \\
&&+\frac{9}{4\mathfrak{r}^{3}}-\frac{15\varepsilon ^{2}}{4\mathfrak{r}^{3}}-%
\frac{15(\rho _{x}^{4}+\rho _{y}^{4}+\rho
_{z}^{4})}{4\mathfrak{r}^{7}}. \label{2}
\end{eqnarray}

The next term we consider ( here $\zeta ^{2}=\rho _{z}^{2}+\varepsilon ^{2}$)%
\begin{eqnarray}
\frac{2\rho _{x}\rho _{y}\pi _{x}\pi _{y}}{\mathfrak{r}^{3}}
&=&\frac{2\rho _{x}\rho _{y}\pi _{x}\pi _{y}}{\emph{(}\rho
_{x}^{2}+\rho _{y}^{2}+\zeta
^{2})^{3/2}}=\frac{1}{\zeta ^{3}}\frac{2\rho _{x}\rho _{y}\pi _{x}\pi _{y}}{%
\emph{(}1+(\rho _{x}^{2}+\rho _{y}^{2})/\zeta ^{2})^{3/2}}  \nonumber \\
&=&\frac{1}{\zeta ^{3}}2\rho _{x}\rho _{y}\pi _{x}\pi
_{y}\sum_{n=0}^{\infty
}\binom{-3/2}{n}\sum_{l=0}^{n}\binom{n}{l}\rho _{x}^{2(n-l)}\rho
_{y}^{2l}.
\end{eqnarray}%
Thus we need
\begin{eqnarray}
&&\left( \pi _{x}\rho _{x}^{2(n-l)+1}\pi _{y}\rho _{y}^{2m+1}\right)
^{W}=\left( \pi _{x}\rho _{x}^{2(n-l)+1})^{W}(\pi _{y}\rho
_{y}^{2l+1}\right) ^{W}  \nonumber \\
&=&\frac{1}{2^{2n+2}}\sum_{m=0}^{2(n-l)+1}\binom{2(n-l)+1}{m}%
\sum_{k=0}^{2l+1}\binom{2l+1}{k}[\rho _{x}^{2(n-l)+1}\pi _{x}-im\rho
_{x}^{2(n-l)}][\rho _{y}^{2l+1}\pi _{y}-ik\rho _{y}^{2l}].  \nonumber \\
&&{}
\end{eqnarray}%
Performing the inner sums, we obtain%
\begin{eqnarray}
&&\left( \frac{2\rho _{x}\rho _{y}\pi _{x}\pi
_{y}}{\mathfrak{r}^{3}}\right)
^{W}  \nonumber \\
&=&\frac{2}{\zeta ^{3}}\sum_{n=0}^{\infty }\binom{-3/2}{n}\sum_{l=0}^{n}%
\binom{n}{l}\frac{1}{\zeta ^{2n}}[\rho _{x}^{2(n-l)}\rho
_{y}^{2l}(\rho
_{x}\pi _{x}\rho _{y}\pi _{y}  \nonumber \\
&&-i\frac{\left( 2(n-l)+1\right) }{2}\rho _{y}\pi _{y}-i\frac{\left(
2l+1\right) }{2}\rho _{x}\pi _{x}-\frac{\left( 2(n-l)+1\right)
\left( 2l+1\right) }{4}].
\end{eqnarray}%
Performing the next inner sum we find%
\begin{eqnarray}
&&\left( \frac{2\rho _{x}\rho _{y}\pi _{x}\pi
_{y}}{\mathfrak{r}^{3}}\right)
^{W}  \nonumber \\
&=&\frac{2}{\zeta ^{3}}\sum_{n=0}^{\infty }\left(
\underset{n}{-3/2}\right) \frac{1}{\zeta ^{2n}}[(\rho _{x}^{2}+\rho
_{y}^{2})^{n}\rho _{x}\pi _{x}\rho _{y}\pi _{y}-i\frac{\rho
_{y}}{2}\frac{d}{d\rho _{x}}\rho _{x}(\rho
_{x}^{2}+\rho _{y}^{2})^{n}\pi _{y}-i\frac{\rho _{x}}{2}\frac{d}{d\rho _{y}}%
\rho _{y}(\rho _{x}^{2}+\rho _{y}^{2})^{n}\pi _{x}  \nonumber \\
&&-\frac{1}{4}\frac{d}{d\rho _{x}}\frac{d}{d\rho _{y}}\rho _{x}\rho
_{y}(\rho _{x}^{2}+\rho _{y}^{2})^{n}]  \nonumber \\
&=&\frac{2}{\mathfrak{r}^{3}}\rho _{x}\rho _{y}\pi _{x}\pi _{y}-i\rho _{y}(%
\frac{1}{\mathfrak{r}^{3}}-\frac{3\rho
_{x}^{2}}{\mathfrak{r}^{5}})\pi
_{y}-i\rho _{x}(\frac{1}{\mathfrak{r}^{3}}-\frac{3\rho _{y}^{2}}{\mathfrak{r}%
^{5}})\pi _{x}-\frac{1}{2}(\frac{1}{\mathfrak{r}^{3}}-\frac{3(\rho
_{x}^{2}+\rho _{y}^{2})}{\mathfrak{r}^{5}}+\frac{15\rho _{x}^{2}\rho _{y}^{2}%
}{\mathfrak{r}^{7}}).
\end{eqnarray}%
Adding the cyclic terms and we obtain%
\begin{eqnarray}
&&\left( \frac{2\rho _{x}\rho _{y}\pi _{x}\pi _{y}+2\rho _{y}\rho
_{z}\pi _{z}\pi _{y}+2\rho _{x}\rho _{z}\pi _{x}\pi
_{z}}{\mathfrak{r}^{3}}\right)
^{W}  \nonumber \\
&=&\frac{2}{\mathfrak{r}^{3}}(\rho _{x}\rho _{y}\pi _{x}\pi
_{y}+\rho _{z}\rho _{y}\pi _{z}\pi _{y}+\rho _{x}\rho _{z}\pi
_{x}\pi _{z})  \nonumber
\\
&&-i(-\frac{1}{\mathfrak{r}^{3}}+\frac{3(\varepsilon ^{2}+\rho _{y}^{2})}{%
\mathfrak{r}^{5}})\rho _{y}\pi _{y}-i(-\frac{1}{\mathfrak{r}^{3}}+\frac{%
3(\varepsilon ^{2}+\rho _{x}^{2})}{\mathfrak{r}^{5}})\rho _{x}\pi _{x}-i(-%
\frac{1}{\mathfrak{r}^{3}}+\frac{3(\varepsilon ^{2}+\rho _{z}^{2})}{%
\mathfrak{r}^{5}})\rho _{z}\pi _{z}  \nonumber \\
&&+\frac{3}{2\mathfrak{r}^{3}}-\frac{3\varepsilon ^{2}}{\mathfrak{r}^{5}}-%
\frac{15(\rho _{x}x^{2}\rho _{y}^{2}+\rho _{y}^{2}\rho _{z}^{2}+\rho
_{z}^{2}\rho _{x}^{2})}{2\mathfrak{r}^{7}}).
\end{eqnarray}

Let us combine this with Eq. (\ref{2}) and in addition to Eq. (\ref{1d}) use%
\begin{equation}
2\pi \delta ^{3}(\vec{\rho})=\underset{\varepsilon \rightarrow 0}{\lim }%
\frac{15\varepsilon ^{4}}{4\mathfrak{r}^{7}}.
\end{equation}%
We find%
\begin{eqnarray}
&&\left( \frac{\rho _{x}^{2}\pi _{x}^{2}+\rho _{y}^{2}\pi
_{y}^{2}+\rho _{z}^{2}\pi _{z}^{2}+2\rho _{x}\rho _{y}\pi _{x}\pi
_{y}+2\rho _{y}\rho
_{z}\pi _{z}\pi _{y}+2\rho _{x}\rho _{z}\pi _{x}\pi _{z}}{\mathfrak{r}^{3}}%
\right) ^{W}  \nonumber \\
&=&\frac{1}{\rho ^{3}}\rho _{i}\rho _{j}\pi _{i}\pi _{j}-i\frac{1}{\rho ^{3}}%
\vec{\rho}\cdot \vec{\pi}-i4\pi \delta ^{3}(\vec{\rho})\vec{\rho}\cdot \vec{%
\pi}+\pi \delta ^{3}(\vec{\rho})  \nonumber \\
&\rightarrow &\frac{1}{\rho ^{3}}\rho _{i}\rho _{j}\pi _{i}\pi _{j}-i\frac{1%
}{\rho ^{3}}\vec{\rho}\cdot \vec{\pi}-\pi \delta ^{3}(\vec{\rho})
\end{eqnarray}%
We note for comparison that
\begin{eqnarray}
\vec{\pi}\cdot \vec{\rho}\frac{1}{\rho ^{3}}\vec{\rho}\cdot
\vec{\pi} &=&\pi
_{i}\rho _{i}\frac{1}{\rho ^{3}}\rho _{j}\pi _{j}  \nonumber \\
&=&\rho _{i}\frac{1}{\mathfrak{r}^{3}}\rho _{j}\pi _{i}\pi _{j}+[\pi
_{i},\rho _{i}\frac{1}{\mathfrak{r}^{3}}\rho _{j}]\pi _{j}  \nonumber \\
&=&\frac{1}{\mathfrak{r}^{3}}\rho _{i}\rho _{j}\pi _{i}\pi _{j}+\frac{i}{%
\mathfrak{r}^{3}}\vec{\rho}\cdot \vec{\pi}-\frac{3i\varepsilon ^{2}}{%
\mathfrak{r}^{5}}\vec{\rho}\cdot \vec{\pi}  \nonumber \\
&\rightarrow &\frac{1}{\rho ^{3}}\rho _{i}\rho _{j}\pi _{i}\pi _{j}+\frac{i}{%
\rho ^{3}}\vec{\rho}\cdot \vec{\pi}.
\end{eqnarray}%
since the delta function in the third line kills the $\vec{\rho}.~$Thus%
\begin{equation}
\left( \vec{\pi}\cdot \vec{\rho}\frac{1}{\rho ^{3}}\vec{\rho}\cdot \vec{\pi}%
\right) ^{W}-\vec{\pi}\cdot \vec{\rho}\frac{1}{\rho
^{3}}\vec{\rho}\cdot
\vec{\pi}=-\frac{2i}{r^{3}}\vec{\rho}\cdot \vec{\pi}-\pi \delta ^{3}(\vec{%
\rho})
\end{equation}

The total Weyl Darwin terms combine to%
\begin{eqnarray}
&&\left( \frac{1}{\mathfrak{r}}\pi
_{x}^{2}+\frac{1}{\mathfrak{r}}\pi _{\rho
}^{2}+\frac{1}{\mathfrak{r}}\pi _{z}^{2}\right) ^{W}+\left(
\frac{\rho _{x}^{2}\pi _{x}^{2}+\rho _{y}^{2}\pi _{\rho }^{2}+\rho
_{z}^{2}\pi _{z}^{2}+2\rho _{x}\rho _{y}\pi _{x}\pi _{y}+2\rho
_{y}\rho _{z}\pi _{z}\pi _{y}+2\rho _{x}\rho _{z}\pi _{x}\pi
_{z}}{\mathfrak{r}^{3}}\right) ^{W}
\nonumber \\
&=&\frac{1}{\rho }\vec{\pi}^{2}+\frac{1}{\rho ^{3}}\rho _{i}\rho
_{j}\pi _{i}\pi _{j}.
\end{eqnarray}%
which gives the same as the total $O(1/c^{2})$ hermitian Darwin
interaction
\begin{equation}
\frac{Q_{1}Q_{1}}{8\pi m_{1}m_{2}c^{2}}[\vec{\pi}\cdot \frac{1}{\rho }\vec{%
\pi}+\vec{\pi}\cdot \vec{\rho}\frac{1}{\rho ^{3}}\vec{\rho}\cdot \vec{\pi}]=%
\frac{Q_{1}Q_{1}}{8\pi m_{1}m_{2}c^{2}}[\frac{1}{\rho }\vec{\pi}^{2}+\frac{1%
}{\rho ^{3}}\rho _{i}\rho _{j}\pi _{i}\pi _{j}].
\end{equation}%
Thus, we find that the difference in two treatments (Weyl and
conventional) of the total quantum $O(1/c^{2})$ Darwin terms is
zero, although the differences in the two treatments of the
individual Darwin terms are not zero!

\subsection{Weyl Quantization of the Classical Darwin Hamiltonian}

From \cite{10} Eqs. (6.35), (6.36), (6.37), for equal masses the
Hamiltonian
is \footnote{%
There is a sign error in Eq. (6.35) which effects Eq. (6.37) of that
reference \cite{10}. \ Here we correct (6.37).}%
\begin{eqnarray}
&&M=\mathcal{P}_{(int)}^{\tau }=2\sqrt{m^{2}+\vec{\pi}_{12}^{2}}+\frac{%
Q_{1}Q_{2}}{4\pi \left\vert \vec{\rho}_{12}\right\vert }+{\frac{Q_{1}Q_{2}}{%
8\pi \left\vert \vec{\rho}_{12}\right\vert }}\times \lbrack \frac{\vec{\pi}%
_{12}^{2}+(\vec{\pi}_{12}\cdot \frac{\vec{\rho}_{12}}{\left\vert \vec{\rho}%
_{12}\right\vert })^{2}}{(m^{2}c^{2}+\vec{\pi}_{12}^{2})}  \nonumber \\
&&+\frac{1}{(m^{2}c^{2}+\vec{\pi}_{12}^{2})[m^{2}c^{2}+(\vec{\pi}_{12}\cdot
\frac{\vec{\rho}_{12}}{\left\vert \vec{\rho}_{12}\right\vert
})^{2}]}\times
(m^{2}c^{2}[3\vec{\pi}_{12}^{2}+(\vec{\pi}_{12}\cdot \frac{\vec{\rho}_{12}}{%
\left\vert \vec{\rho}_{12}\right\vert })^{2}]-  \nonumber \\
&&-[3\vec{\pi}_{12}^{2}+(\vec{\pi}_{12}\cdot \frac{\vec{\rho}_{12}}{%
\left\vert \vec{\rho}_{12}\right\vert })^{2}][m^{2}c^{2}+(\vec{\pi}%
_{12}\cdot \frac{\vec{\rho}_{12}}{\left\vert \vec{\rho}_{12}\right\vert }%
)^{2}]-2\vec{\pi}_{12}^{2}[\vec{\pi}_{12}^{2}-3(\vec{\pi}_{12}\cdot \frac{%
\vec{\rho}_{12}}{\left\vert \vec{\rho}_{12}\right\vert })^{2}]\sqrt{\frac{%
m^{2}c^{2}+\vec{\pi}_{12}^{2}}{m^{2}c^{2}+(\vec{\pi}_{12}\cdot \frac{\vec{%
\rho}_{12}}{\left\vert \vec{\rho}_{12}\right\vert })^{2}}}]  \nonumber \\
&=&2\sqrt{m^{2}+\vec{\pi}_{12}^{2}}+\frac{Q_{1}Q_{2}}{4\pi \left\vert \vec{%
\rho}_{12}\right\vert }  \nonumber \\
&&+{\frac{Q_{1}Q_{2}}{8\pi \left\vert \vec{\rho}_{12}\right\vert }}\frac{1}{%
(m^{2}c^{2}+\vec{\pi}_{12}^{2})[m^{2}c^{2}+(\vec{\pi}_{12}\cdot \frac{\vec{%
\rho}_{12}}{\left\vert \vec{\rho}_{12}\right\vert
})^{2}]}[m^{2}c^{2}\left( \vec{\pi}_{12}^{2}+(\vec{\pi}_{12}\cdot
\frac{\vec{\rho}_{12}}{\left\vert
\vec{\rho}_{12}\right\vert })^{2}\right) -2\vec{\pi}_{12}^{2}(\vec{\pi}%
_{12}\cdot \frac{\vec{\rho}_{12}}{\left\vert \vec{\rho}_{12}\right\vert }%
)^{2}  \nonumber \\
&&-2\vec{\pi}_{12}^{2}(\vec{\pi}_{12}^{2}-3(\vec{\pi}_{12}\cdot \frac{\vec{%
\rho}_{12}}{\left\vert \vec{\rho}_{12}\right\vert })^{2})\sqrt{\frac{%
m^{2}c^{2}+\vec{\pi}_{12}^{2}}{m^{2}c^{2}+(\vec{\pi}_{12}\cdot \frac{\vec{%
\rho}_{12}}{\left\vert \vec{\rho}_{12}\right\vert })^{2}}}].
\label{1}
\end{eqnarray}%
Let us see how we can construct a corresponding \ self-adjoint
quantum operator on position space wave functions using the Weyl -
quantization procedure \cite{41}

\subsubsection{ Weyl-quantization}

Let $K$ be a classical observable and a function of the relative variables $%
\vec{\pi}_{12}$ and $\vec{\rho}_{12}.$ \ The Weyl-quantization $K^{W}$ of $K(%
\vec{\rho}_{12},\vec{\pi}_{12})$ is defined on a well behaved wave function $%
\psi (\vec{\rho}_{12})$ as \footnote{%
In this appendix we use the $\hbar =1$ convention.}
\begin{equation}
K^{W}\psi (\vec{\rho}_{12})=\frac{1}{(2\pi )^{3}}\int \int \exp \left( i(%
\vec{\rho}_{12}-\vec{\rho}_{12}^{\prime })\cdot \vec{\pi}_{12}\right) K(%
\frac{\vec{\rho}_{12}+\vec{\rho}_{12}^{\prime }}{2},\vec{\pi}_{12})\psi (%
\vec{\rho}_{12}^{\prime })d^{3}\rho _{12}d^{3}\pi _{12}.
\end{equation}%
In the case of functions that are dependent only on
$\vec{\rho}_{12}$ we have by doing the $d^{3}\pi _{12}$ integral \
\begin{eqnarray}
K^{W}\psi (\vec{\rho}_{12}) &=&\frac{1}{(2\pi )^{3}}\int \int \exp \left( i(%
\vec{\rho}_{12}-\vec{\rho}_{12}^{\prime })\cdot \vec{\pi}_{12}\right) K(%
\frac{\vec{\rho}_{12}+\vec{\rho}_{12}^{\prime }}{2})\psi (\vec{\rho}%
_{12}^{\prime })d^{3}\rho _{12}d^{3}\pi _{12}  \nonumber \\
&=&K(\vec{\rho}_{12})\psi (\vec{\rho}_{12}).
\end{eqnarray}%
The only term of this form in Eq. (\ref{1}) is the Coulomb term%
\begin{equation}
K_{C}^{W}\psi (\vec{\rho}_{12})=\frac{Q_{1}Q_{2}}{4\pi \left\vert \vec{\rho}%
_{12}\right\vert }\psi (\vec{\rho}_{12}).
\end{equation}

For functions that are dependent only on $\vec{\pi}_{12}$ we have%
\begin{eqnarray}
K^{W}\psi (\vec{\rho}_{12}) &=&\frac{1}{(2\pi )^{3}}\int \int \exp \left( i(%
\vec{\rho}_{12}-\vec{\rho}_{12}^{\prime })\cdot \vec{\pi}_{12}\right) K(\vec{%
\pi}_{12})\psi (\vec{\rho}_{12}^{\prime })d^{3}\rho _{12}d^{3}\pi
_{12}
\nonumber \\
&=&\int \tilde{K}(\vec{\rho}_{12}-\vec{\rho}_{12}^{\prime })\psi (\vec{\rho}%
_{12}^{\prime })d^{3}\rho _{12},  \label{3}
\end{eqnarray}%
in which $\tilde{K}((\vec{\rho}_{12}-\vec{\rho}_{12}^{\prime })$ is
the
Fourier transform of $K(\vec{\pi}_{12}).$%
\begin{equation}
\tilde{K}(\vec{\rho}_{12}-\vec{\rho}_{12}^{\prime })=\frac{1}{(2\pi )^{3}}%
\int \exp \left( i(\vec{\rho}_{12}-\vec{\rho}_{12}^{\prime })\cdot \vec{\pi}%
_{12}\right) K(\vec{\pi}_{12})d^{3}\pi _{12}.
\end{equation}%
The only term like that in the whole Hamiltonian (\ref{1}) is the
kinetic
piece%
\begin{equation}
2\tilde{K}_{T}(\vec{\rho}_{12}-\vec{\rho}_{12}^{\prime
})=\frac{1}{(2\pi )^{3}}\int \exp \left(
i(\vec{\rho}_{12}-\vec{\rho}_{12}^{\prime })\cdot
\vec{\pi}_{12}\right) 2\sqrt{m^{2}+\vec{\pi}_{12}^{2}}d^{3}\pi _{12}
\label{KT}
\end{equation}

\subsubsection{ Lowest Order $1/c^{2}$ Expressions}

From the second part of Eq. (\ref{1}) the first order $1/c^{2}$
Darwin terms
are%
\begin{equation}
D\equiv {\frac{Q_{1}Q_{2}}{8\pi \left\vert \vec{\rho}_{12}\right\vert }}(%
\frac{\vec{\pi}_{12}^{2}+(\vec{\pi}_{12}\cdot \frac{\vec{\rho}_{12}}{%
\left\vert \vec{\rho}_{12}\right\vert })^{2}}{m^{2}c^{2}}).
\label{8}
\end{equation}%
Our Weyl quantized version on a position space wave function is%
\begin{eqnarray}
D^{w}\psi (\vec{\rho}_{12}) &=&\frac{1}{(2\pi
)^{3}}\frac{Q_{1}Q_{2}}{8\pi m^{2}c^{2}}\int \int \exp \left(
i(\vec{\rho}_{12}-\vec{\rho}_{12}^{\prime
})\cdot \vec{\pi}_{12}\right) \frac{2}{\left\vert \vec{\rho}_{12}+\vec{\rho}%
_{12}^{\prime }\right\vert }  \nonumber \\
&&\times \left( \vec{\pi}_{12}^{2}+(\vec{\pi}_{12}\cdot \hat{n}%
_{12})^{2}\right) \psi (\vec{\rho}_{12}^{\prime })d^{3}\rho
_{12}d^{3}\pi _{12}.
\end{eqnarray}%
in which%
\begin{equation}
\hat{n}_{12}=\frac{\vec{\rho}_{12}+\vec{\rho}_{12}^{\prime
}}{\left\vert \vec{\rho}_{12}+\vec{\rho}_{12}^{\prime }\right\vert
}.
\end{equation}%
The first term from integration by parts is%
\begin{eqnarray}
&&\frac{1}{(2\pi )^{3}}\frac{Q_{1}Q_{2}}{8\pi m^{2}c^{2}}\int \int
\exp
\left( i(\vec{\rho}_{12}-\vec{\rho}_{12}^{\prime })\cdot \vec{\pi}%
_{12}\right) \frac{2}{\left\vert
\vec{\rho}_{12}+\vec{\rho}_{12}^{\prime }\right\vert
}\vec{\pi}_{12}^{2}\psi (\vec{\rho}_{12}^{\prime })d^{3}\rho
_{12}d^{3}\pi _{12}  \nonumber \\
&=&-\frac{1}{(2\pi )^{3}}\frac{Q_{1}Q_{2}}{8\pi m^{2}c^{2}}\int \int
[\exp
\left( i(\vec{\rho}_{12}-\vec{\rho}_{12}^{\prime })\cdot \vec{\pi}%
_{12}\right) ]\vec{\partial}^{\prime 2}\frac{2}{\left\vert \vec{\rho}_{12}+%
\vec{\rho}_{12}^{\prime }\right\vert }\psi (\vec{\rho}_{12}^{\prime
})d^{3}\rho _{12}d^{3}\pi _{12}  \nonumber \\
&=&-\frac{Q_{1}Q_{2}}{8\pi m^{2}c^{2}}\vec{\partial}\cdot (-\frac{\vec{\rho}%
_{12}}{\left\vert \vec{\rho}_{12}\right\vert ^{3}}+\frac{1}{\left\vert \vec{%
\rho}_{12}\right\vert }\vec{\partial})\psi (\vec{\rho}_{12}).
\end{eqnarray}%
The second term is%
\begin{eqnarray}
&&\frac{1}{(2\pi )^{3}}\frac{Q_{1}Q_{2}}{8\pi m^{2}c^{2}}\int \int [(\hat{n}%
_{12})_{r}(\hat{n}_{12})_{s}\pi _{12}{}_{r}\pi _{12}{}_{s}\exp \left( i(\vec{%
\rho}_{12}-\vec{\rho}_{12}^{\prime })\cdot \vec{\pi}_{12}\right) ]\frac{2}{%
\left\vert \vec{\rho}_{12}+\vec{\rho}_{12}^{\prime }\right\vert }\psi (\vec{%
\rho}_{12}^{\prime })d^{3}\rho _{12}d^{3}\pi _{12}  \nonumber \\
&=&-\frac{1}{(2\pi )^{3}}\frac{Q_{1}Q_{2}}{8\pi m^{2}c^{2}}\int \int
\exp
\left( i(\vec{\rho}_{12}-\vec{\rho}_{12}^{\prime })\cdot \vec{\pi}%
_{12}\right) \partial _{r}^{\prime }\partial _{s}^{\prime }[(\hat{n}%
_{12})_{r}(\hat{n}_{12})_{s}\frac{2}{\left\vert \vec{\rho}_{12}+\vec{\rho}%
_{12}^{\prime }\right\vert }\psi (\vec{\rho}_{12}^{\prime
})]d^{3}\rho
_{12}d^{3}\pi _{12}  \nonumber \\
&=&-\frac{Q_{1}Q_{2}}{8\pi m^{2}c^{2}}\partial _{r}[\frac{\rho _{12r}}{%
\left\vert \vec{\rho}_{12}\right\vert ^{3}}\psi
(\vec{\rho}_{12})+\frac{\rho _{12r}\rho _{12s}}{\left\vert
\vec{\rho}_{12}\right\vert ^{3}}\partial _{s}\psi (\vec{\rho}_{12})]
\end{eqnarray}%
Thus, the $O(1/c^{2})$ Darwin contribution to the Schrodinger
equation has
the two form%
\begin{equation}
D^{w}\psi (\vec{\rho}_{12})=-\frac{Q_{1}Q_{2}}{8\pi m^{2}c^{2}}\partial _{r}(%
\frac{\delta _{rs}}{\left\vert \vec{\rho}_{12}\right\vert
}+\frac{\rho _{12r}\rho _{12s}}{\left\vert
\vec{\rho}_{12}\right\vert ^{3}})\partial _{s}\psi
(\vec{\rho}_{12}).
\end{equation}

This agrees (in notation of this paper) with known Darwin results (see \cite%
{42},\cite{43} and \cite{44}). \ If we bring the remaining
derivative through we obtain
\begin{equation}
D^{w}\psi (\vec{\rho}_{12})=-\frac{Q_{1}Q_{2}}{8\pi
m^{2}c^{2}}(\frac{\delta
_{rs}}{\left\vert \vec{\rho}_{12}\right\vert }+\frac{\rho _{12r}\rho _{12s}}{%
\left\vert \vec{\rho}_{12}\right\vert ^{3}})\partial _{r}\partial _{s}\psi (%
\vec{\rho}_{12}).
\end{equation}%
These two results arise from the Weyl ordering on the position wave
function of the classical function Eq. (\ref{8}) In operator form
they correspond to the operator forms of
\begin{eqnarray}
&&\frac{Q_{1}Q_{2}}{8\pi m^{2}c^{2}}(\pi _{12}\cdot \frac{1}{\left\vert \vec{%
\rho}_{12}\right\vert }\pi _{12}+\pi _{12}\cdot \frac{\vec{\rho}_{12}\vec{%
\rho}_{12}}{\left\vert \vec{\rho}_{12}\right\vert ^{3}}\cdot \pi
_{12}),
\nonumber \\
&&\frac{Q_{1}Q_{2}}{8\pi m^{2}c^{2}}\frac{1}{\left\vert \vec{\rho}%
_{12}\right\vert }(\delta _{rs}+\hat{\eta}_{r}\hat{\eta}_{s})\pi
_{12}{}_{r}\pi _{12}{}_{s}.
\end{eqnarray}

\subsubsection{Quantization of the Complete Expression Eq. (%
\ref{1}) -The Coulomb Potential as a Local $c\rightarrow \infty $
Limit}

\bigskip

\ Before going on to the Weyl quantization of the exact expression
we
rearrange the first two portions of the Darwin term in the last line of Eq. (%
\ref{1}) to read%
\begin{eqnarray*}
&&\frac{Q_{1}Q_{2}}{8\pi \left\vert \vec{\rho}_{12}\right\vert }\frac{1}{%
\left( m^{2}c^{2}+(\vec{\pi}_{12}\cdot
\frac{\vec{\rho}_{12}}{\left\vert
\vec{\rho}_{12}\right\vert })^{2}\right) (m^{2}c^{2}+\vec{\pi}_{12}^{2})}%
\times \lbrack m^{2}c^{2}\left(
\vec{\pi}_{12}^{2}+(\vec{\pi}_{12}\cdot
\frac{\vec{\rho}_{12}}{\left\vert \vec{\rho}_{12}\right\vert
})^{2}\right)
\\
&&-2(\vec{\pi}_{12}^{2}+m^{2}c^{2})((\pi _{12}\cdot \hat{\eta}%
)^{2}+m^{2}c^{2})+2(\vec{\pi}_{12}^{2}+(\pi _{12}\cdot \hat{\eta}%
)^{2})m^{2}c^{2}+2m^{4}c^{4}] \\
&=&-\frac{Q_{1}Q_{2}}{4\pi \left\vert \vec{\rho}_{12}\right\vert }+\frac{%
Q_{1}Q_{2}}{8\pi \left\vert \vec{\rho}_{12}\right\vert }[\frac{3m^{2}c^{2}}{%
m^{2}c^{2}+(\vec{\pi}_{12}\cdot \frac{\vec{\rho}_{12}}{\left\vert \vec{\rho}%
_{12}\right\vert
})^{2}}+\frac{3m^{2}c^{2}}{(m^{2}c^{2}+\vec{\pi}_{12}^{2})}
\\
&&-\frac{4m^{4}c^{4}}{\ (m^{2}c^{2}+\vec{\pi}_{12}^{2})(m^{2}c^{2}+(\vec{\pi}%
_{12}\cdot \frac{\vec{\rho}_{12}}{\left\vert \vec{\rho}_{12}\right\vert }%
)^{2})}].
\end{eqnarray*}%
Our total classical Hamiltonian then becomes%
\begin{eqnarray}
M &=&2\sqrt{m^{2}c^{4}+\vec{\pi}_{12}^{2}c^{2}}+  \nonumber \\
&&+\frac{Q_{1}Q_{2}}{8\pi \left\vert \vec{\rho}_{12}\right\vert }[\frac{%
3m^{2}c^{2}}{(m^{2}c^{2}+\vec{\pi}_{12}^{2})}+\frac{3m^{2}c^{2}}{m^{2}c^{2}+(%
\vec{\pi}_{12}\cdot \frac{\vec{\rho}_{12}}{\left\vert \vec{\rho}%
_{12}\right\vert })^{2}}-\frac{4m^{4}c^{4}}{\ (m^{2}c^{2}+\vec{\pi}%
_{12}^{2})(m^{2}c^{2}+(\vec{\pi}_{12}\cdot
\frac{\vec{\rho}_{12}}{\left\vert
\vec{\rho}_{12}\right\vert })^{2})}]-  \nonumber \\
&&-{\frac{Q_{1}Q_{2}}{4\pi \left\vert \vec{\rho}_{12}\right\vert }}\frac{2%
\vec{\pi}_{12}^{2}(\vec{\pi}_{12}^{2}-3(\vec{\pi}_{12}\cdot \frac{\vec{\rho}%
_{12}}{\left\vert \vec{\rho}_{12}\right\vert })^{2})\sqrt{\frac{m^{2}c^{2}+%
\vec{\pi}_{12}^{2}}{m^{2}c^{2}+(\vec{\pi}_{12}\cdot \frac{\vec{\rho}_{12}}{%
\left\vert \vec{\rho}_{12}\right\vert })^{2}}}}{\left( m^{2}c^{2}+(\vec{\pi}%
_{12}\cdot \frac{\vec{\rho}_{12}}{\left\vert \vec{\rho}_{12}\right\vert }%
)^{2}\right) (m^{2}c^{2}+\vec{\pi}_{12}^{2})}.  \label{2}
\end{eqnarray}%
Note that in this rearrangement, the local Coulomb potential is
canceled and replaced by momentum dependent terms. Note, however,
that in the non-relativistic limit ($c\rightarrow \infty $) the
momentum dependent potential energy terms in the first line reduces
to the ordinary Coulomb term while the second line vanishes. \
Although Eq. (\ref{1}) has the advantage of seeing the lowest order
expansion more clearly, the above shows that the exact expression
does not have local Coulomb potentials except in the the
non-relativistic limit ($c\rightarrow \infty $).

Consider the simplest part
\begin{equation}
\frac{Q_{1}Q_{2}}{8\pi \left\vert \vec{\rho}_{12}\right\vert }\frac{%
3m^{2}c^{2}}{(m^{2}c^{2}+\vec{\pi}_{12}^{2})}.
\end{equation}%
The corresponding Weyl term would be%
\begin{equation}
K_{1}^{W}\psi (\vec{\rho}_{12})=\frac{3m^{2}c^{2}}{(2\pi )^{3}}\frac{%
Q_{1}Q_{2}}{8\pi }\int \int \exp \left( i(\vec{\rho}_{12}-\vec{\rho}%
_{12}^{\prime })\cdot \vec{\pi}_{12}\right) \frac{2}{\left\vert \vec{\rho}%
_{12}+\vec{\rho}_{12}^{\prime }\right\vert }\frac{1}{(m^{2}c^{2}+\vec{\pi}%
_{12}^{2})}\psi (\vec{\rho}_{12}^{\prime })d^{3}\rho _{12}^{\prime
}d^{3}\pi _{12}.
\end{equation}%
Perform the $\vec{\pi}_{12}$ integral to give%
\begin{equation}
\frac{1}{(2\pi )^{3}}\int d^{3}\pi _{12}\exp (i(\vec{\rho}_{12}-\vec{\rho}%
_{12}^{\prime })\cdot \vec{\pi}_{12})\frac{1}{(m^{2}c^{2}+\vec{\pi}_{12}^{2})%
}=\frac{1}{4\pi }\frac{\exp (-mc\left\vert \vec{\rho}_{12}-\vec{\rho}%
_{12}^{\prime }\right\vert )}{\left\vert \vec{\rho}_{12}-\vec{\rho}%
_{12}^{\prime }\right\vert }
\end{equation}%
and so%
\begin{equation}
K_{1}^{W}\psi (\vec{\rho}_{12})=\frac{3m^{2}c^{2}Q_{1}Q_{2}}{16\pi
^{2}}\int
\frac{1}{\left\vert \vec{\rho}_{12}+\vec{\rho}_{12}^{\prime }\right\vert }%
\frac{\exp (-mc\left\vert \vec{\rho}_{12}-\vec{\rho}_{12}^{\prime
}\right\vert )}{\left\vert \vec{\rho}_{12}-\vec{\rho}_{12}^{\prime
}\right\vert }\psi (\vec{\rho}_{12}^{\prime })d^{3}\rho
_{12}^{\prime }.
\end{equation}%
It is a nonlocal term just as the kinetic energy is. \

Note that we recover the local non-relativistic limit of this
expression by using the form below for the Dirac delta function,
\begin{equation}
\delta ^{3}(\vec{\rho}_{12}^{\prime }-\vec{\rho}_{12})=\lim
(c\rightarrow
\infty )\frac{~m^{2}c^{2}}{4\pi }\frac{\exp (-mc\left\vert \vec{\rho}%
_{12}^{\prime }-\vec{\rho}_{12}\right\vert )}{\left\vert \vec{\rho}%
_{12}^{\prime }-\vec{\rho}_{12}\right\vert }.  \label{4}
\end{equation}%
In that case%
\begin{equation}
K_{1}^{W}\psi (\vec{\rho}_{12})\rightarrow \frac{3Q_{1}Q_{2}}{8\pi
\left\vert \vec{\rho}_{12}\right\vert },  \label{6}
\end{equation}%
which agrees with the expectation from the $c\rightarrow \infty $
limit of the corresponding expression in Eq. (\ref{2}). \

\ More problematic is%
\begin{equation}
\frac{3Q_{1}Q_{2}}{8\pi \left\vert \vec{\rho}_{12}\right\vert }\frac{%
m^{2}c^{2}}{m^{2}c^{2}+(\vec{\pi}_{12}\cdot \frac{\vec{\rho}_{12}}{%
\left\vert \vec{\rho}_{12}\right\vert })^{2}}.
\end{equation}%
Its Weyl ordering is
\begin{eqnarray}
K_{2}^{W}\psi (\vec{\rho}_{12}) &=&\frac{3m^{2}c^{2}}{(2\pi )^{3}}\frac{%
Q_{1}Q_{2}}{8\pi }\int \int \exp \left( i(\vec{\rho}_{12}-\vec{\rho}%
_{12}^{\prime })\cdot \vec{\pi}_{12}\right) \frac{2}{\left\vert \vec{\rho}%
_{12}+\vec{\rho}_{12}^{\prime }\right\vert }\frac{1}{(m^{2}c^{2}+(\vec{\pi}%
_{12}\cdot \hat{n}_{12})^{2})}  \nonumber \\
&&\times \psi (\vec{\rho}_{12}^{\prime })d^{3}\rho _{12}d^{3}\pi
_{12},
\end{eqnarray}

Let us focus on the $\vec{\pi}_{12}$ integral%
\begin{equation}
I=\int \exp \left( i(\vec{\rho}_{12}-\vec{\rho}_{12}^{\prime })\cdot \vec{\pi%
}_{12}\right) \frac{1}{(m^{2}c^{2}+(\hat{n}_{12})^{2})}d^{3}\pi
_{12}.
\end{equation}%
Let us divide $\vec{\pi}_{12}=\vec{\pi}_{12}\cdot (\hat{n}_{12})\hat{n}_{12}+%
\vec{\pi}_{12\perp }.$ \ Then we obtain%
\begin{equation}
I=\int \exp \left( i(\vec{\rho}_{12}-\vec{\rho}_{12}^{\prime })\cdot
\left( \vec{\pi}_{12}\cdot (\hat{n}_{12})\right)
\hat{n}_{12}+\vec{\pi}_{12\perp
}\right) \frac{1}{(m^{2}c^{2}+(\vec{\pi}_{12}\cdot \hat{n}_{12})^{2})}d^{2}%
\vec{\pi}_{12\perp }d\left( \vec{\pi}_{12}\cdot
(\hat{n}_{12})\right) .
\end{equation}%
Perform the $d^{2}\vec{\pi}_{12\perp }$ integral and call $k=\vec{\pi}%
_{12}\cdot (\hat{n}_{12}).$ Then, with%
\begin{equation}
(2\pi )^{2}\delta ^{2}((\vec{\rho}_{12}-\vec{\rho}_{12}^{\prime
})_{\perp })\equiv \int \exp \left(
i(\vec{\rho}_{12}-\vec{\rho}_{12}^{\prime })\cdot \vec{\pi}_{12\perp
}\right) d^{2}\vec{\pi}_{12\perp },  \label{2dlt}
\end{equation}%
we have
\begin{eqnarray}
I &=&(2\pi )^{2}\delta ^{2}((\vec{\rho}_{12}-\vec{\rho}_{12}^{\prime
})_{\perp })\int_{-\infty }^{\infty }\frac{\exp \left( i(\vec{\rho}_{12}-%
\vec{\rho}_{12}^{\prime })\cdot \hat{n}_{12}k\right)
}{m^{2}c^{2}+k^{2}}dk
\nonumber \\
&=&\frac{(2\pi )^{3}\delta
^{2}((\vec{\rho}_{12}-\vec{\rho}_{12}^{\prime
})_{\perp })}{2mc}\exp (-mc\left\vert (\vec{\rho}_{12}-\vec{\rho}%
_{12}^{\prime })\cdot \hat{n}_{12}\right\vert ).
\end{eqnarray}%
and so%
\begin{eqnarray}
K_{2}^{W}\psi (\vec{\rho}_{12}) &=&\frac{3m^{2}c^{2}}{(2\pi )^{3}}\frac{%
Q_{1}Q_{2}}{8\pi }\int \int \exp \left( i(\vec{\rho}_{12}-\vec{\rho}%
_{12}^{\prime })\cdot \vec{\pi}_{12}\right) \frac{2}{\left\vert \vec{\rho}%
_{12}+\vec{\rho}_{12}^{\prime }\right\vert }  \nonumber \\
&&\times \frac{1}{(m^{2}c^{2}+(\vec{\pi}_{12}\cdot \hat{n}_{12})^{2})}\psi (%
\vec{\rho}_{12}^{\prime })d^{3}\rho _{12}d^{3}\pi _{12}  \nonumber \\
&=&3mc\frac{Q_{1}Q_{2}}{8\pi }\int \frac{\exp (-mc\left\vert (\vec{\rho}%
_{12}-\vec{\rho}_{12}^{\prime })\cdot \hat{n}_{12}\right\vert
)}{\left\vert
\vec{\rho}_{12}+\vec{\rho}_{12}^{\prime }\right\vert }\delta ^{2}((\vec{\rho}%
_{12}-\vec{\rho}_{12}^{\prime })_{\perp })\psi
(\vec{\rho}_{12}^{\prime })d^{3}\rho _{12}.
\end{eqnarray}%
We recover the non-relativistic limit by using the one dimensional
expression
for the delta function of%
\begin{equation}
\delta ((\vec{\rho}_{12}-\vec{\rho}_{12}^{\prime })\cdot
\hat{n}_{12})=\lim
(c\rightarrow \infty )\frac{~mc}{2}\exp (-mc\left\vert (\vec{\rho}_{12}-\vec{%
\rho}_{12}^{\prime })\cdot \hat{n}_{12}\right\vert ).  \label{5}
\end{equation}%
Thus in that limit%
\begin{eqnarray}
K_{2}^{W}\psi (\vec{\rho}_{12}) &\rightarrow
&\frac{3Q_{1}Q_{2}}{4\pi }\int \frac{\exp (-mc\left\vert
(\vec{\rho}_{12}-\vec{\rho}_{12}^{\prime })\cdot
\hat{n}_{12}\right\vert )}{\left\vert \vec{\rho}_{12}+\vec{\rho}%
_{12}^{\prime }\right\vert }\delta
((\vec{\rho}_{12}-\vec{\rho}_{12}^{\prime })\cdot
\hat{n}_{12})\delta ^{2}((\vec{\rho}_{12}-\vec{\rho}_{12}^{\prime
})_{\perp })\psi (\vec{\rho}_{12}^{\prime })d^{3}\rho _{12}  \nonumber \\
&\rightarrow &\frac{3Q_{1}Q_{2}}{8\pi \left\vert \vec{\rho}_{12}\right\vert }%
\psi (\vec{\rho}_{12}),  \label{7}
\end{eqnarray}%
as expected.

The next term to Weyl transform in Eq.(\ref{2}) is%
\begin{equation}
-\frac{Q_{1}Q_{2}}{2\pi \left\vert \vec{\rho}_{12}\right\vert }[\frac{%
m^{4}c^{4}}{\ (m^{2}c^{2}+\vec{\pi}_{12}^{2})(m^{2}c^{2}+(\vec{\pi}%
_{12}\cdot \frac{\vec{\rho}_{12}}{\left\vert \vec{\rho}_{12}\right\vert }%
)^{2})}]
\end{equation}%
The corresponding Weyl transform is%
\begin{eqnarray*}
K^{W}\psi (\vec{\rho}_{12}) &=&-\frac{m^{4}c^{4}}{(2\pi )^{3}}\frac{%
Q_{1}Q_{2}}{\pi }\int \int \frac{\exp \left( i(\vec{\rho}_{12}-\vec{\rho}%
_{12}^{\prime })\cdot \vec{\pi}_{12}\right) }{\left\vert \vec{\rho}_{12}+%
\vec{\rho}_{12}^{\prime }\right\vert }\frac{1}{\ (m^{2}+\vec{\pi}%
_{12}^{2})(m^{2}+(\vec{\pi}_{12}\cdot \hat{n}_{12})^{2})} \\
&&\times \psi (\vec{\rho}_{12}^{\prime })d^{3}\rho _{12}d^{3}\pi
_{12},
\end{eqnarray*}%
We focus on the Fourier transform%
\begin{equation}
J=\frac{1}{(2\pi )^{3}}\int \exp \left( i(\vec{\rho}_{12}-\vec{\rho}%
_{12}^{\prime })\cdot \vec{\pi}_{12}\right) \frac{d^{3}\pi _{12}}{\
(m^{2}c^{2}+\vec{\pi}_{12}^{2})(m^{2}c^{2}+(\vec{\pi}_{12}\cdot \hat{n}%
_{12})^{2})}.
\end{equation}%
Let us recall that if
\begin{eqnarray}
\tilde{f}(\pi _{12}) &=&\int \exp \left( -i\pi _{12}\cdot \vec{\rho}%
_{12}\right) f(\vec{\rho}_{12})d^{3}\eta ,  \nonumber \\
\tilde{g}(\pi _{12}) &=&\int \exp \left( -i\pi _{12}\cdot \vec{\rho}%
_{12}^{\prime }\right) g(\vec{\rho}_{12}^{\prime })d^{3}\eta
^{\prime },
\end{eqnarray}%
then we obtain the convolution result of
\begin{eqnarray}
&&\frac{1}{(2\pi )^{3}}\int \exp \left( i(\vec{\rho}_{12}-\vec{\rho}%
_{12}^{\prime })\cdot \pi _{12}\right) \tilde{f}(\pi
_{12})\tilde{g}(\pi
_{12})d^{3}\pi _{12}  \nonumber \\
&=&\int f(\vec{\rho}_{12})g(\vec{\rho}_{12}-\vec{\rho}_{12}^{\prime }-\vec{%
\rho}_{12}^{\prime })d^{3}\eta .
\end{eqnarray}%
Thus, with
\begin{eqnarray}
\frac{1}{\ (m^{2}c^{2}+\vec{\pi}_{12}^{2})} &=&\frac{1}{4\pi }\int
\exp \left( -i\vec{\pi}_{12}\cdot \vec{\rho}_{12}\right) \frac{\exp
(-mc\left\vert \vec{\rho}_{12}\right\vert )}{\left\vert \vec{\rho}%
_{12}\right\vert }d^{3}\eta ,  \nonumber \\
\frac{1}{\ (m^{2}c^{2}+(\vec{\pi}_{12}\cdot \hat{n}_{12})^{2})}
&=&\int \exp \left( -i\vec{\pi}_{12}\cdot \vec{\rho}_{12}\right)
\frac{(2\pi )^{3}\delta _{\perp }^{2}(\vec{\rho}_{12})}{2m}\exp
(-mc\left\vert \vec{\rho}_{12}\cdot \hat{n}_{12}\right\vert ^{\prime
}),
\end{eqnarray}%
we have%
\begin{eqnarray}
J &=&\frac{1}{(2\pi )^{3}}\int \exp \left( i(\vec{\rho}_{12}-\vec{\rho}%
_{12}^{\prime })\cdot \pi _{12}\right) \frac{1}{\ (m^{2}c^{2}+\vec{\pi}%
_{12}^{2})}\frac{1}{\ (m^{2}c^{2}+(\vec{\pi}_{12}\cdot \hat{n}_{12})^{2})}%
d^{3}\pi _{12}  \nonumber \\
&=&\int \frac{\exp (-mc\left\vert \vec{\rho}_{12}\right\vert )}{4\pi
\left\vert \vec{\rho}_{12}\right\vert }\frac{(2\pi )^{3}\delta _{\perp }^{2}(%
\vec{\rho}_{12}-\vec{\rho}_{12}^{\prime }-\vec{\rho}_{12}^{\prime \prime })}{%
2mc}\exp \left( -mc\left\vert \left(
\vec{\rho}_{12}-\vec{\rho}_{12}^{\prime }-\vec{\rho}_{12}^{\prime
\prime }\right) \cdot \hat{n}_{12}\right\vert \right) d^{3}\rho
_{12}^{\prime \prime }.\nonumber \\
&&{}
\end{eqnarray}%
Hence,%
\begin{eqnarray}
&&K_{3}^{W}\psi (\vec{\rho}_{12})  \nonumber \\
&=&-\frac{m^{4}c^{4}}{(2\pi )^{3}}\frac{Q_{1}Q_{2}}{\pi }\int \int \frac{%
\exp \left( i(\vec{\rho}_{12}-\vec{\rho}_{12}^{\prime })\cdot \vec{\pi}%
_{12}\right) }{\left\vert \vec{\rho}_{12}+\vec{\rho}_{12}^{\prime
}\right\vert }\frac{1}{\ (m^{2}c^{2}+\vec{\pi}_{12}^{2})(m^{2}c^{2}+(\vec{\pi%
}_{12}\cdot \hat{n}_{12})^{2})}  \nonumber \\
&&\times \psi (\vec{\rho}_{12}^{\prime })d^{3}\rho _{12}d^{3}\pi
_{12}
\nonumber \\
&=&-\frac{m^{3}c^{3}Q_{1}Q_{2}}{8\pi ^{2}}\int \frac{\exp
(-mc\left\vert
\vec{\rho}_{12}^{\prime \prime }\right\vert )}{\left\vert \vec{\rho}%
_{12}^{\prime \prime }\right\vert })\frac{\delta _{\perp }^{2}(\vec{\rho}%
_{12}-\vec{\rho}_{12}^{\prime }-\vec{\rho}_{12}^{\prime \prime })}{%
\left\vert \vec{\rho}_{12}+\vec{\rho}_{12}^{\prime }\right\vert }
\nonumber
\\
&&\times \exp \left( -mc\left\vert \left( \vec{\rho}_{12}-\vec{\rho}%
_{12}^{\prime }-\vec{\rho}_{12}^{\prime \prime }\right) \cdot \hat{n}%
_{12}\right\vert \right) \psi (\vec{\rho}_{12}^{\prime })d^{3}\rho
_{12}^{\prime }d^{3}\rho _{12}^{\prime \prime }
\end{eqnarray}%
Using the delta function expression \ in Eq. (\ref{4}) and (\ref{5})
we see
that in the $c\rightarrow \infty $ limit the above becomes%
\begin{equation}
K_{3}^{W}\psi (\vec{\rho}_{12})\rightarrow -\frac{Q_{1}Q_{2}}{2\pi
\left\vert \vec{\rho}_{12}\right\vert },
\end{equation}%
which when combined with Eq. (\ref{6}) and (\ref{7}) produces the
correct non-relativistic limit of Eq. (\ref{8}) expected from Eq.
(\ref{1}). This
completes the first portion of the Weyl quantization. \ \footnote{%
We point out that had we chosen not to make the rearrangement of Eq. (\ref{1}%
) , in the quantization, a cancelation of the local Coulomb
potential would still have taken place by the multiple derivatives
(of the nonlocal Yukawa kernels) that come from the higher order
momentum terms in the numerator.}

\bigskip

\subsubsection{ Combined Non-local Weyl Ordered Hamiltonian}

\bigskip
The second portion of the classical Darwin Hamiltonian is

\begin{eqnarray}
&&-{\frac{Q_{1}Q_{2}}{4\pi \left\vert \vec{\rho}_{12}\right\vert }}\frac{%
\vec{\pi}_{12}^{2}[\vec{\pi}_{12}^{2}-3(\vec{\pi}_{12}\cdot \frac{\vec{\rho}%
_{12}}{\left\vert \vec{\rho}_{12}\right\vert })^{2}]\sqrt{\frac{m^{2}c^{2}+%
\vec{\pi}_{12}^{2}}{m^{2}c^{2}+(\vec{\pi}_{12}\cdot \frac{\vec{\rho}_{12}}{%
\left\vert \vec{\rho}_{12}\right\vert })^{2}}}}{(m^{2}c^{2}+\vec{\pi}%
_{12}^{2})[m^{2}c^{2}+(\vec{\pi}_{12}\cdot
\frac{\vec{\rho}_{12}}{\left\vert
\vec{\rho}_{12}\right\vert })^{2}]}  \nonumber \\
&=&-{\frac{Q_{1}Q_{2}}{4\pi \left\vert \vec{\rho}_{12}\right\vert }[}\frac{%
\left( \vec{\pi}_{12}^{2}+m^{2}c^{2}\right) ^{3/2}}{(m^{2}c^{2}+(\vec{\pi}%
_{12}\cdot \frac{\vec{\rho}_{12}}{\left\vert \vec{\rho}_{12}\right\vert }%
)^{2})^{3/2}}-\frac{3\left( \vec{\pi}_{12}^{2}+m^{2}c^{2}\right) ^{1/2}}{%
(m^{2}c^{2}+(\vec{\pi}_{12}\cdot \frac{\vec{\rho}_{12}}{\left\vert \vec{\rho}%
_{12}\right\vert })^{2})^{1/2}}  \nonumber \\
&&+\frac{m^{2}c^{2}\left( \vec{\pi}_{12}^{2}+m^{2}c^{2}\right) ^{1/2}}{%
(m^{2}c^{2}+(\vec{\pi}_{12}\cdot \frac{\vec{\rho}_{12}}{\left\vert \vec{\rho}%
_{12}\right\vert })^{2})^{3/2}}+\frac{3m^{2}c^{2}}{(m^{2}c^{2}+\vec{\pi}%
_{12}^{2})^{1/2}(m^{2}c^{2}+(\vec{\pi}_{12}\cdot \frac{\vec{\rho}_{12}}{%
\left\vert \vec{\rho}_{12}\right\vert })^{2})^{1/2}}  \nonumber \\
&&-\frac{2m^{4}c^{4}}{(m^{2}c^{2}+\vec{\pi}_{12}^{2})^{1/2}(m^{2}c^{2}+(\vec{%
\pi}_{12}\cdot \frac{\vec{\rho}_{12}}{\left\vert \vec{\rho}_{12}\right\vert }%
)^{2})^{3/2}}].
\end{eqnarray}%
Each Weyl transform would involve a convolution. \ The first is%
\begin{eqnarray}
K_{4}^{W}\psi (\vec{\rho}_{12}) &=&-\frac{1}{(2\pi )^{3}}\frac{2Q_{1}Q_{2}}{%
4\pi }\int \int \frac{\exp \left(
i(\vec{\rho}_{12}-\vec{\rho}_{12}^{\prime
})\cdot \vec{\pi}_{12}\right) }{\left\vert \vec{\rho}_{12}+\vec{\rho}%
_{12}^{\prime }\right\vert }  \nonumber \\
&&\times \frac{\left( \vec{\pi}_{12}^{2}+m^{2}c^{2}\right) ^{3/2}}{\
(m^{2}c^{2}+(\vec{\pi}_{12}\cdot \hat{n}_{12})^{2})^{3/2}}\psi (\vec{\rho}%
_{12}^{\prime })d^{3}\rho _{12}d^{3}\pi _{12}.
\end{eqnarray}%
As with $K_{3}^{W}\psi (\vec{\rho}_{12})$ it involves a convolution
\begin{equation}
\frac{1}{(2\pi )^{3}}\int \exp \left( i(\vec{\rho}_{12}-\vec{\rho}%
_{12}^{\prime })\cdot \pi _{12}\right) \tilde{f}(\pi
_{12})\tilde{g}(\pi
_{12})d^{3}\pi _{12}=\int f(\vec{\rho}_{12}^{\prime \prime })g(\vec{\rho}%
_{12}-\vec{\rho}_{12}^{\prime }-\vec{\rho}_{12}^{\prime \prime
})d^{3}\eta .
\end{equation}%
Now defining
\begin{eqnarray}
K_{T1}(\vec{\rho}_{12}) &=&\frac{1}{(2\pi )^{3}}\int \exp (i\pi
_{12}\cdot \vec{\rho}_{12})\left(
\vec{\pi}_{12}^{2}+m^{2}c^{2}\right) ^{3/2}d^{3}\pi
_{12},  \nonumber \\
K_{K1}(\vec{\rho}_{12}) &=&\frac{1}{(2\pi )^{3}}\int \frac{\exp
(i\pi
_{12}\cdot \vec{\rho}_{12})}{(m^{2}c^{2}+(\vec{\pi}_{12}\cdot \hat{n}%
_{12})^{2})^{3/2}}d^{3}\pi _{12},  \label{KTK1}
\end{eqnarray}%
we have%
\begin{equation}
K_{4}^{W}\psi (\vec{\rho}_{12})=-\frac{2Q_{1}Q_{2}}{4\pi }\int \frac{K_{T1}(%
\vec{\rho}_{12}^{\prime \prime })K_{K1}(\vec{\rho}_{12}-\vec{\rho}%
_{12}^{\prime }-\vec{\rho}_{12}^{\prime \prime })}{\left\vert \vec{\rho}%
_{12}+\vec{\rho}_{12}^{\prime }\right\vert }\psi
(\vec{\rho}_{12}^{\prime })d^{3}\rho _{12}^{\prime }d^{3}\rho
_{12}^{\prime \prime }.
\end{equation}

The next portion is%
\begin{eqnarray}
K_{5}^{W}\psi (\vec{\rho}_{12}) &=&\frac{3}{(2\pi )^{3}}\frac{2Q_{1}Q_{2}}{%
4\pi }\int \int \frac{\exp \left(
i(\vec{\rho}_{12}-\vec{\rho}_{12}^{\prime
})\cdot \vec{\pi}_{12}\right) }{\left\vert \vec{\rho}_{12}+\vec{\rho}%
_{12}^{\prime }\right\vert }  \nonumber \\
&&\times \frac{\left( \vec{\pi}_{12}^{2}+m^{2}c^{2}\right) ^{1/2}}{\
(m^{2}c^{2}+(\vec{\pi}_{12}\cdot \hat{n}_{12})^{2})^{1/2}}\psi (\vec{\rho}%
_{12}^{\prime })d^{3}\rho _{12}d^{3}\pi _{12}.
\end{eqnarray}%
Defining%
\begin{equation}
K_{K2}(\vec{\rho}_{12})=\frac{1}{(2\pi )^{3}}\int \frac{\exp (i\vec{\pi}%
_{12}\cdot \vec{\rho}_{12})}{(m^{2}c^{2}+(\vec{\pi}_{12}\cdot \hat{n}%
_{12})^{2})^{1/2}}d^{3}\pi _{12},  \label{K2}
\end{equation}%
we have%
\begin{equation}
K_{5}^{W}\psi (\vec{\rho}_{12})=\frac{6Q_{1}Q_{2}}{4\pi }\int \frac{K_{T}(%
\vec{\rho}_{12}^{\prime \prime })K_{K2}(\vec{\rho}_{12}-\vec{\rho}%
_{12}^{\prime }-\vec{\rho}_{12}^{\prime \prime })}{\left\vert \vec{\rho}%
_{12}+\vec{\rho}_{12}^{\prime }\right\vert }\psi
(\vec{\rho}_{12}^{\prime })d^{3}\rho _{12}^{\prime }d^{3}\rho
_{12}^{\prime \prime }.
\end{equation}

Following this term is%
\begin{eqnarray}
K_{6}^{W}\psi (\vec{\rho}_{12}) &=&-\frac{m^{2}c^{2}}{(2\pi )^{3}}\frac{%
2Q_{1}Q_{2}}{4\pi }\int \int \frac{\exp \left( i(\vec{\rho}_{12}-\vec{\rho}%
_{12}^{\prime })\cdot \vec{\pi}_{12}\right) }{\left\vert \vec{\rho}_{12}+%
\vec{\rho}_{12}^{\prime }\right\vert }  \nonumber \\
&&\times \frac{\left( \vec{\pi}_{12}^{2}+m^{2}c^{2}\right) ^{1/2}}{\
(m^{2}c^{2}+(\vec{\pi}_{12}\cdot \hat{n}_{12})^{2})^{3/2}}\psi (\vec{\rho}%
_{12}^{\prime })d^{3}\rho _{12}d^{3}\pi _{12}
\end{eqnarray}%
and its contribution is%
\begin{equation}
K_{6}^{W}\psi (\vec{\rho}_{12})=-\frac{2m^{2}c^{2}Q_{1}Q_{2}}{4\pi
}\int
\frac{K_{T}(\vec{\rho}_{12}^{\prime \prime })K_{K1}(\vec{\rho}_{12}-\vec{\rho%
}_{12}^{\prime }-\vec{\rho}_{12}^{\prime \prime }}{\left\vert \vec{\rho}%
_{12}+\vec{\rho}_{12}^{\prime }\right\vert }\psi
(\vec{\rho}_{12}^{\prime })d^{3}\rho _{12}^{\prime }d^{3}\rho
_{12}^{\prime }.
\end{equation}

The next term is%
\begin{eqnarray}
K_{7}^{W}\psi (\vec{\rho}_{12}) &=&-\frac{3m^{2}c^{2}}{(2\pi )^{3}}\frac{%
2Q_{1}Q_{2}}{4\pi }\int \int \frac{\exp \left( i(\vec{\rho}_{12}-\vec{\rho}%
_{12}^{\prime })\cdot \vec{\pi}_{12}\right) }{\left\vert \vec{\rho}_{12}+%
\vec{\rho}_{12}^{\prime }\right\vert }  \nonumber \\
&&\times \frac{1}{\ \left( \vec{\pi}_{12}^{2}+m^{2}c^{2}\right)
^{1/2}(m^{2}c^{2}+(\vec{\pi}_{12}\cdot \hat{n}_{12})^{2})^{1/2}}\psi (\vec{%
\rho}_{12}^{\prime })d^{3}\rho _{12}d^{3}\pi _{12}.
\end{eqnarray}%
and with%
\begin{equation}
K_{T2}(\vec{\rho}_{12})=\frac{1}{(2\pi )^{3}}\int \frac{\exp (i\vec{\pi}%
_{12}\cdot \vec{\rho}_{12})}{\left(
\vec{\pi}_{12}^{2}+m^{2}c^{2}\right) ^{1/2}}d^{3}\pi _{12},
\label{KT2}
\end{equation}%
we have%
\begin{equation}
K_{7}^{W}\psi (\vec{\rho}_{12})=-\frac{6m^{2}c^{2}Q_{1}Q_{2}}{4\pi
}\int
\frac{K_{T2}(\vec{\rho}_{12}^{\prime \prime })K_{K2}(\vec{\rho}_{12}-\vec{%
\rho}_{12}^{\prime }-\vec{\rho}_{12}^{\prime \prime })}{\left\vert \vec{\rho}%
_{12}+\vec{\rho}_{12}^{\prime }\right\vert }\psi
(\vec{\rho}_{12}^{\prime })d^{3}\rho _{12}^{\prime }d^{3}\rho
_{12}^{\prime \prime }.
\end{equation}

The final term is%
\begin{eqnarray}
K_{8}^{W}\psi (\vec{\rho}_{12}) &=&\frac{2m^{4}c^{4}}{(2\pi )^{3}}\frac{%
2Q_{1}Q_{2}}{4\pi }\int \int \frac{\exp \left( i(\vec{\rho}_{12}-\vec{\rho}%
_{12}^{\prime })\cdot \vec{\pi}_{12}\right) }{\left\vert \vec{\rho}_{12}+%
\vec{\rho}_{12}^{\prime }\right\vert }  \nonumber \\
&&\times \frac{1}{\ \left( \vec{\pi}_{12}^{2}+m^{2}c^{2}\right)
^{1/2}(m^{2}c^{2}+(\vec{\pi}_{12}\cdot \hat{n}_{12})^{2})^{3/2}}\psi (\vec{%
\rho}_{12}^{\prime })d^{3}\rho _{12}d^{3}\pi _{12},
\end{eqnarray}%
and it contributes%
\begin{equation}
K_{8}^{W}\psi (\vec{\rho}_{12})=-\frac{4m^{4}c^{4}Q_{1}Q_{2}}{4\pi
}\int
\frac{K_{T2}(\vec{\rho}_{12}^{\prime \prime })K_{K1}(\vec{\rho}_{12}-\vec{%
\rho}_{12}^{\prime }-\vec{\rho}_{12}^{\prime \prime })}{\left\vert \vec{\rho}%
_{12}+\vec{\rho}_{12}^{\prime }\right\vert }\psi
(\vec{\rho}_{12}^{\prime })d^{3}\rho _{12}^{\prime \prime }.
\end{equation}%
Although in the $c\rightarrow \infty $ limit each of the Weyl
ordered terms is finite, they cancel altogether.

Altogether, our Weyl order Hamiltonian is%

\begin{equation}
2K_{T}(\vec{\rho}_{12})+\sum_{n=1}^{8}K_{n}^{W}(\vec{\rho}_{12}).
\end{equation}

\vfill\eject


\begin{thebibliography}{99}

\bibitem{1} C.Cohen-Tannoudji, J. Dupont-Roc and G.Grynberg, \textit{
Atom-Photon Interactions. Basic Processes and Applications} (Wiley,
New York, 1992).

\bibitem{2} C.Cohen-Tannoudji, J. Dupont-Roc and G.Grynberg, \textit{
Photons and Atoms. Introduction to Quantum Electrodynamics} (Wiley,
New York, 1989).

\bibitem{3} W.P.Schleich, \textit{Quantum Optics in Phase Space} (Wiley-VCH,
Berlin, 2001).

\bibitem{4}L.Lusanna, {\it The N- and 1-Time Classical
Descriptions of N-Body Relativistic Kinematics and the
Electromagnetic Interaction}, Int.J.Mod.Phys. {\bf A12}, 645 (1997).

\bibitem{5}L.Lusanna, {\it The Chrono-Geometrical Structure of Special and
General Relativity: A Re-Visitation of Canonical Geometrodynamics},
lectures at 42nd Karpacz Winter School of Theoretical Physics:
Current Mathematical Topics in Gravitation and Cosmology, Ladek,
Poland, 6-11 Feb 2006, Int.J.Geom.Methods in Mod.Phys. {\bf 4}, 79
(2007). (gr-qc/0604120).\hfill\break
  D.Alba and L.Lusanna,  \textit{Generalized Radar 4-Coordinates and
Equal-Time Cauchy Surfaces for Arbitrary Accelerated Observers}
(2005), Int.J.Mod.Phys. {\bf D16}, 1149 (2007) (gr-qc/0501090).




\bibitem{6} D.Alba, L.Lusanna and M.Pauri, \textit{New Directions in
Non-Relativistic and Relativistic Rotational and Multipole
Kinematics for N-Body and Continuous Systems} (2005), in
\textit{Atomic and Molecular Clusters: New Research}, ed.Y.L.Ping
(Nova Science, New York, 2006) (hep-th/0505005).\hfill\break D.Alba,
L.Lusanna and M.Pauri, \textit{Centers of Mass and Rotational
Kinematics for the Relativistic N-Body Problem in the Rest-Frame
Instant Form}, J.Math.Phys. \textbf{43}, 1677-1727 (2002)
(hep-th/0102087).\hfill\break D.Alba, L.Lusanna and M.Pauri,
\textit{ Multipolar Expansions for Closed and Open Systems of
Relativistic Particles} , J. Math.Phys. \textbf{46}, 062505, 1-36
(2004) (hep-th/0402181).

\bibitem{7} D.Alba, H.W.Crater and L.Lusanna, \textit{Hamiltonian
Relativistic Two-Body Problem: Center of Mass and Orbit
Reconstruction}, J.Phys. {\bf A40}, 9585 (2007) (gr-qc/0610200).


\bibitem{8}D.Alba, H.W.Crater and L.Lusanna, {\it Towards Relativistic
Atom Physics. I. The Rest-Frame Instant Form of Dynamics and a
Canonical Transformation for a system of Charged Particles plus the
Electro-Magnetic Field}, to appear in Canad.J.Phys. (arXiv:
0806.2383).

\bibitem{9}D.Alba and L.Lusanna,
{\it Charged Particles and the Electro-Magnetic Field in
Non-Inertial Frames: I. Admissible 3+1 Splittings of Minkowski
Spacetime and the Non-Inertial Rest Frames},  Int.J.Geom.Methods in
Physics {\bf 7}, 33 (2010) (0908.0213) and {\it II. Applications:
Rotating Frames, Sagnac Effect, Faraday Rotation, Wrap-up Effect
(0908.0215)}, Int.J.Geom.Methods in Physics, {\bf 7}, 185 (2010).




\bibitem{10} H.W.Crater and L.Lusanna, \textit{The Rest-Frame Darwin
Potential from the Lienard-Wiechert Solution in the Radiation
Gauge}, Ann.Phys. (N.Y.) \textbf{289}, 87 (2001).


\bibitem{11} D.Alba, H.W.Crater and L.Lusanna, \textit{The Semiclassical
Relativistic Darwin Potential for Spinning Particles in the Rest
Frame Instant Form: Two-Body Bound States with Spin 1/2
Constituents}, Int.J.Mod.Phys. \textbf{A16}, 3365-3478 (2001)
(hep-th/0103109).

\bibitem{12} D.Alba and L.Lusanna, \textit{Quantum Mechanics in Noninertial
Frames with a Multitemporal Quantization Scheme: I. Relativistic
Particles}, Int.J.Mod.Phys. \textbf{A21}, 2781 (2006)
(hep-th/0502194).


\bibitem{13}L.Bel, {\it Mecanica Relativista Predictiva}, courso
impartido en el Departamento di Fisica Teorica de la Univrsidad
Autonoma de Barcelona, UAB FT-34 (1977); {\it Dynamique des systèmes
de N particules ponctuelles en relativité restreinte},
Ann.Inst.Henry Poincare' {\bf 12}, 307 (1970); {\it Predictive
Relativistic Mechanics.}, Ann.Inst.Henry Poincare' {\bf 14}, 189
(1971).\hfill\break L.Bel and F.X.Fustero, {\it  Predictive
Relativistic Mechanics of n Particle Systems.}, Ann.Inst.Henry
Poincare' {\bf 25}, 411 (1976).\hfill\break L.Bel and J.Martin, {\it
Hamiltonians and Conservative Systems}, Ann.Inst.Henry Poincare'
{\bf 22}, 173 (1975) and {\it Predictive Relativistic Mechanics of
Systems of N Particles with Spin}, {\bf 33}, 409 (1980).\hfill\break
  Ph.Droz Vincent, \textit{Relativistic
Systems of Interacting Particles}, Phys.Scr. \textbf{2}, 129 (1970);
\textit{ \ \ Hamiltonian Systems of Relativistic Particles}, Rep.
Math. Phys.,\textbf{ 8} ,79 (1975); \textit{Two-Body Relativistic
Systems}, Ann.Inst.H.Poincar \'{e} \textbf{27}, 407 (1977) and
\textit{N-Body Relativistic Systems}, \textbf{32A }, 377 (1980).


\bibitem{14}D.Alba, H.W.Crater and L.Lusanna, {\it Towards Relativistic
Atom Physics. II. Collective and Relative Relativistic Variables for
a System of Charged Particles plus the Electro-Magnetic Field}, to
appear in Canad.J.Phys. (arXiv:0811.0715).



\bibitem{15} G.Longhi and L.Lusanna, \textit{
Bound-State Solutions, Invariant Scalar Products and Conserved
Currents for a Class of Two-Body Relativistic Systems}, Phys.Rev.
\textbf{D34}, 3707 (1986).




\bibitem{16}G.N.Fleming, {\it Covariant Position Operators, Spin
and Locality}, Phys.Rev. {\bf 137B}, 188 (1965); {\it Non-Local
Properties of Stable Particles}, Phys.Rev. {\bf 139B}, 963 (1965);
{\it A Manifestly Covariant Description of Arbitrary Dynamical
Variables in Relativistic Quantum Theory}, J.Math.Phys. {\bf 7},
1959 (1966);  {\it Just How Radical is Hyperplane Dependence?}, in
{\it Perspectives on Quantum Reality}, ed. R.Clifton, p.11 (Kluwer,
Dordrecht, 1996); {\it The Dependence of Lorentz Boost Generators on
the Presence and Nature of Interactions}
(http://philsci-archive.pitt.edu/archive/00000633/); {\it
Observation on Hyperplanes: I. State Reduction and Unitary
Evolution} (2003)
(http://philsci-archive.pitt.edu/archive/00001533/); {\it
Observation on Hyperplanes:II. Dynamical Variables and Localization
Observables} (2004)
(http://philsci-archive.pitt.edu/archive/00002085/). \hfill\break
  G.N.Fleming and H.Bennet, {\it Hyperplane Dependence in
Relativistic Quantum Mechanics}, Found.Phys. {\bf 19}, 231 (1989).





\bibitem{17} J.Butterfield and G.Fleming, {\it Strange Positions}, in {\it
From Physics to Philosophy}, eds. J.Butterfield and C.Pagonis,
pp.108-165 (Cambridge Univ.Press, Cambridge, 1999).




\bibitem{18}G.N.Fleming,  {\it Reeh-Schlieder meets Newton-Wigner},
Philosophy of Science {\bf 67}, S495 (2000).


\bibitem{19} W.H. Zurek, {\it Decoherence, Einselection, and the Quantum Origins of the
Classical},  Rev.Mod.Phys. {\bf 75}, 715-775 (2003).

\bibitem{20} D.Alba, L.Lusanna and M.Pauri, \textit{Dynamical Body Frames,
Orientation-Shape Variables and Canonical Spin Bases for the
Nonrelativistic N-Body Problem} J.Math.Phys. \textbf{43}, 373 (2002)
(hep-th/0011014).\hfill\break R.G.Littlejohn and M.Reinsch,
\textit{Gauge Fields in the Separation of Rotations and Internal
Motions in the N Body Problem}, Rev.Mod.Phys. \textbf{69}, 213
(1997).




\bibitem{21} A.Einstein, B.Podolski and N.Rosen, {\it Can Quantum-Mechanical
Description of Physical Reality be Considered Complete?}, Phys.Rev.
{\bf 47}, 777 (1935).


\bibitem{22}H.R.Brown, {\it Physical Relativity. Space-Time
Structure from a Dynamical Perspective} (Oxford Univ.Press, Oxford,
2005).\hfill\break C.G.Timpson and H.R.Brown, {\it Entanglement and
Relativity}, in {\it Understanding Physical Knowledge}, p.147,
preprint 24. Dip.Filosofia, Univ.Bologna (CLUEB, Bologna,
2002)(quant-ph/0212140)

\bibitem{23}A.Peres, {\it Quantum Theory: Concepts and Methods}
(Kluwer, Dordrecht, 1995).

\bibitem{24}M.Schlosshauer, {\it Decoherence, the Measurement Problem
and Interpretations of Quantum Mechanics}, Rev.Mod.Phys. {\bf 76},
1267 (2004).\hfill\break M.Schlosshauer, {\it Decoherence and the
Quantum-to-Classical Transition} (Springer, Berlin, 2007)

\bibitem{25}N.L.Harshman, {\it Basis States for Relativistic
Dinamically Entangled Particles}, Phys.Rev. {\bf A71}, 022312
(2005); {\it Dynamical Entanglement in Particle Scattering},
Int.J.Mod.Phys. {\bf A20}, 6220 (2005)(quant-ph/0506212); {\it
Poincare' Semigroup Symmetry as an Emergent Property of Unstable
Systems} (hep-ph/0511298); {\it Limits on Entaglement in
Rotationally-Invariant Scattering of Spin Systems}
(quant-ph/0509013).

\bibitem{26}A.Peres and D.R.Terno, {\it Quantum Information and
Relativity Theory}, Rev.Mod.Phys. {\bf 76}, 93
(2004)(quant-ph/0212023).\hfill\break D.R.Terno, {\it Introduction
to Relativistic Quantum Information} (quant-ph/0508049).\hfill\break
O.Hay and A.Peres, {\it Quantum and Classical Descriptions of a
Measuring Apparatus}, Phys.Rev. {\bf A58}, 116 (1998)
(quant-ph/9712044).


\bibitem{27}H.Feshbach and F.Villars, {\it Elementary Relativistic Wave Mechanics
of Spin 0 and Spin 1/2 Particles}, Rev.Mod.Phys. {\bf 30}, 24
(1958).


\bibitem{28} T.A.Debs and M.L.G.Redhead, {\it The 'Jericho Effect' and
Hegerfeldt Non-Locality}, Studies in History and Philosophy of
Science {\bf 34}, 61 (2003).

\bibitem{29}G.C.Hegerfeldt, {\it Remark on Causality and Particle
Localization}, Phys.Rev. {\bf D10}, 3320 (1974).\hfill\break
G.C.Hegerfeldt, {\it Violation of Causality in Relativistic Quantum
Theory?}, Phys.Rev.Lett. {\bf 54}, 2395 (1985).

\bibitem{30}G.C.Hegerfeldt, {\it Instantaneous Spreading and
Einstein Causality in Quantum Theory}, Ann.Phys.Lpz. {\bf 7}, 716
(1998) (quant-ph/9809030); {\it Causality, Particle Localization and
Positivity of the Energy},  in {\it Irreversibility and Causality in
Quantum Theory - Semigroups and Rigged Hilbert Spaces}, eds. A.Bohm,
H.D.Doebner and P.Kielanowski, Lecture Notes in Physics 504, p.238
(Springer , NewYork, 1998) (quant-ph/9806036).

\bibitem{31} H.Halvorson, {\it Reeh-Schlieder Defeats Newton-Wigner: on
Alternative Localization Schemes in Relativistic Quantum FField
Theory}, Philosophy of Science {\bf 68}, 111 (2001).

\bibitem{32}H.Reeh and S.Schlieder, {\it Bemerkungen zur Unitaraquivalenz
von Lorentzinvarianten Feldern}, Nuovo Cimento {\bf 22}, 1051
(1961).


\bibitem{33} D.Alba, \textit{Quantum Mechanics in Noninertial Frames with a
Multitemporal Quantization Scheme: II. Nonrelativistic Particles},
Int.J.Mod.Phys. \textbf{A21}, 3917 (hep-th/0504060).

\bibitem{34}H.Epstein, V.Glaser and A.Jaffe, {\it Nonpositivity of the
Energy Density in Quantized Field Theories}, Nuovo Cimento {\bf 36},
1016 (1965).


\bibitem{35}M.Pauri and G.Prosperi, {\it Canonical Realizations of
the Galilei Group}, J.Math.Phys. {\bf 9}, 1146 (1968).\hfill\break
R.DePietri, L.Lusanna and M.Pauri, {\it Gauging Kinematical and
Internal Symmetry Groups for Extended Systems: the Galilean One-Time
and Two-Times Harmonic Oscillators}, Class.Quantum Grav. {\bf 11},
1417 (1996).




\bibitem{36}G.Longhi, L.Lusanna and J.M.Pons, {\it On the
Many-Time Formulation of Classical Particle Dynamics}, J.Math.Phys.
{\bf 30}, 1893 (1989).


\bibitem{37}E.Corinaldesi and F.Strocchi, {\it Relativistic Wave
Equations} (North-Holland, Amsterdam, 1963).

\bibitem{38}C.J.Borde', {\it Quantum Theory of Atom-Wave Beam Splitters and
Applications to Multidimensional Atomic Gravito-Inertial Sensors},
Gen.Rel.Grav. {\bf 36}, 475 (2004).

\bibitem{30}D.Dominici, J.Gomis and G.Longhi, {\it A Lagrangian for Two
Interacting Relativistic Particles}, Nuovo Cimento {\bf B48}, 152
(1978).



\bibitem{40}C.Lammerzahl, {\it The Pseudo-Differential Operator
Square Root of the Klein-Gordon Equation}, J.Math.Phys. {\bf 34},
3918 (1993).

\bibitem{41}M.Hillery, R.O'Connell, M.Scully and E.Wigner,
{\it Distribution Functions in Physics: Fundamentals}, Phys.Rep.
{\bf 106}, 121 (1984).

\bibitem{42}I.T.Todorov, {\it Quasipotential Equation Corresponding
to the Relativistic Eikonal Approximation}, Phys.Rev. {\bf D3}, 2351
(1971).

\bibitem{43}H.W.Crater and P. Van Alstine, {\it Quantum Constraint
Dynamics for Two Spinless Particles under Vector Interaction},
Phys.Rev. {\bf D30}, 2585 (1984).

\bibitem{44} W.Lucha and F.F.Schoeberl, {\it Facets of the Spinless
Salpeter Equation}, 2004 (arXiv: hep-th/0408184).\hfill\break
R.L.Hall, W.lucha and F.F.Schoeberl, {\it Relativistic N-Boson
Systems Bound by Pair Potentials $V({\vec r}_{ij}) = g({\vec
r}_{ij}^2)$}, J.Math.Phys. {\bf 45}, 3086 (2004).

\bibitem{45}M.Reed and B.Simon, Methods of Modern
Mathematical Physics, Vol.II, Academic Press, NewYork, 1975.


\bibitem{46}T.D.Newton and E.P.Wigner, {\it Localized States for
Elementary Systems}, Rev.Mod.Phys. {\bf 21}, 400 (1949).


\bibitem{47}B.Thaller, {\it The Dirac Equation} (Springer, Berlin,
1992).

\bibitem{48} D.Malament, {\it In Defence of Dogma: why there cannot be a
Relativistic Quantum Mechanics of (Localizable) Particles}, in {\it
Perspectives on Quantum Reality}, ed. R.Clifton, p. 1-10 (Kluwer,
Dordrechet, 1996).

\bibitem{49} H.Halvorson and R.Clifton, {\it No Place for Particles in
Relativistic Quantum Theories?}, Philosophy of Science {\bf 69}, 1
(2002) (quant-ph/0103041).

\bibitem{50}R.Haag, {\it Local Quantum Physics} (Springer, Berlin, 1992).

\bibitem{51}D.Fraser, {\it The Fate of 'Particles' in Quantum Field Theories
with Interactions} (2008),
http://philsci-archive.pitt.edu/archive/00004038/

\bibitem{52}J.Earman and D.Fraser, {\it Haag's Theorem and its Implications
for the Foundations of Quantum Field Theory}, Erkenntnis {\bf 64},
305 (2006) (philsci-archive.pitt.edu/archive/00002673/).




\bibitem{53}I.E.Segal, {\it Quantum Fields and Analysis in the
Solution Manifolds of Differential Equations}, in Proc. of a Conf.
on {\it Theory and Applications of Analysis in Function Spaces},
p.129, eds. W.T.Martin and I.E.Segal (MIT Press, Cambridge, 1964).
\hfill\break I.E.Segal and R.W.Goodman, {\it Anti-Locality of
Certain Lorentz-Invariant Operators}, Journal of Mathematics and
Mechanics {\bf 14}, 629-638 (1965).

\bibitem{54} P.Busch, {\it Unsharp Localization and Causality in Relativistic
Quantum Theory}, J.Phys. {\bf A32}, 6535 (1999).


\bibitem{55}D.Wallace, {\it Emergence of Particles from Bosonic Quantum
Fields}(2001) (quant-ph/0112149).\hfill\break
  D.Wallace, {\it In Defence of Naivete: The Conceptual Status of Lagrangian
  QFT},( 2001) (quant-ph/0112148).

\bibitem{56}J.Larson, {\it Dynamics of the Jaynes-Cummings and Rabi
models: Old Wine in New Bottles}, Phys. Scripta {\bf 76}, 146
(2007).


\bibitem{57}E.P.Wigner, {\it  Die Messung quantenmechanischer
Operatoren}, Z.Phys. {\bf 133}, 101 (1952).\hfill\break H.Araki and
M.M.Yanase, {\it Measurement of Quantum Mechanical Operators},
Phys.Rev. {\bf 120}, 622 (1960).\hfill\break
  G.C.Ghirardi, F.Miglietta, A.Rimini and T.Weber, {\it Limitations
  on Quantum Measurements. I. Determination of the minimal Amount of
  Nonideality and Identification of the Optimal Measuring Apparatuses
   and II. Analysis of a Model Example}, Phys.Rev. {\bf
  D24}, 347 and 353 (1981).\hfill\break
S.Luo, {\it Wigner-Yanase Skew Information and Uncertainty
Relations}, Phys.Rev.Lett. {\bf 91 }, 180403   (2003).

\bibitem{58}D.Marolf, {\it  Mass superselection, Canonical Gauge
Transformations, and Asymptotically Flat Variational Principles},
Clas.Quantum Grav. {\bf 13}, 1871 (1996).\hfill\break J.Hartle,
R.Laflamme and D.Marolf, {\it  Conservation Laws in the Quantum
Mechanics of Closed Systems}, Phys.Rev. {\bf D51}, 7007 (1995).


\bibitem{59}D.Giulini, C.Kiefer and H.D.Zeh, {\it
Symmetries, Superselection Rules and Decoherence}, Phys.Lett. {\bf
A199}, 291 (1995) (gr-qc/9410029).



\bibitem{60}C.Rovelli, {\it Relational Quantum Mechanics},
Int.J.Theor.Phys. {\bf 35}, 1637 (1996(
(quant-ph/9609002).\hfill\break C.Rovelli and M.Smerlak, {\it
Relational EPR} Found.Phys. {\bf 37}, 427 (2007) (quant-ph/0604064).



\bibitem{61}J. Preskill, {\it Quantum Information and Computation}, (1998),
Physics 229 Lecture Notes online at
http://www.theory.caltech.edu/people/preskill/ph229/outline.html




\bibitem{62}O.Pessoa jr., {\it Can the Dechorence Approach Help to Solve the
Measurement Problem?}, Synthese {\bf 113}, 323 (1998).


\bibitem{63} A.Barducci, R.Casalbuoni, D.Dominici and L.Lusanna,
{\it Pseudoclassical Description of Weyl Particles}, Phys.Lett. {\bf
100B}, 126 (1981).\hfill\break A.Barducci and L.Lusanna, {\it The
Photon in Pseudoclassical Mechanics}, Nuovo Cim. {\bf 77A}, 39
(1983); {\it The Massive Photon in Pseudoclassicaal Mechanics},
J.Phys. {\bf A16}, 1993 (1983). \hfill\break J.A.Brooke and
F.E.Schroeck, {\it Localization of the Photon on Phase Space},
J.Math.Phys. {\bf 37}, 5958 (1996).\hfill\break O.Keller, {\it On
the Theory of Spatial Photon Localization}, Phys.Rep. {\bf 411}, 1
(2005).

\bibitem{64} F.Bigazzi and L.Lusanna, {\it Spinning Particles on
Spacelike Hypersurfaces and their Rest Frame Description},
Int.J.Mod.Phys. {\bf A14}, 1429 (1999) (hep-th/9807052).


\bibitem{65}D.Alba, H.W.Crater and L.Lusanna, {\it Massless Particles
plus Matter in the Rest-Frame Instant Form of Dynamics}, in
preparation.




\bibitem{66}A.Peres and D.R.Terno, {\it Relativistic Doppler Effect
in Quantum Communication}, J.Mod.Optics {\bf 50}, 1165
(2003).\hfill\break N.H.Lindner, A.Peres and D.R.Terno, {\it
Wigner's Little Group and Berry's Phase for Massless Particles},
J.Phys. {\bf A36}, L449 (2003) (hep-th/0304017).




\bibitem{67}S.Dimopoulous, P.W.Graham, J.M.Hogan and M.A.Kasevich, {\it General
Relativistic Effects in Atom Interferometry} (arXiv:
0802.4098).\hfill\break S.Dimopoulous, P.W.Graham, J.M.Hogan,
M.A.Kasevich and S.Rajendran, {\it An Atomic Gravitational Wave
Interferometric Sensor (AGIA)} (arXiv: 0806.2125).

\bibitem{68}QUEST - {\it Quantum Entanglement in Space Experiments},
http://www.quantum.at/quest



\bibitem{69}R.Lynch, {\it The Quantum Phase Problem: A Critical
Review}, Phys.Rep. {\bf 256}, 367 (1995).\hfill\break Eds.
W.P.Schleich and S.M.Barnett, \textit{Quantum Phase and Phase
Dependent Measurements}, Physica Scripta vol. \textbf{T48}, (1993).

















\end{thebibliography}
\end{document}